\documentclass[longauth]{aa}

\usepackage{txfonts}
\usepackage{graphicx}	
\usepackage{amsmath}	
\usepackage{amssymb}	
\usepackage[hyperfootnotes=false]{hyperref}
\usepackage{upgreek}    
\usepackage[dvipsnames]{xcolor}
\usepackage[dvipsnames]{xcolor}
\usepackage[para,online,flushleft]{threeparttable}
\usepackage{array,multirow,booktabs}
\usepackage{bigstrut}
\usepackage{float}
\usepackage{longtable}
\usepackage{tablefootnote}
\begin{document}

   \title{The GAPS Programme at TNG XL: \\
   A puffy and warm Neptune-sized planet and an outer Neptune-mass candidate orbiting the solar-type star TOI-1422}
   \titlerunning{The GAPS Programme at TNG XL}
   \authorrunning{L. Naponiello}


   \author{L.~Naponiello\inst{1,2}
          \and L.~Mancini\inst{1,3,4}
          \and M.~Damasso\inst{3}
          \and A.~S.~Bonomo\inst{3}
          \and A.~Sozzetti\inst{3}
          \and D.~Nardiello\inst{6,7}
          \and K.~Biazzo\inst{5}
          \and R.~G.~Stognone\inst{8}
          \and J.~Lillo-Box\inst{9}
          \and A.~F.~Lanza\inst{14}
          \and E.~Poretti\inst{18,19}
          \and J.~J.~Lissauer\inst{35,36}
          \and L.~Zeng\inst{10,11}
          \and A.~Bieryla\inst{32}
          \and G.~Hébrard\inst{21,22}
\and M.~Basilicata\inst{1}
\and S.~Benatti\inst{15}
\and A.~Bignamini\inst{12}
\and F.~Borsa\inst{18}
\and R.~Claudi\inst{7}
\and R.~Cosentino\inst{19}
\and E.~Covino\inst{13}
\and A.~de~Gurtubai\inst{19}
\and X.~Delfosse\inst{24}
\and S.~Desidera\inst{7}
\and D.~Dragomir\inst{33}
\and J.~D.~Eastman\inst{32}
\and Z.~Essack\inst{30,31}
\and A.~F.~M.~Fiorenzano\inst{19}
\and P.~Giacobbe\inst{3}
\and A.~Harutyunyan\inst{19}
\and N.~Heidari\inst{25,26,23}
\and C.~Hellier\inst{39}
\and J.~M.~Jenkins\inst{34} 
\and C.~Knapic\inst{12}
\and P.-C.~König\inst{20,21}
\and D.~W.~Latham\inst{32}
\and A.~Magazzù\inst{37}
\and A.~Maggio\inst{15}
\and J.~Maldonado\inst{15}
\and G.~Micela\inst{15}
\and E.~Molinari\inst{16}
\and M.~Molinaro\inst{12}
\and E.~H.~Morgan\inst{28} 
\and C.~Moutou\inst{27}
\and V.~Nascimbeni\inst{7}
\and E.~Pace\inst{2}
\and I.~Pagano\inst{14}
\and M.~Pedani\inst{19}
\and G.~Piotto\inst{17}
\and M.~Pinamonti\inst{3}
\and E.~V.~Quintana\inst{29}
\and M.~Rainer\inst{18}
\and G.~R.~Ricker\inst{31}
\and S.~Seager\inst{28,30,40} 
\and J.~D.~Twicken\inst{34,35} 
\and R.~Vanderspek\inst{28}
\and J.~N.~Winn\inst{38} 
}

\institute{
Department of Physics, University of Rome ``Tor Vergata'', Via della Ricerca Scientifica 1, 00133, Rome, Italy \\
\email{luca.naponiello@unifi.it}
\and
Department of Physics and Astronomy, University of Florence, Largo Enrico Fermi 5, 50125 Firenze
\and
INAF -- Turin Astrophysical Observatory, via Osservatorio 20, 10025, Pino Torinese, Italy
\and
Max Planck Institute for Astronomy, K\"{o}nigstuhl 17, 69117 -- Heidelberg, Germany
\and
INAF -- Osservatorio  Astronomico  di  Roma,  via  Frascati  33,  00040, Monte Porzio Catone (RM), Italy
\and
Aix-Marseille Université, CNRS, CNES, LAM, Marseille, France
\and
INAF -- Osservatorio  Astronomico  di  Padova,  Vicolo  dell’Osservatorio 5, 35122, Padova, Italy 
\and
Dipartimento di Fisica, Universit\'a degli Studi di Torino, via Pietro Giuria 1, 10125, Torino, Italy
\and
Centro de Astrobiolog\'{i}a, Depto. de Astrof\'{i}sica,
ESAC campus, 28692, Villanueva de la Ca\~{n}ada (Madrid), Spain
\and
Center for Astrophysics \textbar \ Harvard \& Smithsonian, 60 Garden Street, Cambridge, MA 02138, USA
\and
Department of Earth and Planetary Sciences, Harvard University, 20 Oxford Street, Cambridge, MA 02138, USA
\and
INAF -- Osservatorio Astronomico di Trieste, via Tiepolo 11, 34143, Trieste, Italy
\and
INAF -- Osservatorio Astronomico di Capodimonte, Salita Moiariello 16, 80131, Naples, Italy
\and
INAF -- Osservatorio Astrofisico di Catania, via S. Sofia 78, 95123, Catania, Italy
\and
INAF -- Osservatorio Astronomico di Palermo, Piazza del Parlamento, 1, 90134, Palermo, Italy
\and
INAF -- Osservatorio Astronomico di Cagliari, via della Scienza 5, 09047, Selargius (CA), Italy
\and
Dipartimento di Fisica e Astronomia ``Galileo Galilei'' – Universit\`{a} di Padova, Vicolo dell’Osservatorio 2, 35122, Padova, Italy
\and
INAF -- Osservatorio Astronomico di Brera, Via E. Bianchi 46, 23807, Merate (LC), Italy
\and
Fundaci\'{o}n Galileo Galilei - INAF, Rambla Jos\'{e} Ana Fernandez
P\'{e}rez 7, 38712 Bre\~{n}a Baja, TF, Spain
\and
European Southern Observatory, Karl-Schwarzschild-Straße 2, D-85748 Garching, Germany
\and
Institut d'Astrophysique de Paris, CNRS, UMR 7095 \& Sorbonne Université, UPMC Paris 6, 98bis Bd Arago, 75014 Paris, France
\and
Observatoire de Haute-Provence, CNRS, Universit\'{e} d’Aix-Marseille, 04870 Saint-Michel-l'Observatoire, France
\and
Aix-Marseille Université, CNRS, CNES, LAM, Marseille, France
\and
Université Grenoble Alpes, CNRS, IPAG, 38000 Grenoble, France
\and
Department of Physics, Shahid Beheshti University, Tehran, Iran
\and
Laboratoire J.-L. Lagrange, OCA, Université de Nice-Sophia Antipolis, CNRS, Campus Valrose, 06108 Nice Cedex 2, France
\and
IRAP, Université de Toulouse, CNRS, UPS, CNES, F-31400 Toulouse, France
\and
Department of Physics and Kavli Institute for Astrophysics and Space Research, Massachusetts Institute of Technology, Cambridge, MA 02139, USA
\and
NASA Goddard Space Flight Center, 8800 Greenbelt Road, Greenbelt, MD 20771, USA
\and
Department of Earth, Atmospheric and Planetary Sciences, Massachusetts Institute of Technology, Cambridge, MA 02139, USA
\and
Kavli Institute for Astrophysics and Space Research, Massachusetts Institute of Technology, Cambridge, MA 02139, USA
\and
Harvard-Smithsonian Center for Astrophysics, 60 Garden Street, Cambridge, MA 02138, USA
\and
Department of Physics and Astronomy, University of New Mexico, 210 Yale Blvd NE, Albuquerque, NM 87106, USA
\and
NASA Ames Research Center, Moffett Field, CA 94035, USA
\and
SETI Institute, Mountain View, CA 94043, USA
\and
Space Science \& Astrobiology Division MS 245-3
\and
INAF-Telescopio Nazionale Galileo, Apartado 565, E-38700, Santa Cruz de La Palma, Spain
\and
Department of Astrophysical Sciences, Princeton University, Princeton, NJ 08544, USA
\and
Astrophysics Group, Keele University, Staffordshire, ST5 5BG, UK
\and
Department of Aeronautics and Astronautics, MIT, 77 Massachusetts Avenue, Cambridge, MA 02139, USA
\and
INAF – Osservatorio Astrofisico di Arcetri, Largo E. Fermi 5, 50125, Florence, Italy
}


 
  \abstract
{Neptunes represent one of the main types of exoplanets and have chemical-physical characteristics halfway between rocky and gas giant planets.
Therefore, their characterization is important for understanding and constraining both the formation mechanisms and the evolution patterns of planets.
}
{We investigate the exoplanet candidate TOI-1422\,b, which was discovered by the \emph{TESS} space telescope around the high proper-motion G2\,V star TOI-1422 ($V=10.6$\,mag), 155\,pc away, with the primary goal of confirming its planetary nature and characterising its properties.}
{We monitored TOI-1422 with the HARPS-N spectrograph for 1.5 years to precisely quantify its radial velocity variation. We analyze these radial velocity measurements jointly with TESS photometry and check for blended companions through high-spatial resolution images using the AstraLux instrument.}
{We estimate that the parent star has a radius and a mass of $R_{\star}=1.019_{-0.013}^{+0.014}\,R_{\sun}$, $M_{\star}=0.981_{-0.065}^{+0.062}\,M_{\sun}$, respectively. Our analysis confirms the planetary nature of TOI-1422\,b and also suggests the presence of a Neptune-mass planet on a more distant orbit, the candidate TOI-1422\,c, which is not detected in \emph{TESS} light curves. The inner planet, TOI-1422\,b, orbits on a period $P_{\rm b}=12.9972\pm0.0006$\,days and has an equilibrium temperature $T_{\rm eq, b}=867\pm17$\,K. With a radius of $R_{\rm b}=3.96^{+0.13}_{-0.11}\,R_{\oplus}$, a mass of $M_{\rm b}=9.0^{+2.3}_{-2.0}\,M_{\oplus}$ and, consequently, a density of $\rho_{\rm b}=0.795^{+0.290}_{-0.235}$\,g\,cm$^{-3}$, it can be considered a warm Neptune-size planet. Compared to other exoplanets of similar mass range, TOI-1422\,b is among the most inflated ones and we expect this planet to have an extensive gaseous envelope that surrounds a core with a mass fraction around $10\%-25\%$ of the total mass of the planet. The outer non-transiting planet candidate, TOI-1422\,c, has an orbital period of $P_{\rm c}=29.29^{+0.21}_{-0.20}$\,days, a minimum mass, $M_{\rm c}\sin{i}$, of $11.1^{+2.6}_{-2.3}\,M_{\oplus}$, an equilibrium temperature of $T_{\rm eq, c}=661\pm13$\,K and, therefore, if confirmed, it could be considered as another warm Neptune.} 
   {}

   \keywords{planetary systems -- 
             techniques: photometric -- 
             techniques: spectroscopic --
             techniques: radial velocities
             stars: individual: TOI-1422 --
             method: data analysis
            }

   \maketitle
%

\section{Introduction}

Exoplanetary science has expanded quickly from the simple detection of new worlds to their in-depth characterization. The latter is especially feasible for planets orbiting bright stars on a plane almost aligned to our line of sight, so that their radius and mass can both be derived by transit photometry and radial velocity measurements, respectively. 
The population of known transiting planets has increased significantly in the last two decades mainly thanks to dedicated ground-based surveys, which were then followed by surveys from space that resulted to be much more efficient considering the total number of discoveries.

So far, the \emph{Kepler} and the K2 space missions \citep{Borucki2010,Howell2014} have had a very important impact on the exoplanet field by discovering thousands of confirmed and candidate planets, many of which are not amenable to radial velocity (RV) follow up due to the faintness of their host stars. The Transiting Exoplanet Survey Satellite (\emph{TESS}) \citep{Ricker2014}, currently at the end of its first extended mission with a second one already proposed, was designed to target nearby and bright stars on a great portion of the sky (around 85\% sky coverage during the primary mission alone) because such stars are easier to follow-up by means of RV and result in refined measurements of their own exoplanet masses, atmospheres, sizes, and therefore densities. 

The possibility to use a large exoplanet sample, like that of \emph{Kepler}, which is based on homogeneous data and has a minimal pollution from false positives ($<10\%$, \citealt{2013ApJ...766...81F}), has allowed us to distinguish between several distinct exoplanet regimes \citep{2014ApJ...783L...6W,2014Natur.509..593B,2019PNAS..116.9723Z}: the terrestrial-like planets ($R_{\rm p}<1.7 \, R_{\oplus}$), the gas dwarf planets with rocky cores and hydrogen–helium envelopes, the H$_2$O-dominated ices/fluids water worlds (both of the latter two classes have $1.7 \, R_{\oplus} <R_{\rm p}<3.9 \, R_{\oplus}$) and the ice or gas giant planets ($R_{\rm p}>3.9 \, R_{\oplus}$).

Planet occurrence around main-sequence stars has been early investigated thanks to Doppler surveys (e.g., \citealp{2008PASP..120..531C,2012ApJ...753..160W}). In particular, the Keck 
Eta-Earth survey \citep{2010Sci...330..653H} and the CORALIE+HARPS survey \citep{2011arXiv1109.2497M} first explored the domain of low-mass ($3-30\,M_{\oplus}$) close-in ($P_{\rm orb}\sim 50$\,days) planets. These planets resulted an order of magnitude more common than giant planets.

Other studies for determining the occurrence rates of planets, based on the \emph{Kepler} sample, agree in determining that for planets with less than 1-year orbital period, their mean number per star is higher within the radius range $1 \, R_{\oplus} <R_{\rm p}<4 \, R_{\oplus}$ compared to the range $4 \, R_{\oplus} <R_{\rm p}<16 \, R_{\oplus}$ or similar \citep{2012ApJS..201...15H,2013ApJ...766...81F,2013ApJ...770...69P}. The subsequently and gradually refinement of parent-star properties (especially thanks to high-resolution stellar spectra) revealed a clear bimodality of the radius distribution of close-in ($P<100$\,days), small-size ($R_{\rm p}<4.0 \, R_{\oplus}$) planets orbiting bright, main-sequence solar-type stars \citep{2017AJ....154..107P,2018AJ....156..264F,Vaneylen2018}\footnote{For a possible explanation of Fulton's gap see \citealt{Modirrousta2020}}. These two quite distinct populations were identified as ``super-Earths'' ($R_{\rm p} < 1.5 \, R_{\oplus}$) and ``sub-Neptunes'' ($R_{\rm p} = 2-3 \, R_{\oplus}$), which are also represented in the intermediate region ($R_{\rm p} = 1.5-2 \, R_{\oplus}$) with fewer planets. However, it is better to stress that, since we do not know for sure what they are made of, the space of physical parameters ($R_{\rm p}$, $M_{\rm p}$), for which the previous terms apply, are not strictly defined. 

The advantage in studying transiting planets is the possibility, in many cases, to measure both the planetary radius and mass and, therefore, determine their density and bulk composition.
Knowing the structural properties, one should be able to distinguish
among the various scenarios of exoplanet formation and evolution. Unfortunately, theoretical models (e.g., \citealp{2019A&A...624A.109B,2020A&A...638A..41T}) tell us that the mass-radius relationships for small planets present degeneracy due to the vastness of possible different compositions and amounts of rock, ice and gas, especially in the transition between rocky super-Earths and Neptune-like planets (e.g., \citealp{2009ApJ...690.1056M,2018ApJ...866...49L}).
A detailed investigation of the mass-radius relation for small planets can be useful for throwing light on several open questions, such as the diversity of planet core masses and compositions, or those regarding the place where they form (in situ or beyond the snowline) and the existence of the radius gap. 
We refer the reader to the recent review by \citealp{2022ASSL..466..143B} for an exhaustive discussion on this topic.

It is, therefore, clear how RV follow-up observations and in general planetary mass measurements play an important role in this understanding process and why today there is a tremendous effort in this field by many teams (e.g., KESPRINT: \citealp{2018A&A...619L..10G}; HARPS-N consortium: \citealp{2020AJ....160....3C}; NCORES: \citealp{2020Natur.583...39A}; TESS-Keck Survey: \citealp{2022AJ....163..297C}; GAPS: \citealp{2021A&A...645A..71C}) in order to confirm TESS small-planet candidates.

Probing the chemical composition of the atmosphere of a large number of sub-Neptune planets would also be helpful to unravel the skein. Various techniques (like high-resolution spectroscopy, transmission and emission spectroscopy) have been implemented and applied successfully using the HST (Hubble Space Telescope) instruments or the high-resolution spectrographs mounted on large-class ground-based telescopes (e.g., CRIRES: \citealp{2010Natur.465.1049S}; HARPS: \citealp{2015A&A...577A..62W}; LDSS3C: \citealp{2018AJ....156...42D}; GIANO: \citealp{2018A&A...615A..16B}; CARMENES: \citealp{2019A&A...628A...9C}; HARPS-N: \citealp{2020ApJ...894L..27P}; ESPRESSO: \citealp{2021A&A...645A..24B}). Unfortunately, these techniques for probing the planetary atmospheres are currently effectively applicable only to giant planets, as we know few sub-Neptune planets for which the transmission-spectrum signal can be detected with a sufficient signal-to-noise ratio for allowing to discriminate among different atmospheric models. The featureless transmission spectra of GJ\,436\,b \citep{2014Natur.505...66K} and GJ\,1214\,b \citep{2014Natur.505...69K} are emblematic.

The situation should improve soon thanks to the JWST (James Webb Space Telescope) \citep{Barstow2015}, which is about to go into operation, and with the next generation of space-based and large ground-based telescopes (Ariel: \citealp{Tinetti2021};  ELT: \citealp{Ramsay2021}; TMT: \citealt{Skidmore2015}). In the meantime, it is important that we continue to work for uncovering new exoplanets, especially those of small size ($R_{\rm p}<5 \,R_{\oplus}$) that orbit bright ($V<11$\,mag) main-sequence dwarf stars. This is currently possible thanks to the great number of planet candidates (more than 5000) that \emph{TESS} is discovering at the present time. The recent detection of water vapour in the atmosphere of the super-Neptune TOI-674\,b with the HST \citep{2022arXiv220104197B} is a successful example of this effort.

On the 6$^{\rm th}$ of November 2019, the \emph{TESS} target star TIC\,333473672 was officially named TOI-1422 (\emph{TESS} Object of Interest; \citealt{Guerrero2021}), following the Data Validation Report Summary (DRS) produced by the \emph{TESS} Science Processing Operations Center (SPOC) \citep{Jenkins2016} pipeline at NASA Ames Research Center through the Transiting Planet Search (TPS; \citealt{jenkins2002,jenkins2010,Jenkins2020b}) and Data Validation (DV; \citealt{Twicken2018}, \citealt{Li2019}) modules. In particular, TOI-1422\,01 was flagged as a potential planet with an orbital period of $13.0020\,\pm\,0.0040$~days, a transit depth of $1422\,\pm\,94$~ppm (parts per million) and a corresponding radius of $3.85\,\pm\,0.90\,R_{\oplus}$, which is compatible with Neptune's radius. The candidate passed all SPOC DV diagnostic tests; furthermore, all TIC (version 8) objects other than the target star were excluded as sources of the transit signal through the difference image centroid offsets \citep{Twicken2018}.

The long-term, multi-program Global Architecture of Planetary Systems (GAPS) \citep{Covino2013,Poretti2015} exploits Doppler measurements taken with the HARPS-N (High Accuracy Radial velocity Planet Searcher for the Northern hemisphere) \citep{Cosentino2012} instrument at the Telescopio Nazionale Galileo (TNG) in La Palma (Spain). This high-resolution spectrograph (resolving power $R\approx115\,000$) delivers the highest RV precision ($\sim1$ m s$^{-1}$) currently achievable in the northern hemisphere. One of the aims of the GAPS programme is to confirm and obtain an accurate mass determination of planets having an intermediate-mass between super-Earths and super-Neptunes; for this reason, TOI-1422 was selected for RV follow-up observations, which started in June 2020. 

In the present work, we report the results of our measurements and analysis that allowed us to confirm TOI-1422 as a new planetary system. The paper is organized as follows: Sect.\,\ref{sec:data_reduction} contains the details of the instruments and the photometric and RV measurements. The results of our analyses are presented in Sect.\,\ref{sec:analysis} and discussed in Sect.\,\ref{sec:discussion}, before finally addressing the conclusions in Sect.\,\ref{sec:conclusion}.

\section{Observations and data reduction}
\label{sec:data_reduction}
\subsection{TESS photometry}
\label{sec:photometry}
\begin{figure}
\centering
\includegraphics[width=0.45\textwidth]{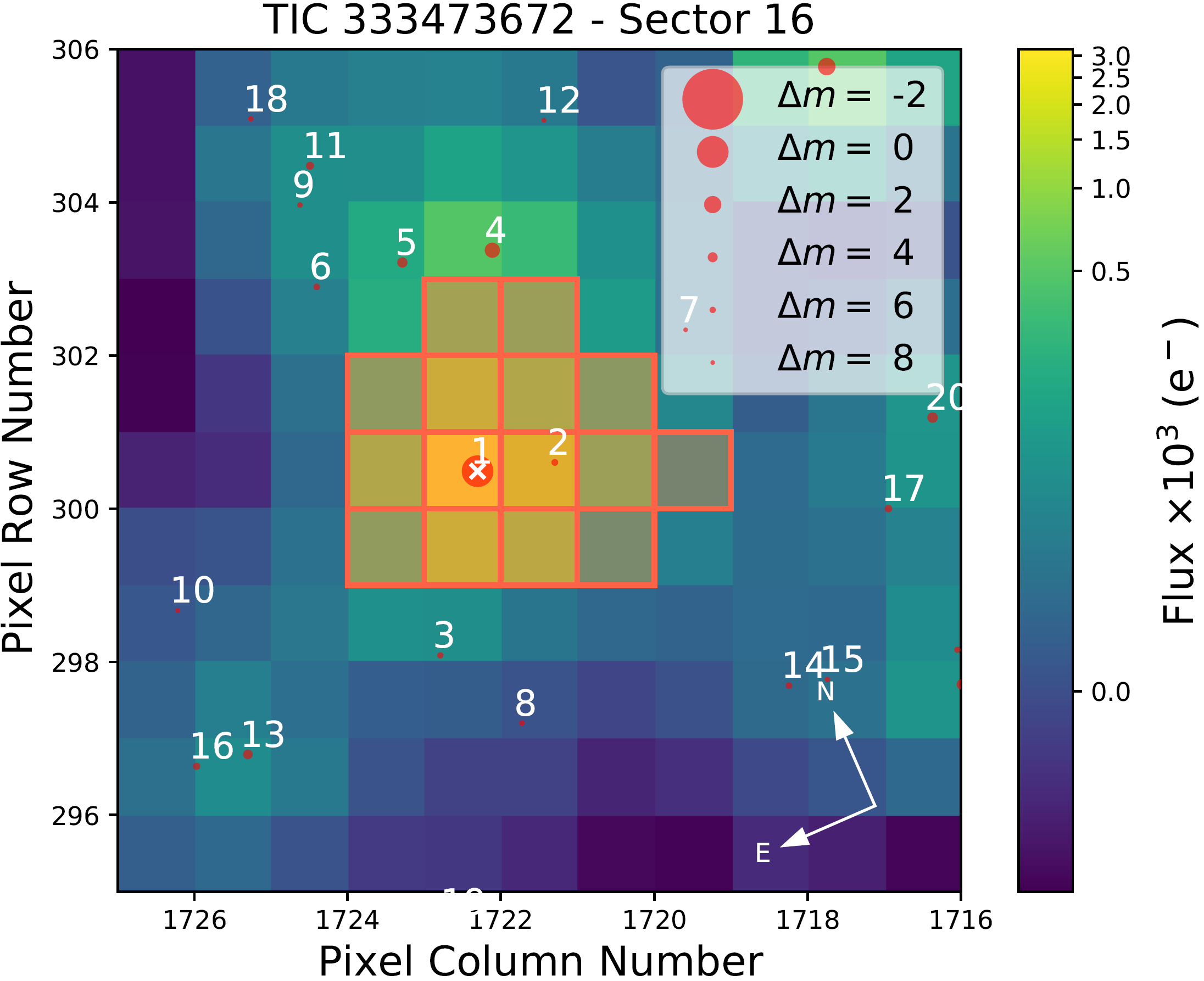}
\caption{The TPF (target pixel file) from the \emph{TESS} observation Sector 16, made with \texttt{tpfplotter} \citep{Aller2020} and centred on TOI-1422, which is marked 
with a white cross. The SPOC pipeline aperture is shown by shaded red squares, and the {\it Gaia} satellite eDR3 catalogue \citep{Brown2018,Prusti2016} is also overlaid with symbol sizes proportional to the magnitude difference with TOI-1422. The difference image centroid locates the source of the transits within $1.89 \pm 5$ arcsec of the target star’s location, as reported by the TicOffset for the multisector DV report for this system.}
\label{fig:star}
\end{figure}

%
Since late July 2018, \emph{TESS} observed more than $200\,000$ stars with its four wide-field optical CCD cameras ($24 \times 96$\,degrees), each having a focal ratio of $f/1.4$ and a broad-band filter range between 600 and 1000\,nm. The pre-selected target TIC\,333473672 was observed in Sectors 16 and 17 between 2019-Sep-11 and 2019-Nov-02, and a first of a total of four transiting events were recorded on 2019-Sep-19. The two-minute cadence photometry of TOI-1422 from \emph{TESS} spans a total of $\thickapprox50$~days and, to analyze it, we used the Presearch Data Conditioning Simple Aperture Photometry (PDC-SAP; \citealt{Stumpe2012,Stumpe2014}, \citealt{Smith2012}) light curve, which is provided by the \emph{TESS} SPOC pipeline and retrieved through the Python package \texttt{lightkurve} \citep{lightkurve} from the Mikulski Archive for Space Telescopes (MAST). We jointly fitted the transit model and a Gaussian Process (GP) using a simple (approximate) Matern kernel, which was implemented in the Python modelling tool \texttt{juliet}\footnote{\texttt{\href{https://juliet.readthedocs.io/en/latest/}{https://juliet.readthedocs.io}}} \citep{Espinoza2019} via \texttt{celerite} \citep{foreman}, of the form:
\begin{equation}
    k(\tau_{i,j})=\sigma_{\rm GP}^2\,M(\tau_{i,j},\,\rho)+(\sigma_i^2+\sigma_w^2)\,\delta_{i,j} \,,
\end{equation}
where $\sigma_i$ is the errorbar of the $i$-th data point, $\sigma_{\rm GP}$ the amplitude of the GP in parts per million (ppm), $\sigma_w$ an added jitter term (in ppm), $\delta_{i,j}$ the Kronecker's delta, $k(\tau_{i,j})$ the element $i$,$j$ of the covariance matrix as a function of $\tau_{i,j}=|t_i-t_j|$, with $t_i$ and $t_j$ being the $i$,$j$ GP regressors (i.e. the observing times), while 
\begin{equation}
    M(\tau_{i,j},\,\rho)=(1+1/\epsilon)\,e^{-[1-\epsilon]\sqrt{3}\tau_{i,j}/\rho}+(1-1/\epsilon)\,e^{-[1+\epsilon]\sqrt{3}\tau_{i,j}/\rho}
\label{eq:matern}\end{equation}
is the kernel with its characteristic time-scale $\rho$. The parameter $\epsilon$ controls the quality of the approximation since, in the limit $\epsilon\to0$, the Eq.~(\ref{eq:matern}) becomes the Matern-3/2 function.
In \texttt{juliet}, the possible polluting sources inside \emph{TESS} aperture\footnote{\texttt{tpfplotter} is a python package developed by J. Lillo-Box and publicly available on www.github.com/jlillo/tpfplotter.} (Fig.~\ref{fig:star}), which might result in a smaller transit depth compared to the real one, are taken into account with a dilution factor ($D$) that, in this case, has been neglected because the PDC-SAP photometry is already corrected for dilution from other objects contained within the aperture using the Create Optimal Apertures (COA) module \citep{Bryson2010,Bryson2020}\footnote{Since the release of the light curve products from Year 2, the SPOC background estimation algorithm has been updated due to an over-correction bias which was significant for dim and/or crowded targets. For this particular TOI, we estimated this over-correction to be negligible for the planetary radius estimation as it is significantly smaller than the transit depth uncertainty.}. For the purpose of efficiently sample the whole plausible zone in the $(b,k)$ plane, where $b$ is the impact parameter and $k$ is the planet-to-star radius ratio, we used the $(r_1,r_2)$ parametrization described in \cite{Espinoza2018}. This is the same approach that we adopted for the modelling of the transits in the joint analysis with the radial velocities (see Sect.~\ref{sec:joint_analysis}). Moreover, here we make use of the limb-darkening parametrizations of \citealt{Kipping2013} for two-parameter limb-darkening laws ($q_1,q_2 \to u_1,u_2$). 

\begin{figure*}
\centering
\includegraphics[width=0.8\textwidth]{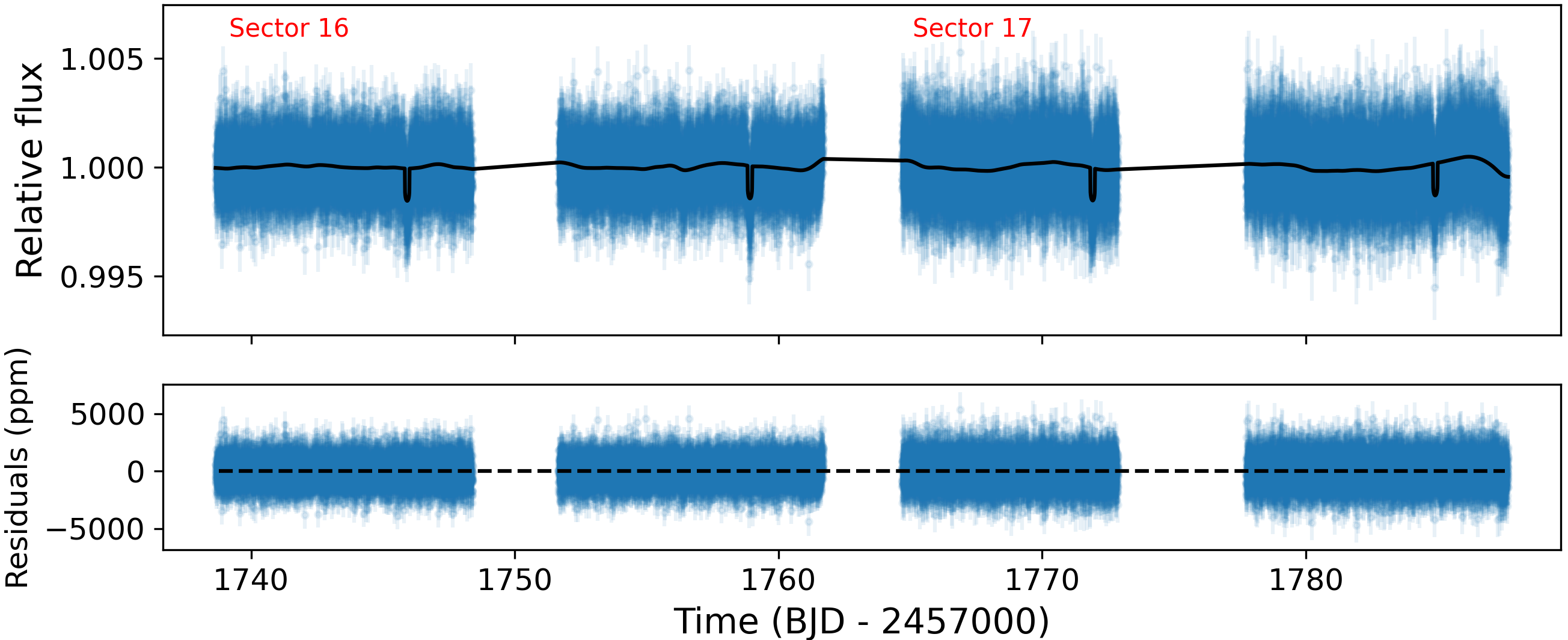}
\caption{{\it Top panel:} PDC-SAP light curve of TOI-1422  (blue points with error bars) as collected by \emph{TESS} in Sectors~16 and 17 with a 2-minute cadence. The black line represents the best-fit model obtained through GP detrending, as detailed in Sect. \ref{sec:photometry}. {\it Bottom panel:} residuals of the best-fitting model in parts per million.}
\label{fig:lc_trend}
\end{figure*}

The PDC-SAP light curve of TOI-1422 and its detrending are plotted in Fig.~\ref{fig:lc_trend}. We also analyzed the SPOC SAP photometry \citep{Twicken2010,Morris2020}, which presents a small long-term variability that might be due to systematics, but no other feature or modulation can be discerned within the experimental uncertainties, aside from a possible single extra transit event which is discussed at the end of Sect.~\ref{sec:results}, and a steep flux drop at the end of both SAP and PDC-SAP light curves which is probably due to high levels of background noise.

\subsection{High-spatial resolution imaging - AstraLux}\label{sec:astralux} 

\begin{figure}
\centering
\includegraphics[width=0.5\textwidth]{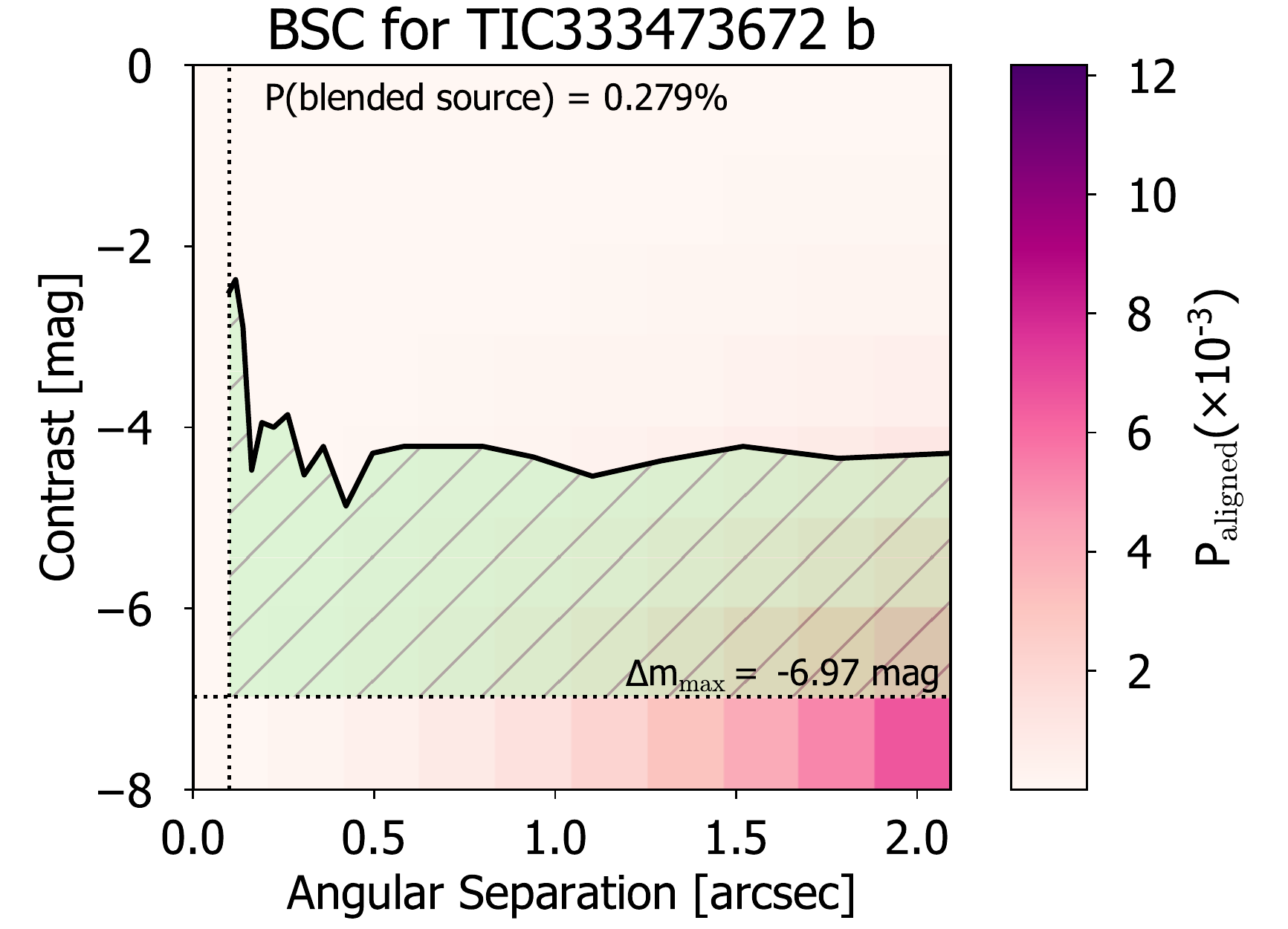}
\caption{Blended Source Confidence (BSC) curve from the AstraLux SDSSz image (solid black line). The colour on each angular separation and contrast bin represent the probability of a source aligned at the location of the target, based on TRILEGAL model. The dotted horizontal line shows the maximum contrast of a blended binary that is capable of imitating the planet's transit depth. The green region represents the regime that is not explored by the high-spatial resolution image. The BSC curve corresponds to the integration of $P_{\rm aligned}$ over this region.}
\label{fig:astralux}
\end{figure}

We observed TOI-1422 with the AstraLux high-spatial resolution camera \citep{hormuth08}, located at the 2.2\,m telescope of the Calar Alto Observatory (CAHA, Almer\'ia, Spain) using the lucky-imaging
technique. This technique obtains diffraction-limited images by acquiring thousands of short-exposure frames and selecting the ones with the highest Strehl ratio \citep{strehl1902} to finally combine
them into a co-added high-spatial resolution image. We observed this target on the night of 29$^{\rm th}$ of September 2021 under good weather conditions with a mean seeing of 1\,arcsec and obtained 50\,000 frames
with 20\,ms exposure time in the Sloan Digital Sky Survey $z$ filter (SDSSz), with a field-of-view windowed to $6\times6$ arcsec. The datacube was reduced by the instrument pipeline \citep{hormuth08} and
we selected the best-quality $10\%$ frames to produce the final high-resolution image. We obtained the sensitivity limits of the co-added image by using our own developed {\sc astrasens} package\footnote{\url{https://github.com/jlillo/astrasens}} with the procedure described in \cite{lillo-box12,lillo-box14b}. The 5$\sigma$ sensitivity curve is shown in Fig.~\ref{fig:astralux}. We can discard sources down to 0.2\,arcsec with a magnitude contrast of $\Delta Z < 4$~mag, corresponding to a maximum contamination level of 2.5\%. By using this high-spatial resolution image, we also estimate the
probability of an undetected blended source. This probability (fully described in \citealt{lillo-box14b}) is called the blended source confidence (BSC). We use a python implementation of this approach
(\texttt{bsc}, by J. Lillo-Box) which uses the TRILEGAL\footnote{\url{http://stev.oapd.inaf.it/cgi-bin/trilegal}} galactic model \citep[v1.6;][]{girardi12} to retrieve a simulated source population of the region around the corresponding
target\footnote{This is done in python by using the astrobase implementation by \citet{astrobase}.}. This simulated population is used to compute the density of stars around the target position (radius $r=1^{\circ}$) and derive the probability of chance-alignment at a given contrast magnitude and separation. When applied to the TOI-1422 location, we use a maximum contrast magnitude
of $\Delta m_{\rm b,max} = 6.97$~mag in the SDSSz passband, corresponding to the maximum contrast of a blended eclipsing binary that could mimic the observed transit depth of planet b ($\sim$ 1000
ppm). Thanks to our high-resolution image, we estimate the probability of an undetected blended source to be 0.28\%. The probability of such an undetected source to be an appropriate eclipsing binary is thus even
lower and consequently, we can assume that the transit signal is not due to a blended eclipsing binary.

\subsection{HARPS-N radial velocities}
\label{sec:rv}
Between June 2020 and January 2022, we collected a total of 112 RV measurements of TOI-1422 with HARPS-N (Table~\ref{tab:TERRA}). The RVs were calculated using the \textsc{TERRA} pipeline \citep{AngladaEscud2012}, version 1.8, through the YABI workflow interface \citep{YABI}, which is maintained by the Italian center for Astronomical Archive (IA2). \textsc{TERRA} is an algorithm based on the template matching technique, and for the RVs retrieval is preferred, in this paper, over the standard Data Reduction Software (DRS) pipeline, which returned a slightly lower overall RV precision\footnote{For a comparison of the performances of TERRA vs DRS see \citealt{perger2017}} on this target. We used the RVs calculated using all the spectral orders and from the full sample of RVs we removed four points following Chauvenet's criterion. TERRA RVs have an average measurement error of $2.6$\,m\,s$^{-1}$, a root mean square error of $4.5$\,m\,s$^{-1}$ and a signal-to-noise ratio S/N\,$\approx$\,35 measured at a reference wavelength of 5500 $\AA$. A long linear trend is evident in HARPS-N RV data, as we discuss in Sect.~\ref{sec:rv_trend}.\newline

TOI-1422 was also observed with the SOPHIE instrument, a stabilized \'echelle spectrograph mounted at the 193-cm Telescope of Observatoire de Haute-Provence in France (\citealt{Perruchot2008}, \citealt{Bouchy2013}), however, for signals of low semi-amplitudes such as those we discuss in this work, the RV measurements of SOPHIE, due to higher uncertainties compared to HARPS-N, do not increase the significance of the results presented in Sect.~\ref{sec:conclusion} and therefore have not been utilized.



\section{System characterization}
\label{sec:analysis}
\subsection{Parent star}
\label{sec:star}
From the co-added spectrum built out of individual HARPS-N spectra extracted with the standard DRS pipeline, we derived the following atmospheric parameters of the planet-host star TOI-1422: effective temperature $T_{\rm eff}$, surface gravity $\log g$, microturbulence velocity $\xi$, iron abundance [Fe/H], and rotational velocity $\upsilon\sin{i_{\star}}$. For $T_{\rm eff}$, $\log g$, $\xi$, and [Fe/H] we applied a method based on equivalent widths of iron lines taken from \cite{Biazzoetal2015} and the spectral analysis package MOOG (\citealt{Sneden1973}; version 2017). The \cite{CastelliKurucz2003} grid of model atmospheres was adopted. $T_{\rm eff}$ and $\xi$ were derived by imposing that the abundance of \ion{Fe}{i} is not dependent on the line excitation potentials and the reduced equivalent widths (i.e. $EW/\lambda$), respectively, while $\log g$ was obtained by imposing the \ion{Fe}{i}/\ion{Fe}{ii} ionization equilibrium condition. The $\upsilon\sin{i_{\star}}$ was measured with the same MOOG code, by applying the spectral synthesis of three regions around 5400, 6200, and 6700\,\AA, and adopting the same grid of model atmosphere after fixing the macroturbulence velocity to the value of $3.4$\,km\,s$^{-1}$ from the relationship by \cite{Doyleetal2014}. From these results, the star can be classified as a G2\,V dwarf with a low projected rotation velocity $\upsilon\sin{i_{\star}}$ of $1.9\pm0.8$\,km\,s$^{-1}$, implying a maximum rotation period of $27^{+19}_{-8}$~d at $1\sigma$. In analogy, using an empirical relation based on the FWHM\footnote{This relation was calibrated using a set of well-aligned transiting exoplanet systems, for which we could infer $\upsilon\sin{i_{\star}}$ as equal to their equatorial velocities. We estimate the equatorial velocities from the stellar radii and rotational period and correlate these values directly to the FWHMs.} derived by the HARPS-N DRS, we find $\upsilon\sin{i_{\star}}\sim2.2$\,km\,s$^{-1}$.

The field of TOI-1422 was also observed in 2004, 2006 and 2007 during the WASP transit-search survey \citep{2006PASP..118.1407P}. A total of 20\,000 photometric data points were obtained by observing the field every $\sim$\,15 min on clear nights, over spans of $\sim$\,120 days in each year. We searched the data for any rotational modulation using the methods from \citet{2011PASP..123..547M} and found no significant periodicity between 1 and 100 days, with a 95\%-confidence upper limit on the amplitude of 2 mmag. The TESS light curve shows no modulation as well (Sect.~\ref{sec:photometry}), confirming that the star is rather magnetically quiet over a period of $\sim$\,100 days.

\begin{figure}
   \centering
   \includegraphics[width=0.49\textwidth]{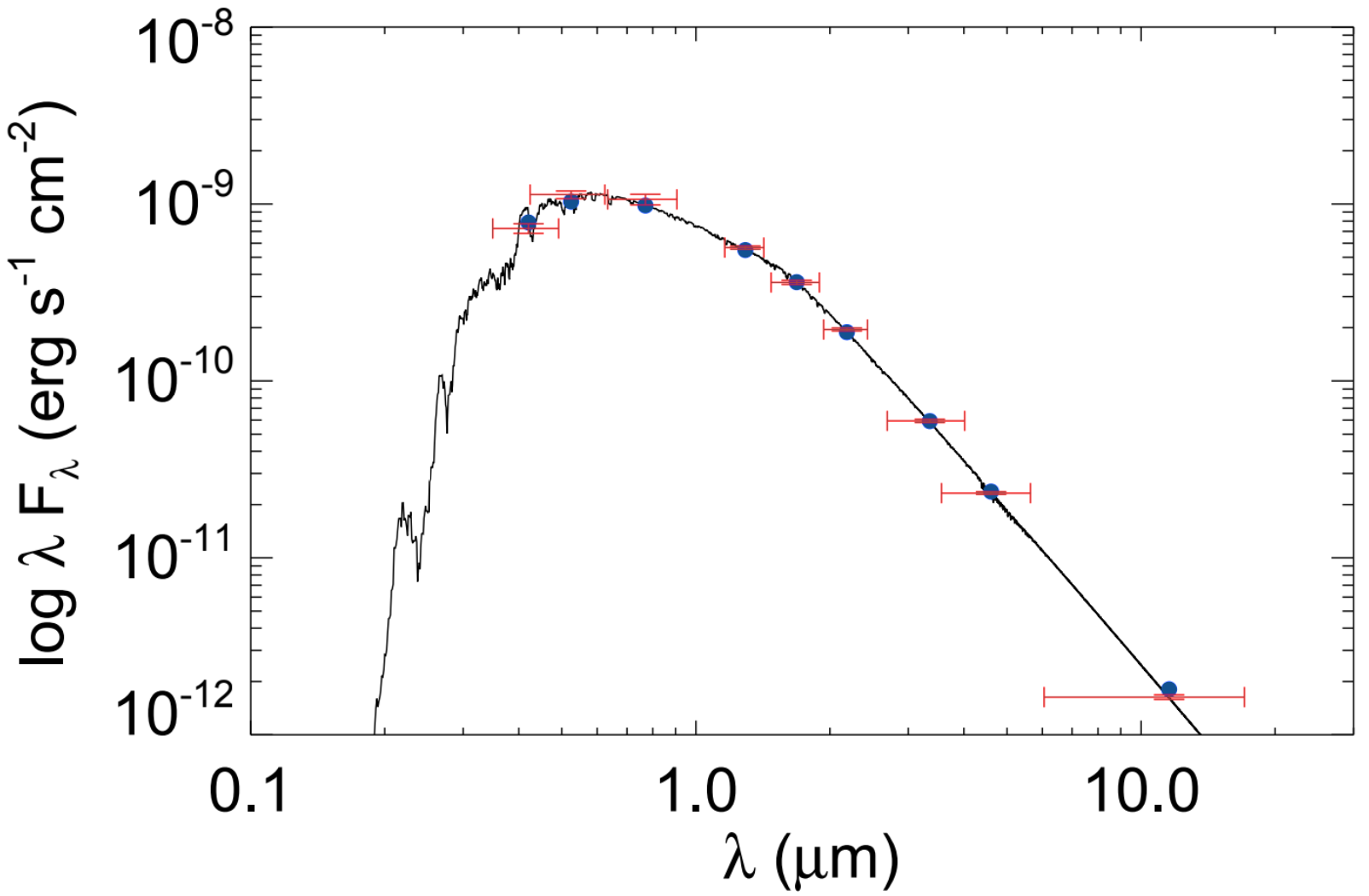}
      \caption{Spectral energy distribution computed for TOI-1422, where the black curve is the most likely atmospheric stellar model and the blue dots correspond to the model fluxes over each passband. The horizontal and vertical red error bars represent, respectively, the effective width of the passbands and the reported photometric measurement uncertainties (refer to the magnitudes in Table \ref{tab:star}).}
         \label{fig:sed}
   \end{figure}

Moreover, the spectrum of TOI-1422 clearly shows a lithium line at $\lambda=6707.8$\,\AA. We, therefore, estimated the lithium abundance $\log A{\rm (Li)}_{\rm NLTE}$ by measuring the lithium $EW$ and considering our $T_{\rm eff}$, $\log g$, $\xi$, and [Fe/H] previously derived together with the NLTE corrections by \citealt{lindetal2009}. The value of the lithium abundance is listed in Table\,\ref{tab:star} and its position in a $\log A{\rm (Li)}$-$T_{\rm eff}$ diagram is compatible with the M67 open cluster \textit{advanced age }($\sim 4.5$\,Gyr; see \citealt{Pasquinietal2008}) in agreement with the star's low-activity level. The physical parameters of TOI-1422 are also displayed in Table \ref{tab:star} and were determined with the EXOFASTv2 Bayesian code \citep{2017ascl.soft10003E, 2019arXiv190709480E}, by fitting the stellar Spectral Energy Distribution (SED) and by employing the MESA Isochrones and Stellar Tracks \citep{2016ApJS..222....8D} to constrain more precisely the stellar mass. In addition, in the table we report the stellar magnitudes used for the SED modelling, while the SED best fit is shown in Fig.~\ref{fig:sed}. Gaussian priors were imposed on the {\it Gaia} eDR3 parallax \citep{2021A&A...649A...1G} as well as on the $T_{\rm eff}$ and $\rm [Fe/H]$ as derived above from the analysis of the HARPS-N spectra. An upper limit was set on the $V$-band extinction, $A_{\rm V}$, from reddening maps \citep{1998ApJ...500..525S, 2011ApJ...737..103S}.

%
%

\begin{table}
\centering %
\caption{TOI-1422 parameters.} %
\label{tab:star} %
\resizebox{\hsize}{!}{
\begin{tabular}{l c c  c}
\hline %
\hline  \\[-8pt]
Parameter & Unit & Value & Source \\
\hline  \\[-6pt]
\multicolumn{1}{l}{\textbf{Cross-identifications}} \\ [2pt] %
TOI \dotfill     & \dotfill& TOI-1422      & TOI catalogue \\
TIC\,ID \dotfill & \dotfill&  333473672    & TIC \\
Tycho \dotfill   & \dotfill&  3235-00524-1 & Tycho \\
2MASS\,ID \dotfill & \dotfill& J23365789+3938218 & 2MASS \\
Gaia\,ID  \dotfill & \dotfill& 1920333449169516288 & Gaia~eDR3 \\ [6pt] %
\multicolumn{1}{l}{\textbf{Astrometric properties}} \\ [2pt] %
R.A. \dotfill & J2016 & 354.240817  & Gaia~eDR3 \\
Dec  \dotfill & J2016 & +39.639275  & Gaia~eDR3 \\
Parallax \dotfill & mas & $6.4418 \pm 0.0138$ & Gaia~eDR3 \\
$\mu_{\alpha}$ \dotfill & mas\,yr$^{-1}$  & $-67.564\pm0.015$ & Gaia~eDR3 \\
$\mu_{\delta}$ \dotfill & mas\,yr$^{-1}$  & $-31.180\pm0.011$ & Gaia~eDR3 \\
Distance \dotfill & pc  & $154.56^{+0.037}_{-0.027}$ & VizieR \\ [6pt] %
\multicolumn{1}{l}{\textbf{Photometric properties}} \\ [2pt] %
$B_{\rm T}$ \dotfill & mag & $11.31\pm0.07$ & Tycho \\ 
$V_{\rm T}$ \dotfill & mag & $10.62\pm 0.05$ & Tycho \\ 
$J$ \dotfill & mag   & $9.585\pm0.022$ & 2MASS \\
$H$ \dotfill & mag   & $9.275\pm0.030$ & 2MASS \\
$K_{\rm S}$ \dotfill  & mag & $9.190\pm0.022$ & 2MASS \\ 
$i'$ \dotfill & mag & $10.311\pm0.075$ & APASS \\
$W1$ \dotfill & mag & $9.161\pm0.023$ & AllWISE \\
$W2$ \dotfill & mag & $9.201\pm0.020$ & AllWISE \\     
$W3$ \dotfill & mag & $9.161\pm0.033$ & AllWISE \\ 
$A_V$ \dotfill & mag & $<0.077$ & This work \\ [6pt] %
\multicolumn{1}{l}{\textbf{Stellar parameters}} \\ [2pt] %
$L_{\star}$ \dotfill & $L_{\sun}$ & $1.116\pm0.037$ & This work \\
$M_{\star}$ \dotfill & $M_{\sun}$ & $0.981^{+0.062}_{-0.065}$ & This work \\
$R_{\star}$ \dotfill & $R_{\sun}$ & $1.019^{+0.014}_{-0.013}$ & This work \\
$T_{\rm eff}$ \dotfill & K & $5840\pm62$ & This work \\
$\log g_{\star}$ \dotfill & cgs & $4.41\pm0.11$ & This work \\
$\xi$ \dotfill & km\,s$^{-1}$ & $0.89\pm0.07$ & This work \\
$\rm{[Fe/H]}$ \dotfill & dex & $-0.09\pm0.07$ & This work \\
Spectral type$^{(a)}$ \dotfill & & G2\,V & This work \\
$\rho_{\star}$\dotfill & g\,cm$^{-3}$ & $1.3\pm0.1$ & This work \\
$\upsilon \sin{i_{\star}}$\dotfill & km\,s$^{-1}$ & $1.7\pm0.4$ & This work \\
$\log A{\rm (Li)_{\rm NLTE}}$\dotfill &  & $1.97\pm0.05$ & This work \\
$\log R^{\prime}_{\rm HK}$\dotfill & dex &  $-4.95\pm0.03$ & This work \\
Age\dotfill & Gyr & $5.1^{+3.9}_{-3.1}$ & This work \\
\hline %
\end{tabular}
}
\tablebib{TESS Primary Mission TOI catalogue \citep{Guerrero2021}; TIC \citep{Stassun2018,Stassun2019}; Tycho \citep{hog}; 2MASS \citep{Skrutskie2006}; Gaia eDR3 \citep{Brown2021}; AllWISE \citep{allwise2013}; APASS \citep{henden2015}; VizieR Online Data catalogue \citep{Bailer2021}.}
\tablefoot{
\tablefoottext{a}{Spectral type defined according to the stellar spectral classification of \citealt{Gray2009}}.}
\end{table}

  \begin{figure*}[t]
    \centering
    \includegraphics[width=0.8\textwidth]{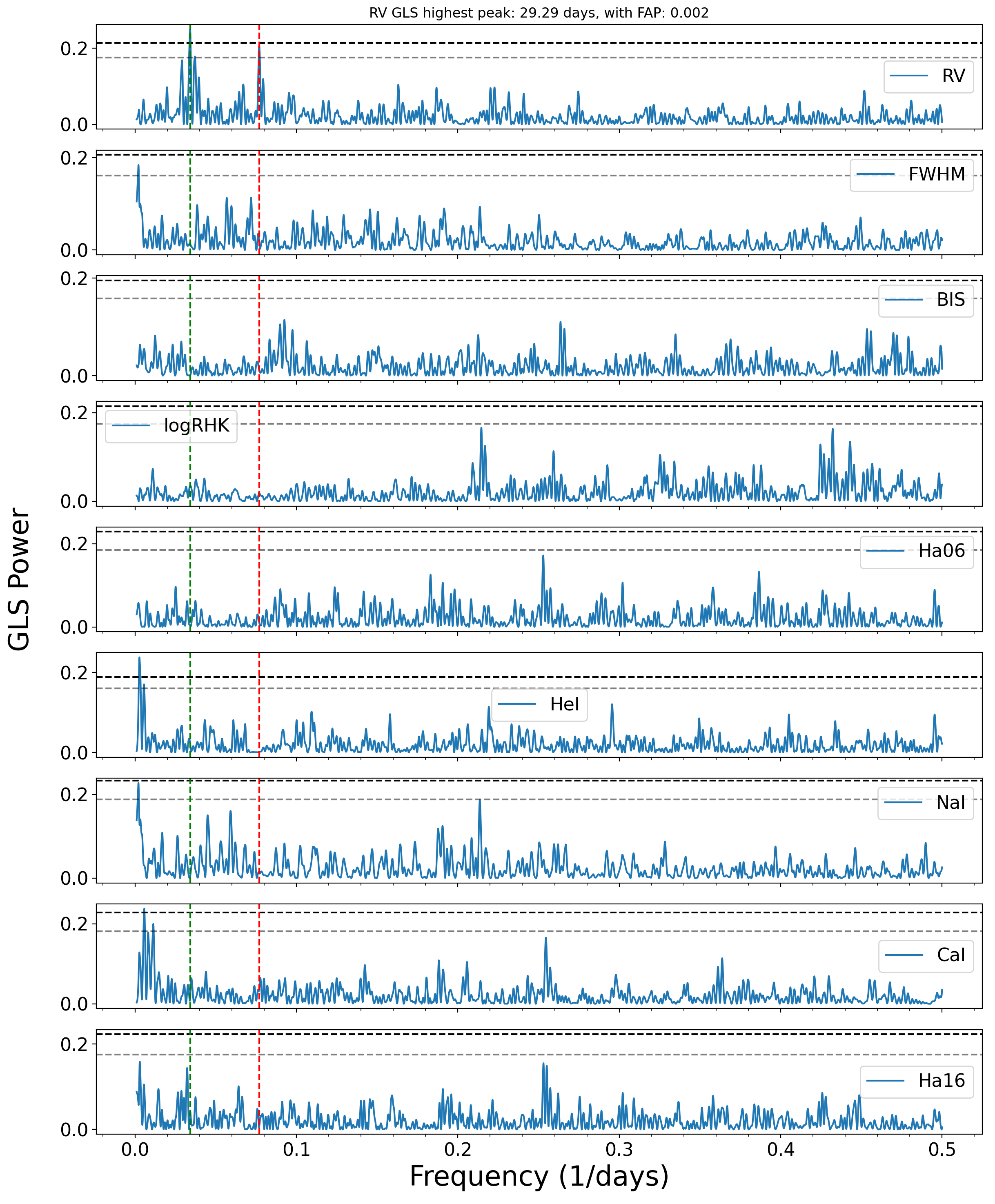}
    \caption{GLS periodogram of HARPS-N radial velocities and of various activity indexes specified in the labels, after the removal of a linear trend (Sect.~\ref{sec:rv_trend}). The main peak of the RV GLS and that of the TOI-1422\,b period are highlighted with a green and red dashed line, respectively. They do not overlap with any of the peaks from the indexes, which in general do not suggest any clear stellar rotation period. The period corresponding to the highest peak in the RV GLS periodogram, and its FAP, are written on the top of the first panel, while the horizontal dashed lines remark the $10\%$ and $1\%$ confidence levels (evaluated with the bootstrap method), respectively. The three peaks surrounding the RVs main frequency can all be explained as aliases of the 29-day signal due to the two highest frequencies of the window function (190 and 390.3 days, as shown in Fig.~\ref{fig:aliases} and Fig.~\ref{fig:window}).}
    \label{fig:GLS_activity}
    \end{figure*}

\subsection{RV and activity indexes periodogram analysis}
\label{sec:rv_trend}

We computed the Generalised Lomb Scargle (GLS) periodogram for the HARPS-N RVs and different stellar activity indexes\footnote{The Full-Width-at-Half-Maximum (FWHM) and the Bisector inverse span (BIS) are calculated using the cross correlation function (CCF) derived by the DRS pipeline. We also analysed the chromospheric $\log{R^{\prime}_{\rm HK}}$ index, and additional activity diagnostics derived from the spectroscopic lines He\,{\tiny I}, Na\,{\tiny I}, Ca\,{\tiny I}, H$\alpha$06 and H$\alpha$16 as defined in the code \texttt{ACTIN} ({\tt https://github.com/gomesdasilva/ACTIN} v.1.3.9, \citealt{actin}, which has been used for the calculation. In particular, the two H-alpha indices have 1.6 and $0.6\,\AA$ band-pass width, respectively.} using the Python package \texttt{astropy} v.4.3.1 \citep{Price2018}. The periodogram of the RVs shows the main peak around 29-days, other than a significant peak at 13 days (TOI-1422\,b transiting period), after correcting for a linear trend of $\sim4$\,m\,s$^{-1}$\,yr$^{-1}$, observed in HARPS-N data. No index shows signs of 29-days periodicity, but a linear trend is also present in the FWHM and $\log R^{\prime}_{\rm HK}$ (see Fig.~\ref{fig:fwhmrhk}), with the former correlating the most with the RVs, unveiling a moderate Spearman coefficient \citep{Spearman1904} of $0.41$ ($p$-value $0.01\%$). Therefore, in order to explain the nature of the main peak in the RVs, we present the GLS periodogram of these coefficients posterior to the removal of their linear trends in Fig.~\ref{fig:GLS_activity} (see Fig.~\ref{fig:aliases} for a closer look at the RVs panel), but again no trace of the 29-days signal is found. We also perform a GP regression analysis, using a quasi-periodic model, of the $\log R^{\prime}_{\rm HK}$ index corrected for the linear trend over the time series, and find no evidence of any particular periodic modulation in the posterior distribution of the periodic time-scale hyper-parameter. In short, there is no evidence pointing to a specific periodic rotation of the star TOI-1422, other than the tentative estimation from $\upsilon \sin{i_{\star}}$.\newline

\noindent A query from the Gaia eDR3 archive returns astrometric excess noise and renormalized unit weight error (RUWE)  values of 80 $\mu$as and 1.09, respectively, for TOI-1422. The star is thus astrometrically quiet. The analysis of Sect. 2.2 rules out the existence of obvious sub-arcsec stellar companions, and no co-moving objects are present in Gaia eDR3 data in a 600 arcsec radius. The linear trend seen in the RV data along with a few activity indexes can therefore be explained by a long star magnetic activity, rather than by the presence of a companion\footnote{In this case, at a projected separation of 0.1 arcsec ($\sim 15$\,au at the distance of TOI-1422), the lower limit of the AstraLux imaging data, a maximum RV slope of the magnitude measured in this work would be produced by a companion of $\sim 30\, M_{\rm Jup}$ (i.e. either a very low-mass star or a massive sub-stellar companion).}. 

\subsection{RV and photometry joint analysis}
\label{sec:joint_analysis}
\begin{figure}
\centering
\includegraphics[width=0.49\textwidth]{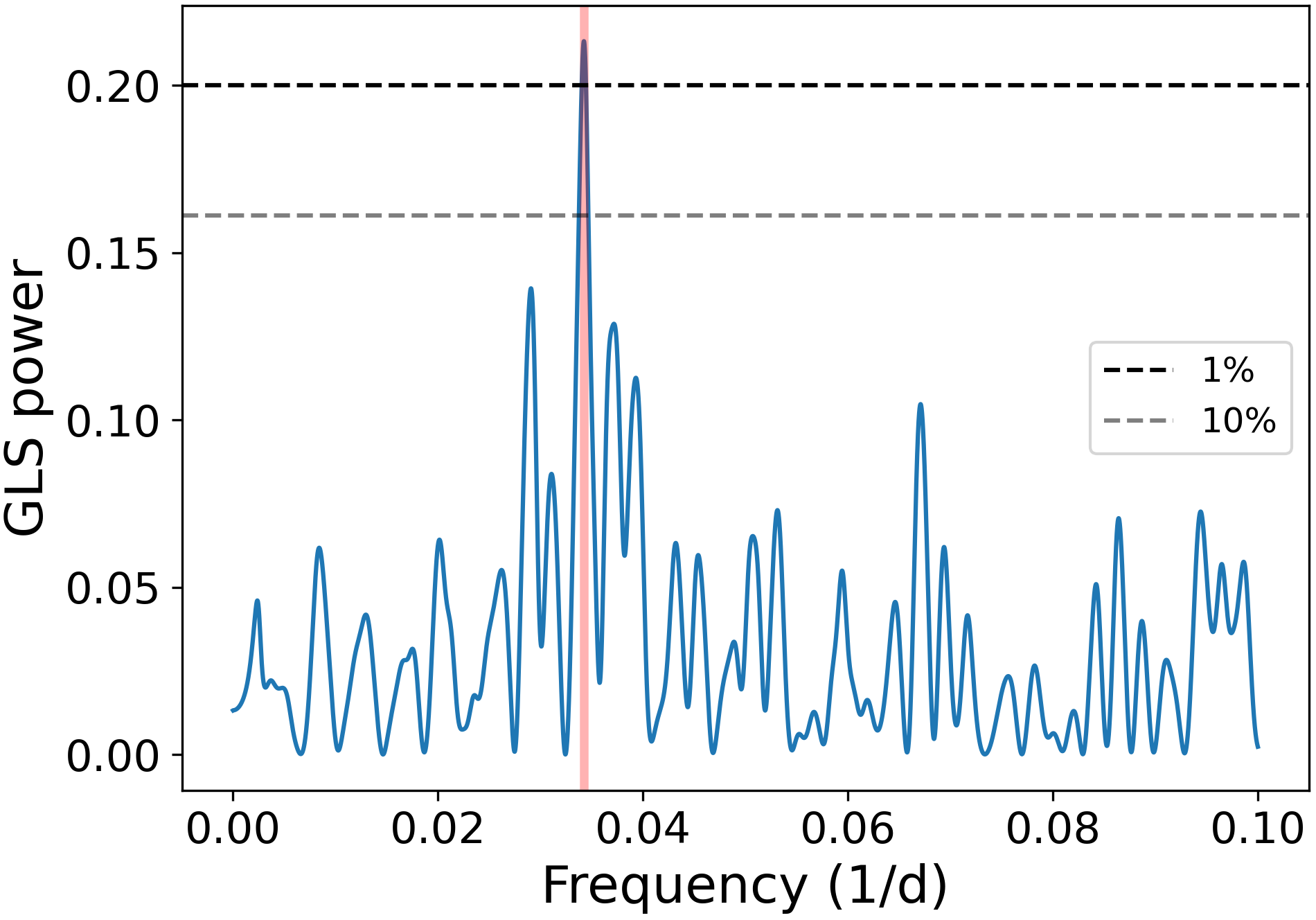}
\caption{GLS periodogram of the transiting one-planet model RV residuals. The main peak is highlighted in red and corresponds to a period of 29.2 days, with a False Alarm Probability (FAP) of 0.45\% (evaluated with the bootstrap method), while the horizontal dashed lines show the 10\% and 1\% confidence levels.}
\label{fig:GLS_1p_res}
\end{figure}

\begin{figure*}
    \centering
    \includegraphics[width=0.9\textwidth]{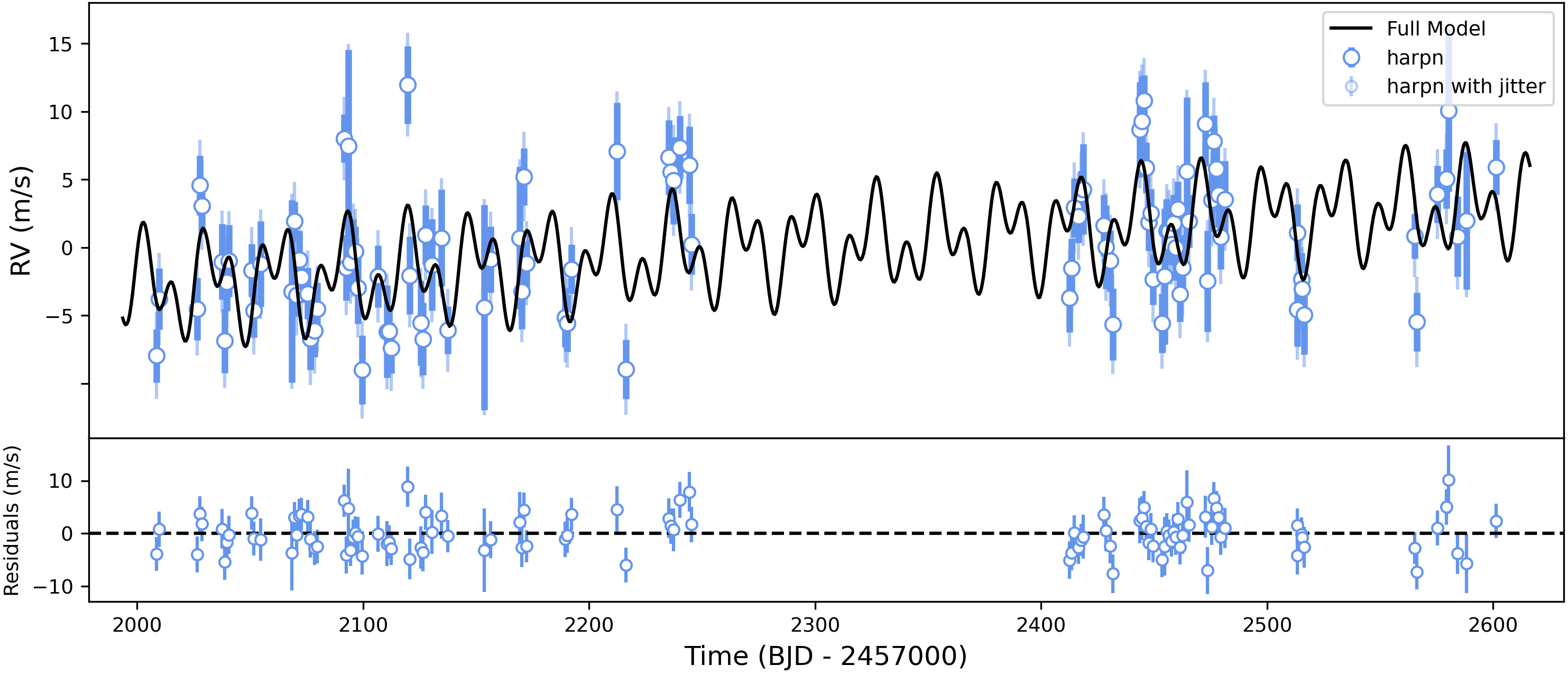}
    \caption{The RV measurements of TOI-1422 versus time are shown on the top panel, while their residuals over the model fit are in the bottom panel. The circles with blue error bars are the RV data taken with HARPS-N. The large and small error bars indicate $\sigma_t$ and $\sigma_w$ (the added jitter term), respectively. In the top panel, the black line represents the two-planet model fit.}
    \label{fig:RV_2p}
\end{figure*}

A joint transit and RV analysis has been carried out with \texttt{juliet}, which employs different Python tools: \texttt{batman}\footnote{\texttt{\href{https://lweb.cfa.harvard.edu/~lkreidberg/batman/}{https://lweb.cfa.harvard.edu/\char`\~lkreidberg/batman}}} \citep{Kreidberg2015} for the modelling of transits, \texttt{RadVel}\footnote{\texttt{\href{https://radvel.readthedocs.io/en/latest/}{https://radvel.readthedocs.io}}} \citep{Fulton2018} for the modelling of RVs and stochastic processes, which are treated as GPs with the packages \texttt{george}\footnote{\texttt{\href{https://george.readthedocs.io/en/latest/}{https://george.readthedocs.io}}} \citep{Ambikasaran2014} and \texttt{celerite}\footnote{\texttt{\href{https://celerite.readthedocs.io}{https://celerite.readthedocs.io}}}. The radial velocity model that we used in \texttt{juliet} is the following:
\begin{equation}
M(t)=K(t)+\epsilon(t)+\overline{\mu}+A\,t+B \, ,
\end{equation}
where $\epsilon(t)$ is a noise model for the HARPS-N instrument here assumed to be white-gaussian noise, i.e. $\epsilon(t)\thickapprox N(0,\sigma(t)^2+\sigma_w^2)$, with $\sigma(t)^2$ being the formal uncertainty of the RV point at time $t$, $\sigma_w^2$ being an added jitter term and $N(\mu,\sigma^2)$ denoting a normal distribution with mean $\mu$ and variance $\sigma^2$. $K(t)$ is the Keplerian model of the RV star perturbations due to the orbiting planet, $\overline{\mu}$ is the systemic velocity linked to the instrument, and the coefficients $A$, $B$ (also referred to as RV slope and RV intercept) represent an additional linear trend used for modelling non-Keplerian signals with a period longer than the observation span. For a total number of data points $N$, we assumed the model likelihood to follow the likelihood of a $N$-dimensional multivariate Gaussian:
\begin{equation}
ln\,p(\vec{y}|\vec{\theta})=-\frac{1}{2}[N\,ln\,2\pi+ln\,|\Sigma|+\vec{r}^T\Sigma^{-1}\vec{r}] \,,
\end{equation}
where $\vec{y}$ and $\vec{\theta}$ are vectors containing respectively all the RV data points and instrumental parameters, while $\vec{r}$ is the residual vector given by
\begin{equation}
r\left(t_i\right)=y\left(t_i\right)-M\left(t\right) \, .
\end{equation}
The elements of the covariance matrix $\Sigma$ are:
\begin{equation}
\Sigma(t_i,t_j)=k(\mu_i,\mu_j)+(\sigma_w^2+\sigma_t^2)\delta_{t_i,t_j} \, ,
\end{equation}
with $k$ equal to any GP kernel model or zero for a pure white-noise one. In order to estimate the Bayesian posteriors and evidence, $\mathcal{Z}$, of different models, we used a Dynamic Nested Sampling package \texttt{dynesty} \citep{Speagle2019}, which adaptively allocates samples based on a posterior structure and, at the same time, estimates evidence and sampling from multi-modal distributions. In general, dynamic nested sampling algorithms sample a dynamic number of live points from the prior ``volume'' and sequentially replace the point with the lowest likelihood with a new one, while updating the Bayesian evidence by the difference $\Delta\mathcal{Z}$. Usually, the stopping criterion is a defined value of $\Delta\mathcal{Z}$, below which the algorithm is said to have converged ($\Delta\mathcal{Z}\thickapprox0.5$). However, here we used the default criterion described in Sect. 3.4 of \cite{Speagle2019}.\newline

\begin{figure}
\centering
\includegraphics[width=0.5\textwidth]{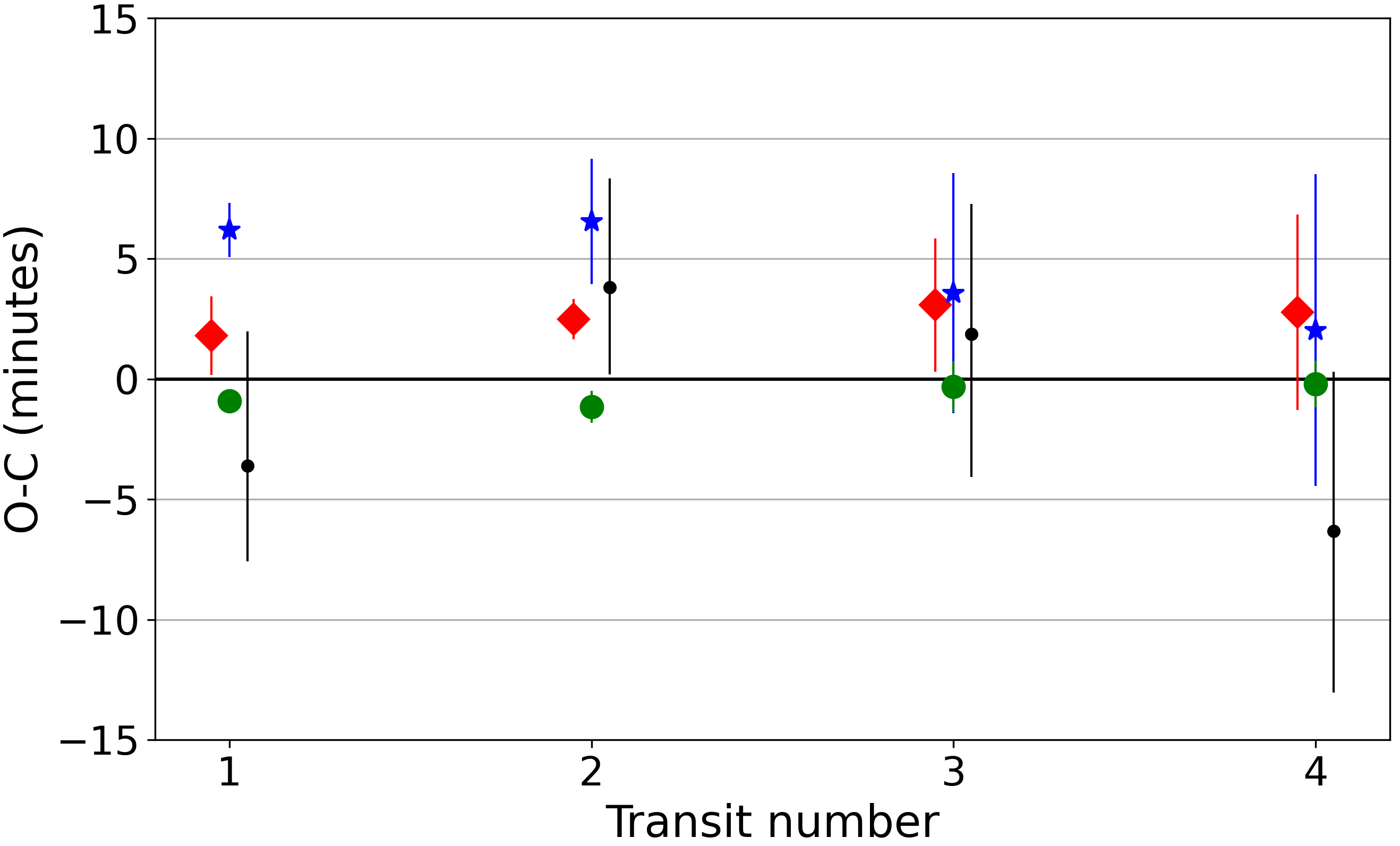}
\caption{The residuals for the mid-transit timings of TOI-1422\,b versus a linear ephemeris, with 1-$\sigma$ error bars, are plotted in black. The green circles, red diamonds and blue stars represent TTV predictions respectively in the case of null, average or maximum eccentricities, with the error bars showing the uncertainty due to $T_{0,c}$ (see Table \ref{tab:2p}). The points have been slightly shifted on the x-axis to allow for more visibility.}
\label{fig:o-c}
\end{figure}

In order to reveal the transiting object suggested by the TESS light curve, we first run the RV and photometry joint analysis with a simple one-planet model using as transit-related priors the parameters in the DVR produced by the SPOC pipeline both with a fixed null and uniformly-sampled eccentricity via the parametrization $S_1=\sqrt{e}\sin{\omega}$,\, $S_2=\sqrt{e}\cos{\omega}$, that is described in \cite{Eastman2012}. All the priors are defined in Table \ref{tab:prior1p}. In particular, we have set Gaussian priors on both the limb-darkening coefficients (from \citealt{Claret2017}) and the star mean density $\rho_{\star}$ (from Sect. \ref{sec:star}), which is implemented here instead of the scaled semi-major axis, $a/R_{\star}$, because the latter can be recovered using Kepler's third law using only the period of the respective planet, which is a direct result of any \texttt{juliet} run. In this way, from the single value of $\rho_{\star}$ we can evenly derive $a/R_{\star}$ in the case of multiple planets.

The best one-planet RV model fit is found with $e=0$ ($\Delta\ln{\mathcal{Z}}^{e_b=0}_{e_b\neq0}=0.7$), but the scatter of the residuals is higher than the average photon-noise uncertainties for this kind of star. In fact, the same peak of 29-days, which was found in the RVs GLS periodogram, is also distinctly found in the residuals of the transiting one-planet model (see Fig.~\ref{fig:GLS_1p_res}). Consequently, we proceed to test two-planet models, whose priors are summed in Table \ref{tab:prior2p}. Since they have comparable statistical significance ($\Delta\ln{\mathcal{Z}}^{e_{b,c}=0}_{e_{b,c}\neq0}=0.4$) for the rest of the paper we use the results of the eccentric model. The two-planet eccentric model is plotted on top of the RVs in Fig. \ref{fig:RV_2p}, along with its residuals. TOI-1422\,b RV semi-amplitude and orbital period result to be $K_{\rm b}=2.47^{+0.50}_{-0.46}$ m/s and $P_{\rm b}=12.9972\pm0.0006$ days, respectively. While the second planet, candidate TOI-1422\,c, has RV semi-amplitude $K_{\rm c}=2.36^{+0.42}_{-0.40}$ m/s, orbital period $P_{\rm c}=29.29^{+0.21}_{-0.20}$ days and $T_{0,{\rm c}}=2458776.6\pm4.6$ BJD (see the posteriors in Fig.~\ref{fig:corner2p} and Table~\ref{tab:posterior2p}). The eccentricities turn out to be $e_{\rm b}=0.04^{+0.05}_{-0.03}$ and $e_{\rm c}=0.14^{+0.17}_{-0.10}$, but it is important to note that when they are fixed to zero, the orbital parameters of TOI-1422\,b and TOI-1422\,c remain, within 1-$\sigma$, compatible with those of the eccentric model.

\subsection{Results}
\label{sec:results}
\noindent TOI-1422\,c's orbital period explains both the main peak found in the residuals of the one-planet model (Fig.~\ref{fig:GLS_1p_res}) and in the RV GLS periodogram (Fig.~\ref{fig:GLS_activity}); it is also in 9:4 orbital resonance with the first planet. The difference between the Bayesian evidence of the two-planet eccentric model and the one-planet model ($\Delta\ln{\mathcal{Z}}^{2p}_{1p}=5.1$) is barely above the \textit{very strong} evidence threshold defined in \citet{Kass1995} ($\Delta\ln{\mathcal{Z}}>5$), so even if the existence of candidate planet c remains unproven, we believe the two-planet model is currently the better one to explain the 29-days signal observed in the RVs, due to the lack of evidence for star activity.

\begin{figure}
\centering
\includegraphics[width=0.49\textwidth]{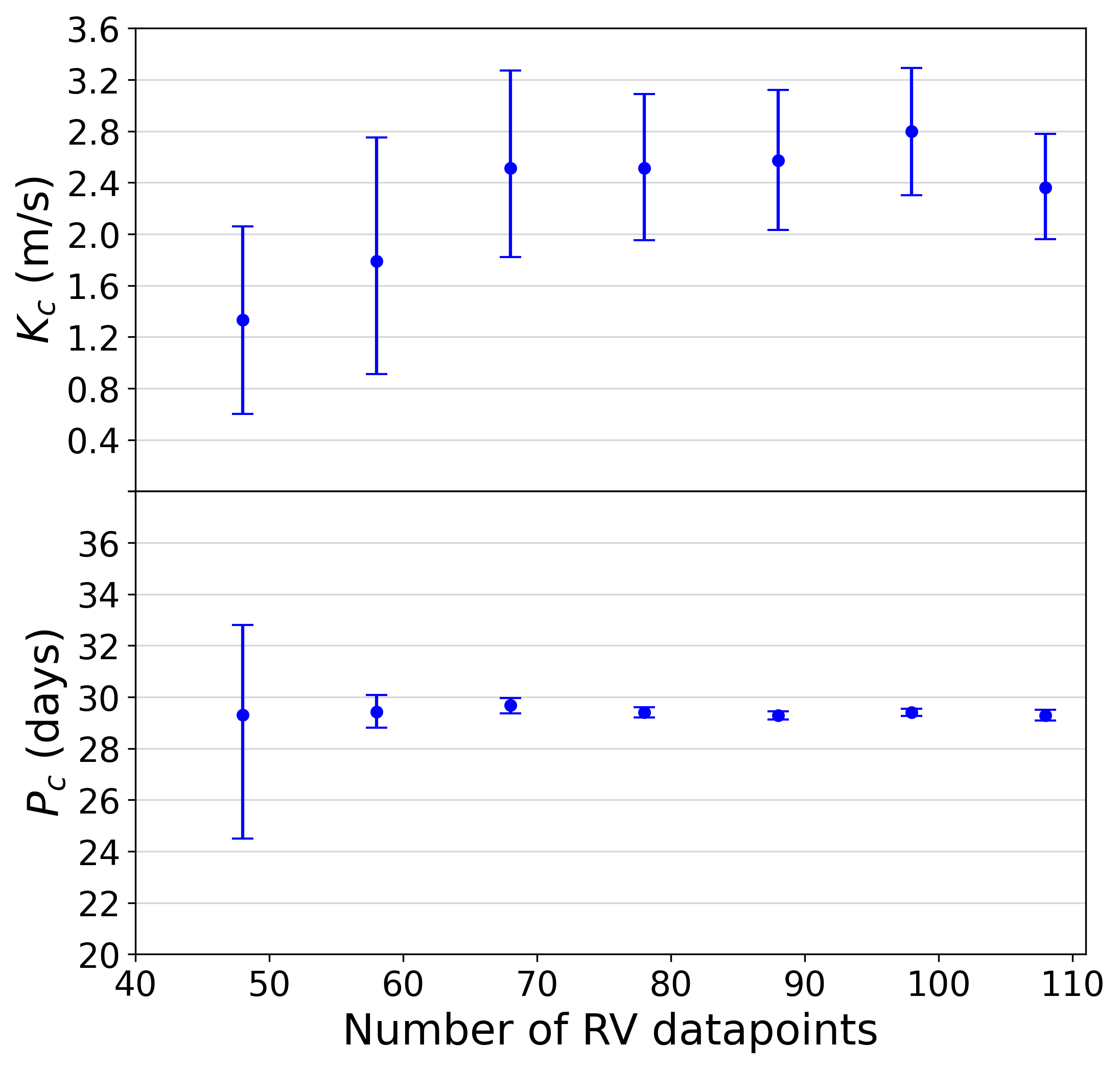}
\caption{RV semi-amplitude $K$ and orbital period $P$, along with their 1-$\sigma$ error bars, for candidate planet c as functions of the number of data points used for the two-planet eccentric model analysis with \texttt{juliet}.}
\label{fig:progressive}
\end{figure}

Furthermore, the two-planet analysis has been replicated with different numbers of data points in order to understand how and if new measurements were impacting the significance of the second planet detection and, as shown in Fig.~\ref{fig:progressive}, both the RV semi-amplitude and the period seem to stabilise after $\approx60$ measurements, which matches the beginning of the second observation season, while the significance of the 29-days peak also grows (Fig.~\ref{fig:power_obs}). It is noteworthy to mention that the GLS periodogram of the residuals of the two-planet model does not show peaks below 50\% FAP and hence does not suggest the presence of additional detectable signals.\newline 

\begin{figure}
\centering
\includegraphics[width=0.49\textwidth]{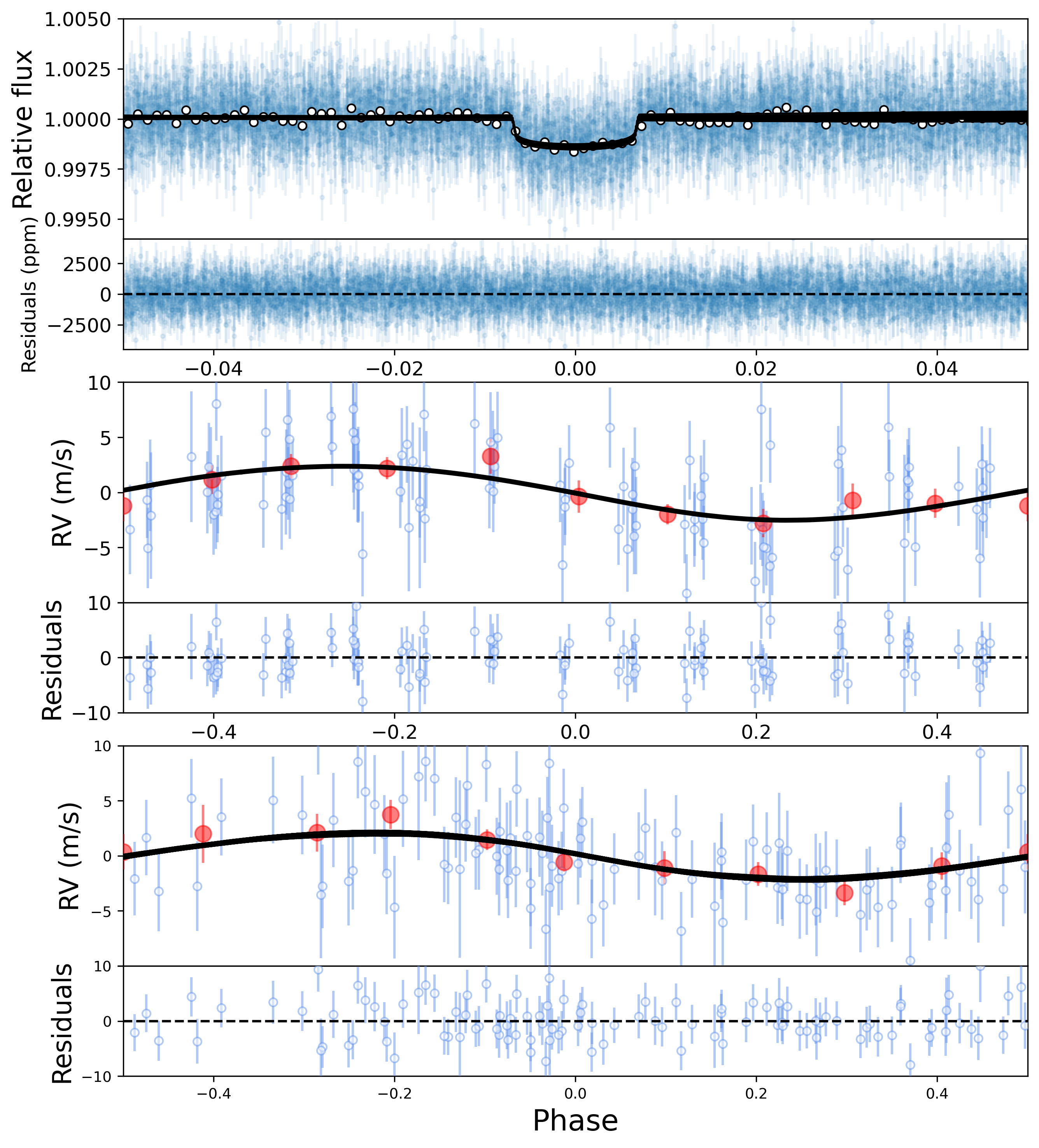}
\caption{{\bf Top panel:} The {\it TESS} phase-folded light curve of TOI-1422\,b transits, compared to the best-fitting model. {\bf Bottom panels:} HARPS-N RV data of TOI-1422 phase-folded to the period of planet b (middle) and candidate c (bottom), along with their residuals over the model. The red circles represent the average value of phased RV data points.}
         \label{fig:phase2p}
   \end{figure}

\noindent A phase-folded plot of both the transit and the radial velocities is shown in Fig.~\ref{fig:phase2p} for the eccentric two-planet model. The radius for TOI-1422\,b was calculated with the transformations provided by \citet{Espinoza2018} and, using the stellar radius of Sect.~\ref{sec:star}, its revised value turns out to be $R_{\rm b}=3.96^{+0.13}_{-0.11}\,R_{\oplus}$. Using the stellar radius from Table~\ref{tab:star}, we derived the mass of both objects to be $M_{\rm b}=9.0^{+2.3}_{-2.0}\,M_{\oplus}$ and $M_{\rm c}\sin{i_c}=11.1^{+2.6}_{-2.3}\,M_{\oplus}$. Their final parameters are reported in Table~\ref{tab:2p}. An independent joint analysis of the HARPS-N radial velocities and \textit{TESS} photometry, after the transits have been normalised through a local linear fitting, was also performed with a DE-MCMC method \citep{Eastman2012,2019arXiv190709480E}, following the same implementation as in \citet{2014A&A...572A...2B,2015A&A...575A..85B}. The obtained results are consistent, within 1-$\upsigma$, with those reported in Table \ref{tab:2p}.\newline

\noindent In order to evaluate possible transit time variations (TTVs) due to the influence of candidate TOI-1422\,c over TOI-1422\,b, we plot the four mid-transit times minus their expected values (based on the two-planet eccentric model) in Fig.~\ref{fig:o-c} along with different TTV predictions made with the code described in \citealt{Agol2016}. Unfortunately, there are not enough transits to draw any conclusion as all the delays are compatible with zero within 1-$\sigma$, therefore further precise monitoring of TOI-1422\,b transits is here encouraged in order to confirm the existence of TOI-1422\,c and, overall, better characterise the planetary system.
\begin{figure*}
\centering
\includegraphics[width=0.9\textwidth]{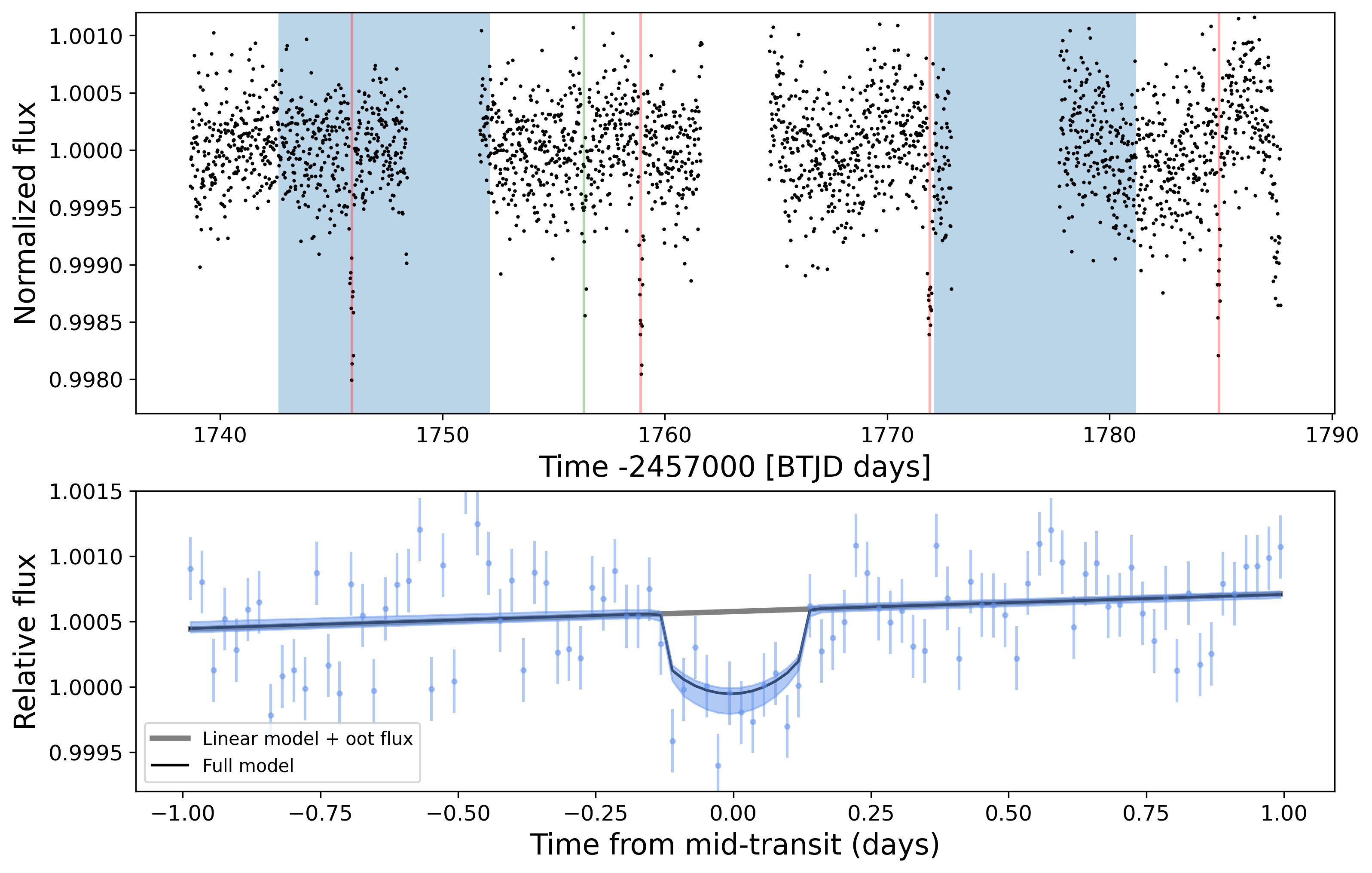}
\caption{{\it Top panel:} PDC-SAP light curve with TOI-1422\,b transits highlighted in red and the expected TOI-1422\,c transits, with their uncertainties, highlighted in blue. A single planetary-transit event has also been marked with a green vertical line and is discussed at the end of Sect.~\ref{sec:results}). {\it Bottom panel:} The single transit-like event as seen in the PATHOS light curve and the corresponding fit.}
\label{fig:ste}
\end{figure*}
\begin{table}
\centering
\caption{Best fit median values, with upper and lower 68\% credibility bands as errors, of the fitted and derived parameters for TOI-1422\,b and TOI-1422\,c, as extracted from the posterior distribution of the two-planet eccentric model (Table~\ref{tab:posterior2p} and Fig.~\ref{fig:corner2p}).}\label{tab:2p}
\renewcommand{\arraystretch}{1.2}
\resizebox{\hsize}{!}{\begin{tabular}{lcc}
    \toprule
     & TOI-1422\,b &  TOI-1422\,c \\ [2pt]
\hline \\[-6pt]%
\multicolumn{1}{l}{\textbf{Transit and orbital parameters}} \\[2pt]
$K$ (m\,s$^{-1}$)\dotfill & $2.47^{+0.50}_{-0.46}$ & $2.36^{+0.42}_{-0.40}$ \\
$P_{\rm orb}$ (days)\dotfill & $12.9972\pm0.0006$ & $29.29^{+0.21}_{-0.20}$ \\
$T_{\rm 0}$ (BJD)\dotfill & $2\,458\,745.9205^{+0.0012}_{-0.0011}$ & $2\,458\,776.6^{+4.6}_{-4.5}$ \\
$T_{\rm 14}$ (hours)\dotfill & $4.52\pm0.16$ & -- \\
$R_{\rm p}/R_{\star}$\dotfill & $0.0356^{+0.0007}_{-0.0005}$ & -- \\ 
$b$\dotfill & $0.19^{+0.11}_{-0.10}$ & --\\ 
$i$ (deg)\dotfill &  $89.52^{+0.26}_{-0.28}$ & -- \\ 
$a/R_{\star}$\dotfill & $22.72^{+0.31}_{-0.40}$ & $39.05^{+0.50}_{-0.73}$ \\
$q_1$\dotfill & $0.28^{+0.11}_{-0.08}$ & -- \\
$q_2$\dotfill & $0.30^{+0.05}_{-0.05}$ & -- \\
$\sqrt{e}\sin\omega$\dotfill & $0.018^{+0.108}_{-0.095}$ & $0.120^{+0.221}_{-0.233}$ \\
$\sqrt{e}\cos\omega$\dotfill & $-0.149^{+0.153}_{-0.128}$ & $-0.070^{+0.349}_{-0.304}$ \\[2pt]
\multicolumn{1}{l}{\textbf{Derived parameters}} \\[2pt]
$M_{\rm p}$ ($M_{\oplus}$)\dotfill & $9.0^{+2.3}_{-2.0}$ & -- \\
$M_{\rm p}\sin{i}$ ($M_{\oplus}$)\dotfill & --  & $11.1^{+2.6}_{-2.3}$ \\[2pt]
$R_{\rm p}$ ($R_{\oplus}$)\dotfill & $3.96^{+0.13}_{-0.11}$ & --\\
$\rho_{\rm p}$ (g\,cm$^{-3}$)\dotfill & $0.795^{+0.290}_{-0.235}$ & -- \\ [2pt]
$\log{g_{p}}$ (cgs)\dotfill & $2.75^{+0.08}_{-0.14}$ & -- \\
$a$ (AU)\dotfill & $0.108\pm0.003$ & $0.185\pm0.006$ \\ 
$T_{\rm eq}^{(\ddagger)}$ (K)\dotfill & $867\pm17$ & $661\pm13$ \\
$u_1$\dotfill & $0.32^{+0.12}_{-0.10}$ & -- \\
$u_2$\dotfill & $0.21^{+0.10}_{-0.08}$ & -- \\
$e$\dotfill & $0.04^{+0.05}_{-0.03}$ & $0.14^{+0.17}_{-0.10}$ \\ [2pt]
$\omega$ (deg)\dotfill & $153^{+20}_{-56}$ & $99^{+63}_{-64}$ \\ [2pt]
    \bottomrule
\end{tabular}
}
\tablefoot{$^{(\ddagger)}$ This is the equilibrium temperature for a zero Bond albedo and uniform heat redistribution to the night side.}
\end{table}

\subsection{Other transit events}
In the search for TOI-1422\,c transits, we found a possible single transit-like event around 2\,458\,756.35 BTJD days, as shown in Fig.~\ref{fig:ste}, which can not be related to either TOI-1422\,b or TOI-1422\,c. We have fitted this potential transit using the light curve from the pipeline PATHOS \citep{Nardiello2019} and retrieved a possible radius $R_{\rm d}=2.82_{-0.05}^{+0.38}\,R_{\oplus}$, which is compatible with the transit depth observed in the PDC-SAP and SAP light curves as well. The duration of the transit suggests an orbital period longer than TOI-1422\,c's but this is very uncertain, while the lack of other transits in the TESS light curve suggests an orbital period between 17-22, or longer than, 35 days, thus incompatible with TOI-1422\,c's. PATHOS is a PSF-based approach to TESS data which allows the minimization of dilution effects in crowded environments, and here it is utilized to extract high-precision photometry of TOI-1422 to independently confirm the presence of this transit even after the application of a different neighbour-subtraction technique. Neither the single transit nor TOI-1422\,b transits show correlation with the X,Y pixels and the sky background signal (Fig.~\ref{fig:systematics}), and the single transit depth also does not change with different photometric apertures (Fig.~\ref{fig:singledepth}). Nevertheless, the 3-planet model for the joint transit-RV analysis is not statistically significant and the lack of other transits makes the suggestion of another candidate impossible to justify.\newline

\noindent However, no transit compatible with the expected $T_{0,{\rm c}}$ and $P_{\rm c}$ evaluated with the RV and photometry joint analysis, was found in the SPOC (both SAP and PDC-SAP) light curves, even though a small part of the supposed transiting window has been missed by \emph{TESS}. When we take into account both the time-span of \emph{TESS} light curve and TOI-1422\,c expected (non-grazing) transit duration, the probability that such transits would have been missed can be estimated to be around 1\% and 7\%, with respectively 1$\sigma$ and 3$\sigma$ uncertainty on $T_{0,c}$. Other than misaligned orbits, another possible explanation for the lack of TOI-1422\,c transits is that despite its mass, which is greater than planet b's, its size could be much smaller (like the high-density sub-Neptune, BD+20594b of \citealt{Espinoza_2016}), as any object with a radius approximately below 2.8$R_{\oplus}$ might be disguised in the light curve noise (as proven by the, so far undetected and uncertain, single transit-like event). Ultimately, it remains unknown if candidate planet c is transiting or not, so further high-precision long photometric follow-up observations will be important to clear up this possibility along with the nature of the single transit event. The new \emph{TESS} observations of this target, during Sector~57, are definitely welcome as they might shed some light on both matters.

\section{Discussion}
\label{sec:discussion}
\subsection{Orbital resonance}
As we have seen, candidate c is within 1-$\sigma$, in 9:4 orbital resonance with planet b. This is likely coincidental since the resonance is 5th-order and thus very weak unless one of the planets is quite eccentric\footnote{We note that, even with the $e$'s suggested by the eccentric fits, which are unusually high compared to most multi-transiting planetary systems according to \citealt{Xie2016}, the 9:4 would not be as strong as a first-order resonance.} or the mutual inclination is high. The exact 9:4 (or 2.25) resonance is within uncertainty perhaps only because the uncertainty of TOI-1422\,c's orbital period is large compared to the tight period uncertainties of transiting planets. As a matter of fact, period ratios a bit above 2 have been found within many exoplanetary systems \citep{Winn2015}, but it is also possible that the 9:4 resonance is actually the result of a resonant chain of 3 planets in 1st-order 3:2 resonances among each other, with the middle one yet to be seen. If that is the case, since the period ratios of Kepler planets near first order resonances are usually slightly wide of resonance, the likely orbital period for this unknown exoplanet would be slightly more than 19.5 days, and thus compatible with the observed single transit discussed in  Sect.~\ref{sec:results}. Given this orbital period and assuming that an RV semi-amplitude roughly up to 2\,m\,s$^{-1}$ might be hidden in the residuals of the two-planet model, this middle object should not have a mass higher than $\approx8\,M_{\oplus}$, or a density higher than $\approx2$ g\,cm$^{-3}$.
%
\subsection{Mass-radius diagram and internal structure of planet b}
\begin{figure}
\centering
\includegraphics[width=\hsize]{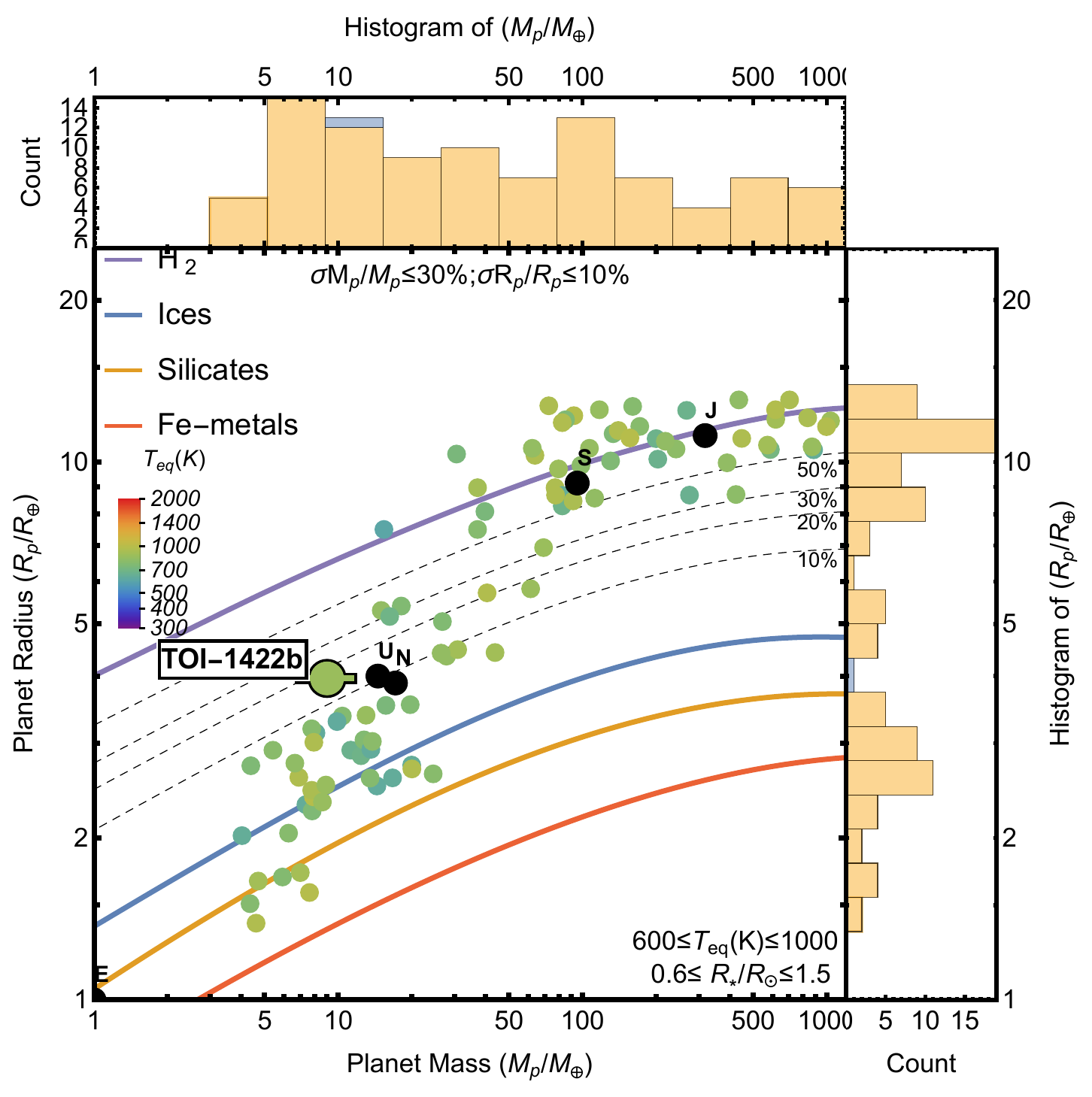}
\caption{Planetary masses and radii of the known transiting exoplanets (values taken from the Transiting Extrasolar Planet catalogueue, TEPCat, which is available at {\tt http://www.astro.keele.ac.uk/jkt/tepcat/} catalogue) \citep{Southworth2010,2011MNRAS.417.2166S} with equilibrium temperature $T_{\rm eq}$ between 600 and 1000\,K and host star radius between 0.6 and 1.5 $R_{\sun}$. Different lines correspond to different mass fractions of relatively cold hydrogen envelopes. The \textit{ice giants} of the Solar System are displayed in black-filled circles. TOI-1422\,b is on the low-density envelope of planets with precise mass/radius estimations ($\sigma M_p/M_p\leq30\%$; $\sigma R_p/R_p\leq10\%$), one of the reasons that make it potentially valuable for transit spectroscopy.}
\label{fig:planetplot}
\end{figure}

\begin{figure}
\centering
\includegraphics[width=\hsize]{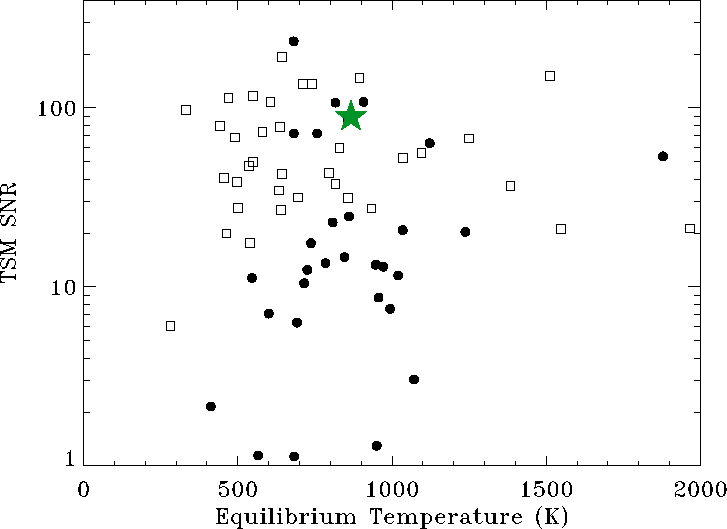}
\caption{Transmission spectroscopy observations TSM values with JWST
over the equilibrium temperature for planets with a measured mass in the radius range $2.75<R_{\oplus}<4.0$, including TOI-1422\,b (green star). Black
filled dots and empty squares identify the sample of planets around stars
with $T_{eff}>5400\,K$ and $T_{eff}<5400\,K$, respectively.}
\label{fig:TSM}
\end{figure}

TOI-1422\,b is among one of the puffier planets with a density of $\sim0.8$\,g\,cm$^{-3}$, which is close to Saturn's and, therefore, lower than most exoplanets in this mass range. It lies towards the upper-left corner of the mass-radius diagram (Fig.~\ref{fig:planetplot}), making it very similar to Kepler-36\,c \citep{Vissapragada2020} and especially to Kepler-11\,e \citep{Lissauer2013}, which even shares the same kind of host star but is on a longer orbit. On one hand, it has a similar radius compared to Neptune and Uranus in our solar system, whereas, on the other hand, its mass is only about $50\%-60\%$ that of our ice giants.  Thus, an extensive gaseous envelope, surrounding a massive core, is expected to be found in TOI-1422\,b. More precisely, the mass fraction of this envelope is expected to be around $10\%-25\%$ of the total mass of the planet (using the equations of state from \citealt{Becker2014}), suggesting that the atmosphere has not been blown away by the stellar wind. The nature of this extensive envelope as well as its core requires further investigation. For this purpose, we assess the expected S/N of the \textit{JWST}/NIRISS measurements\footnote{From a 10-hour observing program assuming a cloud-free, solar-metallicity, $H_2$-dominated atmosphere.} of TOI-1422\,b transits compared to planets of similar sizes, by evaluating the transmission spectroscopy metric (TSM) defined in \citealt{Kempton2018}:
\begin{equation}
   \mbox{TSM}=\mbox{(Scale factor)}\times\frac{R_{\oplus}^3\,T_{\rm eq}}{M_{\oplus}\,R_{\star}^2}\times10^{-0.2\,J}\,,
\end{equation}
where the scale factor is a dimensionless normalization constant, equal to 1.28 for planets with $2.75<R_{\oplus}<4.0$, and $J$ is the apparent magnitude of the host star in the $J$ band (a filter that is near the middle of the NIRISS bandpass). As a result (Fig.~\ref{fig:TSM}), TOI-1422\,b ranks fourth\footnote{Following TOI-561\,c \citep{Lacedelli2020,Lacedelli2022}, HD\,136352\,c \citep{Kane2020,Delrez2021} and HD\,63935\,b \citep{Scarsdale2021}} among Neptunes ($2.75<R_{\oplus}<4.0$) orbiting G-F dwarfs ($T_{eff}>5400$K), but being the one with the lowest density, it is definitely an interesting candidate for atmospheric characterization by the \textit{JWST}.


\section{Conclusions}
\label{sec:conclusion}

In this paper, we have confirmed the planetary nature of the {\it TESS} transiting planet TOI-1422\,b, which turns out to be a low-density and warm Neptune-size planet orbiting around an astrometrically, and overall magnetically, quiet G2\,V star. Therefore, TOI-1422\,b is the latest addition to the low-populated range of exoplanets with the size of Neptune but Saturn-like density. In order to well constrain the mass of TOI-1422\,b, a long RV monitoring with more than a hundred observations was necessary with the HARPS-N instrument at the TNG in La Palma, which resulted in fully characterized orbital and physical parameters of this new planetary system. On top of that, our RV measurements also suggest the presence in the system of a, possibly non-transiting, heavier candidate planet TOI-1422\,c in a weak 9:4 orbital resonance with its inner brother, which will require further study to validate.

\begin{acknowledgements}
We acknowledge the use of public TESS data from pipelines at the TESS Science Office and at the TESS Science Processing Operations Center. Resources supporting this work were provided by the NASA High-End Computing (HEC) Program through the NASA Advanced Supercomputing (NAS) Division at Ames Research Center for the production of the SPOC data products. The research has made use of the SIMBAD database, operated at CDS, Strasbourg, France, NASA's Astrophysics Data System and the NASA Exoplanet Archive, which is operated by the California Institute of Technology, under contract with the National Aeronautics and Space Administration under the Exoplanet Exploration Program. This research was also partly supported by the Department of Energy under awards DE-NA0003904 and DE-FOA0002633 (to S.B.J., principal investigator, and collaborator Li Zeng) with Harvard University and by the Sandia Z Fundamental Science Program. This research represents the authors’ views and not those of the Department of Energy. The work is based on observations made with the Italian Telescopio Nazionale Galileo (TNG) operated on the island of La Palma by the Fundacion Galileo Galilei of the INAF (Istituto Nazionale di Astrofisica) at the Spanish Observatorio del Roque de los Muchachos of the Instituto de Astrofisica de Canarias. This work has also made use of data from the European Space Agency (ESA) mission {\it Gaia} (\url{https://www.cosmos.esa.int/gaia}), processed by the {\it Gaia} Data Processing and Analysis Consortium (DPAC, \url{https://www.cosmos.esa.int/web/gaia/dpac/consortium}). Funding for the DPAC
has been provided by national institutions, in particular the institutions participating in the {\it Gaia} Multilateral Agreement. L.\,M. acknowledges support from the ``Fondi di Ricerca Scientifica d'Ateneo 2021'' of the University of Rome ``Tor Vergata''. J.L-B. acknowledges financial support received from ``la Caixa'' Foundation (ID 100010434) and from the European Unions Horizon 2020 research and innovation programme under the Marie Slodowska-Curie grant agreement No 847648, with fellowship code LCF/BQ/PI20/11760023. This research has also been partly funded by the Spanish State Research Agency (AEI) Projects No. PID2019-107061GB-C61 and No. MDM-2017-0737 Unidad de Excelencia ``Mar\'ia de Maeztu''- Centro de Astrobiolog\'ia (INTA-CSIC). We acknowledge financial contribution from the agreement ASI-INAF n.2018-16-HH.0. D.D. acknowledges support from the \emph{TESS} Guest Investigator Program grants 80NSSC21K0108 and 80NSSC22K0185. L.N. acknowledges the support of the ARIEL ASI-INAF agreement 2021-5-HH.0.

\end{acknowledgements}

\bibliographystyle{aa}
\bibliography{TOI-1422.bib}

\begin{thebibliography}{129}
\expandafter\ifx\csname natexlab\endcsname\relax\def\natexlab#1{#1}\fi

\bibitem[{Agol \& Deck(2016)}]{Agol2016}
Agol, E. \& Deck, K. 2016, The Astrophysical Journal, 818, 177

\bibitem[{{Aller} {et~al.}(2020){Aller}, {Lillo-Box}, {Jones}, {Miranda}, \&
  {Barcel{\'o} Forteza}}]{Aller2020}
{Aller}, A., {Lillo-Box}, J., {Jones}, D., {Miranda}, L.~F., \& {Barcel{\'o}
  Forteza}, S. 2020, \aap, 635, A128

\bibitem[{{Ambikasaran} {et~al.}(2015){Ambikasaran}, {Foreman-Mackey},
  {Greengard}, {Hogg}, \& {O'Neil}}]{Ambikasaran2014}
{Ambikasaran}, S., {Foreman-Mackey}, D., {Greengard}, L., {Hogg}, D.~W., \&
  {O'Neil}, M. 2015, IEEE Transactions on Pattern Analysis and Machine
  Intelligence, 38, 252

\bibitem[{{Anglada-Escud{\'e}} \& {Butler}(2012)}]{AngladaEscud2012}
{Anglada-Escud{\'e}}, G. \& {Butler}, R.~P. 2012, \apjs, 200, 15

\bibitem[{{Armstrong} {et~al.}(2020){Armstrong}, {Lopez}, {Adibekyan}, {Booth},
  {Bryant}, {Collins}, {Deleuil}, {Emsenhuber}, {Huang}, {King}, {Lillo-Box},
  {Lissauer}, {Matthews}, {Mousis}, {Nielsen}, {Osborn}, {Otegi}, {Santos},
  {Sousa}, {Stassun}, {Veras}, {Ziegler}, {Acton}, {Almenara}, {Anderson},
  {Barrado}, {Barros}, {Bayliss}, {Belardi}, {Bouchy}, {Brice{\~n}o}, {Brogi},
  {Brown}, {Burleigh}, {Casewell}, {Chaushev}, {Ciardi}, {Collins},
  {Col{\'o}n}, {Cooke}, {Crossfield}, {D{\'\i}az}, {Delgado Mena}, {Demangeon},
  {Dorn}, {Dumusque}, {Eigm{\"u}ller}, {Fausnaugh}, {Figueira}, {Gan},
  {Gandhi}, {Gill}, {Gonzales}, {Goad}, {G{\"u}nther}, {Helled}, {Hojjatpanah},
  {Howell}, {Jackman}, {Jenkins}, {Jenkins}, {Jensen}, {Kennedy}, {Latham},
  {Law}, {Lendl}, {Lozovsky}, {Mann}, {Moyano}, {McCormac}, {Meru},
  {Mordasini}, {Osborn}, {Pollacco}, {Queloz}, {Raynard}, {Ricker}, {Rowden},
  {Santerne}, {Schlieder}, {Seager}, {Sha}, {Tan}, {Tilbrook}, {Ting}, {Udry},
  {Vanderspek}, {Watson}, {West}, {Wilson}, {Winn}, {Wheatley}, {Villasenor},
  {Vines}, \& {Zhan}}]{2020Natur.583...39A}
{Armstrong}, D.~J., {Lopez}, T.~A., {Adibekyan}, V., {et~al.} 2020, \nat, 583,
  39

\bibitem[{{Bailer-Jones} {et~al.}(2021){Bailer-Jones}, {Rybizki}, {Fouesneau},
  {Demleitner}, \& {Andrae}}]{Bailer2021}
{Bailer-Jones}, C.~A.~L., {Rybizki}, J., {Fouesneau}, M., {Demleitner}, M., \&
  {Andrae}, R. 2021, VizieR Online Data Catalog, I/352

\bibitem[{Barstow {et~al.}(2015)Barstow, Aigrain, Irwin, Kendrew, \&
  Fletcher}]{Barstow2015}
Barstow, J.~K., Aigrain, S., Irwin, P. G.~J., Kendrew, S., \& Fletcher, L.~N.
  2015, Monthly Notices of the Royal Astronomical Society, 448

\bibitem[{Becker {et~al.}(2014)Becker, Lorenzen, Fortney, Nettelmann,
  Schöttler, \& Redmer}]{Becker2014}
Becker, A., Lorenzen, W., Fortney, J.~J., {et~al.} 2014, The Astrophysical
  Journal Supplement Series, 215, 21

\bibitem[{Bhatti {et~al.}(2020)Bhatti, Bouma, Joshua, John, \&
  Price-Whelan}]{astrobase}
Bhatti, W., Bouma, L., Joshua, John, \& Price-Whelan, A. 2020,
  waqasbhatti/astrobase: astrobase v0.5.0

\bibitem[{{Biazzo} {et~al.}(2022){Biazzo}, {Bozza}, {Mancini}, \&
  {Sozzetti}}]{2022ASSL..466..143B}
{Biazzo}, K., {Bozza}, V., {Mancini}, L., \& {Sozzetti}, A. 2022, in
  Astrophysics and Space Science Library, Vol. 466, Demographics of
  Exoplanetary Systems, Lecture Notes of the 3rd Advanced School on
  Exoplanetary Science, ed. K.~{Biazzo}, V.~{Bozza}, L.~{Mancini}, \&
  A.~{Sozzetti}, 143--234

\bibitem[{{Biazzo} {et~al.}(2015){Biazzo}, {Gratton}, {Desidera}, {Lucatello},
  {Sozzetti}, {Bonomo}, {Damasso}, {Gandolfi}, {Affer}, {Boccato}, {Borsa},
  {Claudi}, {Cosentino}, {Covino}, {Knapic}, {Lanza}, {Maldonado}, {Marzari},
  {Micela}, {Molaro}, {Pagano}, {Pedani}, {Pillitteri}, {Piotto}, {Poretti},
  {Rainer}, {Santos}, {Scandariato}, \& {Zanmar Sanchez}}]{Biazzoetal2015}
{Biazzo}, K., {Gratton}, R., {Desidera}, S., {et~al.} 2015, \aap, 583, A135

\bibitem[{{Bitsch} {et~al.}(2019){Bitsch}, {Raymond}, \&
  {Izidoro}}]{2019A&A...624A.109B}
{Bitsch}, B., {Raymond}, S.~N., \& {Izidoro}, A. 2019, \aap, 624, A109

\bibitem[{{Bonomo} {et~al.}(2014){Bonomo}, {Sozzetti}, {Lovis}, {Malavolta},
  {Rice}, {Buchhave}, {Sasselov}, {Cameron}, {Latham}, {Molinari}, {Pepe},
  {Udry}, {Affer}, {Charbonneau}, {Cosentino}, {Dressing}, {Dumusque},
  {Figueira}, {Fiorenzano}, {Gettel}, {Harutyunyan}, {Haywood}, {Horne},
  {Lopez-Morales}, {Mayor}, {Micela}, {Motalebi}, {Nascimbeni}, {Phillips},
  {Piotto}, {Pollacco}, {Queloz}, {S{\'e}gransan}, {Szentgyorgyi}, \&
  {Watson}}]{2014A&A...572A...2B}
{Bonomo}, A.~S., {Sozzetti}, A., {Lovis}, C., {et~al.} 2014, \aap, 572, A2

\bibitem[{{Bonomo} {et~al.}(2015){Bonomo}, {Sozzetti}, {Santerne}, {Deleuil},
  {Almenara}, {Bruno}, {D{\'\i}az}, {H{\'e}brard}, \&
  {Moutou}}]{2015A&A...575A..85B}
{Bonomo}, A.~S., {Sozzetti}, A., {Santerne}, A., {et~al.} 2015, \aap, 575, A85

\bibitem[{{Borsa} {et~al.}(2021){Borsa}, {Allart}, {Casasayas-Barris},
  {Tabernero}, {Zapatero Osorio}, {Cristiani}, {Pepe}, {Rebolo}, {Santos},
  {Adibekyan}, {Bourrier}, {Demangeon}, {Ehrenreich}, {Pall{\'e}}, {Sousa},
  {Lillo-Box}, {Lovis}, {Micela}, {Oshagh}, {Poretti}, {Sozzetti}, {Allende
  Prieto}, {Alibert}, {Amate}, {Benz}, {Bouchy}, {Cabral}, {Dekker},
  {D'Odorico}, {Di Marcantonio}, {Figueira}, {Genova Santos}, {Gonz{\'a}lez
  Hern{\'a}ndez}, {Lo Curto}, {Manescau}, {Martins}, {M{\'e}gevand}, {Mehner},
  {Molaro}, {Nunes}, {Riva}, {Su{\'a}rez Mascare{\~n}o}, {Udry}, \&
  {Zerbi}}]{2021A&A...645A..24B}
{Borsa}, F., {Allart}, R., {Casasayas-Barris}, N., {et~al.} 2021, \aap, 645,
  A24

\bibitem[{Borucki {et~al.}(2010)Borucki, Koch, Basri, Batalha, Brown, Caldwell,
  Caldwell, Christensen-Dalsgaard, Cochran, DeVore, Dunham, Dupree, Gautier,
  Geary, Gilliland, Gould, Howell, Jenkins, Kondo, Latham, Marcy, Meibom,
  Kjeldsen, Lissauer, Monet, Morrison, Sasselov, Tarter, Boss, Brownlee, Owen,
  Buzasi, Charbonneau, Doyle, Fortney, Ford, Holman, Seager, Steffen, Welsh,
  Rowe, Anderson, Buchhave, Ciardi, Walkowicz, Sherry, Horch, Isaacson,
  Everett, Fischer, Torres, Johnson, Endl, MacQueen, Bryson, Dotson, Haas,
  Kolodziejczak, Cleve, Chandrasekaran, Twicken, Quintana, Clarke, Allen, Li,
  Wu, Tenenbaum, Verner, Bruhweiler, Barnes, \& Prsa}]{Borucki2010}
Borucki, W.~J., Koch, D., Basri, G., {et~al.} 2010, Science, 327

\bibitem[{{Bouchy, F.} {et~al.}(2013){Bouchy, F.}, {D\'{\i}az, R. F.},
  {H\'ebrard, G.}, {Arnold, L.}, {Boisse, I.}, {Delfosse, X.}, {Perruchot, S.},
  \& {Santerne, A.}}]{Bouchy2013}
{Bouchy, F.}, {D\'{\i}az, R. F.}, {H\'ebrard, G.}, {et~al.} 2013, A\&A, 549,
  A49

\bibitem[{{Brande} {et~al.}(2022){Brande}, {Crossfield}, {Kreidberg},
  {Oklop{\v{c}}i{\'c}}, {Polanski}, {Barman}, {Benneke}, {Christiansen},
  {Dragomir}, {Foreman-Mackey}, {Fortney}, {Greene}, {Howard}, {Knutson},
  {Lothringer}, {Mikal-Evans}, \& {Morley}}]{2022arXiv220104197B}
{Brande}, J., {Crossfield}, I. J.~M., {Kreidberg}, L., {et~al.} 2022, arXiv
  e-prints, arXiv:2201.04197

\bibitem[{{Brogi} {et~al.}(2018){Brogi}, {Giacobbe}, {Guilluy}, {de Kok},
  {Sozzetti}, {Mancini}, \& {Bonomo}}]{2018A&A...615A..16B}
{Brogi}, M., {Giacobbe}, P., {Guilluy}, G., {et~al.} 2018, \aap, 615, A16

\bibitem[{Brown {et~al.}(2018)Brown, Vallenari, Prusti, de~Bruijne, Babusiaux,
  Bailer-Jones, Biermann, Evans, Eyer, Jansen, Jordi, Klioner, Lammers,
  Lindegren, Luri, Mignard, Panem, Pourbaix, Randich, Sartoretti, Siddiqui,
  Soubiran, van Leeuwen, Walton, Arenou, Bastian, Cropper, Drimmel, Katz,
  Lattanzi, Bakker, Cacciari, Castañeda, Chaoul, Cheek, Angeli, Fabricius,
  Guerra, Holl, Masana, Messineo, Mowlavi, Nienartowicz, Panuzzo, Portell,
  Riello, Seabroke, Tanga, Thévenin, Gracia-Abril, Comoretto,
  Garcia-Reinaldos, Teyssier, Altmann, Andrae, Audard, Bellas-Velidis, Benson,
  Berthier, Blomme, Burgess, Busso, Carry, Cellino, Clementini, Clotet,
  Creevey, Davidson, Ridder, Delchambre, Dell’Oro, Ducourant,
  Fernández-Hernández, Fouesneau, Frémat, Galluccio, García-Torres,
  González-Núñez, González-Vidal, Gosset, Guy, Halbwachs, Hambly, Harrison,
  Hernández, Hestroffer, Hodgkin, Hutton, Jasniewicz, Jean-Antoine-Piccolo,
  Jordan, Korn, Krone-Martins, Lanzafame, Lebzelter, Löffler, Manteiga,
  Marrese, Martín-Fleitas, Moitinho, Mora, Muinonen, Osinde, Pancino, Pauwels,
  Petit, Recio-Blanco, Richards, Rimoldini, Robin, Sarro, Siopis, Smith,
  Sozzetti, Süveges, Torra, van Reeven, Abbas, Aramburu, Accart, Aerts,
  Altavilla, Álvarez, Alvarez, Alves, Anderson, Andrei, Varela, Antiche,
  Antoja, Arcay, Astraatmadja, Bach, Baker, Balaguer-Núñez, Balm, Barache,
  Barata, Barbato, Barblan, Barklem, Barrado, Barros, Barstow, Muñoz,
  Bassilana, Becciani, Bellazzini, Berihuete, Bertone, Bianchi, Bienaymé,
  Blanco-Cuaresma, Boch, Boeche, Bombrun, Borrachero, Bossini, Bouquillon,
  Bourda, Bragaglia, Bramante, Breddels, Bressan, Brouillet, Brüsemeister,
  Brugaletta, Bucciarelli, Burlacu, Busonero, Butkevich, Buzzi, Caffau,
  Cancelliere, Cannizzaro, Cantat-Gaudin, Carballo, Carlucci, Carrasco,
  Casamiquela, Castellani, Castro-Ginard, Charlot, Chemin, Chiavassa, Cocozza,
  Costigan, Cowell, Crifo, Crosta, Crowley, Cuypers†, Dafonte, Damerdji,
  Dapergolas, David, David, de~Laverny, Luise, March, de~Martino, de~Souza,
  de~Torres, Debosscher, del Pozo, Delbo, Delgado, Delgado, Matteo, Diakite,
  Diener, Distefano, Dolding, Drazinos, Durán, Edvardsson, Enke, Eriksson,
  Esquej, Bontemps, Fabre, Fabrizio, Faigler, Falcão, Casas, Federici,
  Fedorets, Fernique, Figueras, Filippi, Findeisen, Fonti, Fraile, Fraser,
  Frézouls, Gai, Galleti, Garabato, García-Sedano, Garofalo, Garralda, Gavel,
  Gavras, Gerssen, Geyer, Giacobbe, Gilmore, Girona, Giuffrida, Glass, Gomes,
  Granvik, Gueguen, Guerrier, Guiraud, Gutiérrez-Sánchez, Haigron,
  Hatzidimitriou, Hauser, Haywood, Heiter, Helmi, Heu, Hilger, Hobbs, Hofmann,
  Holland, Huckle, Hypki, Icardi, Janßen, de~Fombelle, Jonker, Juhász, Julbe,
  Karampelas, Kewley, Klar, Kochoska, Kohley, Kolenberg, Kontizas, Kontizas,
  Koposov, Kordopatis, Kostrzewa-Rutkowska, Koubsky, Lambert, Lanza, Lasne,
  Lavigne, Fustec, Poncin-Lafitte, Lebreton, Leccia, Leclerc, Lecoeur-Taibi,
  Lenhardt, Leroux, Liao, Licata, Lindstrøm, Lister, Livanou, Lobel, López,
  Managau, Mann, Mantelet, Marchal, Marchant, Marconi, Marinoni, Marschalkó,
  Marshall, Martino, Marton, Mary, Massari, Matijevič, Mazeh, McMillan,
  Messina, Michalik, Millar, Molina, Molinaro, Molnár, Montegriffo, Mor,
  Morbidelli, Morel, Morris, Mulone, Muraveva, Musella, Nelemans, Nicastro,
  Noval, O’Mullane, Ordénovic, Ordóñez-Blanco, Osborne, Pagani, Pagano,
  Pailler, Palacin, Palaversa, Panahi, Pawlak, Piersimoni, Pineau, Plachy,
  Plum, Poggio, Poujoulet, Prša, Pulone, Racero, Ragaini, Rambaux,
  Ramos-Lerate, Regibo, Reylé, Riclet, Ripepi, Riva, Rivard, Rixon, Roegiers,
  Roelens, Romero-Gómez, Rowell, Royer, Ruiz-Dern, Sadowski, Sellés,
  Sahlmann, Salgado, Salguero, Sanna, Santana-Ros, Sarasso, Savietto,
  Schultheis, Sciacca, Segol, Segovia, Ségransan, Shih, Siltala, Silva, Smart,
  Smith, Solano, Solitro, Sordo, Nieto, Souchay, Spagna, Spoto, Stampa, Steele,
  Steidelmüller, Stephenson, Stoev, Suess, Surdej, Szabados, Szegedi-Elek,
  Tapiador, Taris, Tauran, Taylor, Teixeira, Terrett, Teyssandier, Thuillot,
  Titarenko, Clotet, Turon, Ulla, Utrilla, Uzzi, Vaillant, Valentini, Valette,
  van Elteren, Hemelryck, van Leeuwen, Vaschetto, Vecchiato, Veljanoski, Viala,
  Vicente, Vogt, von Essen, Voss, Votruba, Voutsinas, Walmsley, Weiler, Wertz,
  Wevers, Wyrzykowski, Yoldas, Žerjal, Ziaeepour, Zorec, Zschocke, Zucker,
  Zurbach, \& Zwitter}]{Brown2018}
Brown, A. G.~A., Vallenari, A., Prusti, T., {et~al.} 2018, \aap, 616, A1

\bibitem[{Brown {et~al.}(2021)Brown, Vallenari, Prusti, de~Bruijne, Babusiaux,
  Biermann, Creevey, Evans, Eyer, Hutton, Jansen, Jordi, Klioner, Lammers,
  Lindegren, Luri, Mignard, Panem, Pourbaix, Randich, Sartoretti, Soubiran,
  Walton, Arenou, Bailer-Jones, Bastian, Cropper, Drimmel, Katz, Lattanzi, van
  Leeuwen, Bakker, Cacciari, Castañeda, Angeli, Ducourant, Fabricius,
  Fouesneau, Frémat, Guerra, Guerrier, Guiraud, Piccolo, Masana, Messineo,
  Mowlavi, Nicolas, Nienartowicz, Pailler, Panuzzo, Riclet, Roux, Seabroke,
  Sordo, Tanga, Thévenin, Gracia-Abril, Portell, Teyssier, Altmann, Andrae,
  Bellas-Velidis, Benson, Berthier, Blomme, Brugaletta, Burgess, Busso, Carry,
  Cellino, Cheek, Clementini, Damerdji, Davidson, Delchambre, Dell’Oro,
  Fernández-Hernández, Galluccio, García-Lario, Garcia-Reinaldos,
  González-Núñez, Gosset, Haigron, Halbwachs, Hambly, Harrison,
  Hatzidimitriou, Heiter, Hernández, Hestroffer, Hodgkin, Holl, Janßen,
  de~Fombelle, Jordan, Krone-Martins, Lanzafame, Löffler, Lorca, Manteiga,
  Marchal, Marrese, Moitinho, Mora, Muinonen, Osborne, Pancino, Pauwels, Petit,
  Recio-Blanco, Richards, Riello, Rimoldini, Robin, Roegiers, Rybizki, Sarro,
  Siopis, Smith, Sozzetti, Ulla, Utrilla, van Leeuwen, van Reeven, Abbas,
  Aramburu, Accart, Aerts, Aguado, Ajaj, Altavilla, Álvarez, Álvarez
  Cid-Fuentes, Alves, Anderson, Varela, Antoja, Audard, Baines, Baker,
  Balaguer-Núñez, Balbinot, Balog, Barache, Barbato, Barros, Barstow,
  Bartolomé, Bassilana, Bauchet, Baudesson-Stella, Becciani, Bellazzini,
  Bernet, Bertone, Bianchi, Blanco-Cuaresma, Boch, Bombrun, Bossini,
  Bouquillon, Bragaglia, Bramante, Breedt, Bressan, Brouillet, Bucciarelli,
  Burlacu, Busonero, Butkevich, Buzzi, Caffau, Cancelliere, Cánovas,
  Cantat-Gaudin, Carballo, Carlucci, Carnerero, Carrasco, Casamiquela,
  Castellani, Castro-Ginard, Sampol, Chaoul, Charlot, Chemin, Chiavassa, Cioni,
  Comoretto, Cooper, Cornez, Cowell, Crifo, Crosta, Crowley, Dafonte,
  Dapergolas, David, David, de~Laverny, Luise, March, Ridder, de~Souza,
  de~Teodoro, de~Torres, del Peloso, del Pozo, Delbo, Delgado, Delgado,
  Delisle, Matteo, Diakite, Diener, Distefano, Dolding, Eappachen, Edvardsson,
  Enke, Esquej, Fabre, Fabrizio, Faigler, Fedorets, Fernique, Fienga, Figueras,
  Fouron, Fragkoudi, Fraile, Franke, Gai, Garabato, Garcia-Gutierrez,
  García-Torres, Garofalo, Gavras, Gerlach, Geyer, Giacobbe, Gilmore, Girona,
  Giuffrida, Gomel, Gomez, Gonzalez-Santamaria, González-Vidal, Granvik,
  Gutiérrez-Sánchez, Guy, Hauser, Haywood, Helmi, Hidalgo, Hilger, Hładczuk,
  Hobbs, Holland, Huckle, Jasniewicz, Jonker, Campillo, Julbe, Karbevska,
  Kervella, Khanna, Kochoska, Kontizas, Kordopatis, Korn, Kostrzewa-Rutkowska,
  Kruszyńska, Lambert, Lanza, Lasne, Campion, Fustec, Lebreton, Lebzelter,
  Leccia, Leclerc, Lecoeur-Taibi, Liao, Licata, Lindstrøm, Lister, Livanou,
  Lobel, Pardo, Managau, Mann, Marchant, Marconi, Santos, Marinoni, Marocco,
  Marshall, Polo, Martín-Fleitas, Masip, Massari, Mastrobuono-Battisti, Mazeh,
  McMillan, Messina, Michalik, Millar, Mints, Molina, Molinaro, Molnár,
  Montegriffo, Mor, Morbidelli, Morel, Morris, Mulone, Munoz, Muraveva, Murphy,
  Musella, Noval, Ordénovic, Orrù, Osinde, Pagani, Pagano, Palaversa,
  Palicio, Panahi, Pawlak, Esteller, Penttilä, Piersimoni, Pineau, Plachy,
  Plum, Poggio, Poretti, Poujoulet, Prša, Pulone, Racero, Ragaini, Rainer,
  Raiteri, Rambaux, Ramos, Ramos-Lerate, Fiorentin, Regibo, Reylé, Ripepi,
  Riva, Rixon, Robichon, Robin, Roelens, Rohrbasser, Romero-Gómez, Rowell,
  Royer, Rybicki, Sadowski, Sellés, Sahlmann, Salgado, Salguero, Samaras,
  Gimenez, Sanna, Santoveña, Sarasso, Schultheis, Sciacca, Segol, Segovia,
  Ségransan, Semeux, Shahaf, Siddiqui, Siebert, Siltala, Slezak, Smart,
  Solano, Solitro, Souami, Souchay, Spagna, Spoto, Steele, Steidelmüller,
  Stephenson, Süveges, Szabados, Szegedi-Elek, Taris, Tauran, Taylor,
  Teixeira, Thuillot, Tonello, Torra, Torra, Turon, Unger, Vaillant, van
  Dillen, Vanel, Vecchiato, Viala, Vicente, Voutsinas, Weiler, Wevers,
  Wyrzykowski, Yoldas, Yvard, Zhao, Zorec, Zucker, Zurbach, \&
  Zwitter}]{Brown2021}
Brown, A. G.~A., Vallenari, A., Prusti, T., {et~al.} 2021, \aap, 650, C3

\bibitem[{{Bryson} {et~al.}(2020){Bryson}, {Jenkins}, {Klaus}, {Cote},
  {Quintana}, {Campbell}, {Zamudio}, {Chandrasekaran}, {Caldwell}, {Van Cleve},
  \& {Haas}}]{Bryson2020}
{Bryson}, S.~T., {Jenkins}, J.~M., {Klaus}, T.~C., {et~al.} 2020, {Kepler Data
  Processing Handbook: Target and Aperture Definitions: Selecting Pixels for
  Kepler Downlink}, Kepler Science Document KSCI-19081-003

\bibitem[{{Bryson} {et~al.}(2010){Bryson}, {Jenkins}, {Klaus}, {Cote},
  {Quintana}, {Hall}, {Ibrahim}, {Chandrasekaran}, {Caldwell}, {Van Cleve}, \&
  {Haas}}]{Bryson2010}
{Bryson}, S.~T., {Jenkins}, J.~M., {Klaus}, T.~C., {et~al.} 2010, in Society of
  Photo-Optical Instrumentation Engineers (SPIE) Conference Series, Vol. 7740,
  Software and Cyberinfrastructure for Astronomy, ed. N.~M. {Radziwill} \&
  A.~{Bridger}, 77401D

\bibitem[{{Buchhave} {et~al.}(2014){Buchhave}, {Bizzarro}, {Latham},
  {Sasselov}, {Cochran}, {Endl}, {Isaacson}, {Juncher}, \&
  {Marcy}}]{2014Natur.509..593B}
{Buchhave}, L.~A., {Bizzarro}, M., {Latham}, D.~W., {et~al.} 2014, \nat, 509,
  593

\bibitem[{{Carleo} {et~al.}(2021){Carleo}, {Desidera}, {Nardiello},
  {Malavolta}, {Lanza}, {Livingston}, {Locci}, {Marzari}, {Messina}, {Turrini},
  {Baratella}, {Borsa}, {D'Orazi}, {Nascimbeni}, {Pinamonti}, {Rainer}, {Alei},
  {Bignamini}, {Gratton}, {Micela}, {Montalto}, {Sozzetti}, {Squicciarini},
  {Affer}, {Benatti}, {Biazzo}, {Bonomo}, {Claudi}, {Cosentino}, {Covino},
  {Damasso}, {Esposito}, {Fiorenzano}, {Frustagli}, {Giacobbe}, {Harutyunyan},
  {Leto}, {Magazz{\`u}}, {Maggio}, {Mainella}, {Maldonado}, {Mallonn},
  {Mancini}, {Molinari}, {Molinaro}, {Pagano}, {Pedani}, {Piotto}, {Poretti},
  {Redfield}, \& {Scandariato}}]{2021A&A...645A..71C}
{Carleo}, I., {Desidera}, S., {Nardiello}, D., {et~al.} 2021, \aap, 645, A71

\bibitem[{{Casasayas-Barris} {et~al.}(2019){Casasayas-Barris}, {Pall{\'e}},
  {Yan}, {Chen}, {Kohl}, {Stangret}, {Parviainen}, {Helling}, {Watanabe},
  {Czesla}, {Fukui}, {Monta{\~n}{\'e}s-Rodr{\'\i}guez}, {Nagel}, {Narita},
  {Nortmann}, {Nowak}, {Schmitt}, \& {Zapatero Osorio}}]{2019A&A...628A...9C}
{Casasayas-Barris}, N., {Pall{\'e}}, E., {Yan}, F., {et~al.} 2019, \aap, 628,
  A9

\bibitem[{{Castelli} \& {Kurucz}(2003)}]{CastelliKurucz2003}
{Castelli}, F. \& {Kurucz}, R.~L. 2003, in Modelling of Stellar Atmospheres,
  ed. N.~{Piskunov}, W.~W. {Weiss}, \& D.~F. {Gray}, Vol. 210, A20

\bibitem[{{Chontos} {et~al.}(2022){Chontos}, {Murphy}, {MacDougall},
  {Fetherolf}, {Van Zandt}, {Rubenzahl}, {Beard}, {Huber}, {Batalha},
  {Crossfield}, {Dressing}, {Fulton}, {Howard}, {Isaacson}, {Kane}, {Petigura},
  {Robertson}, {Roy}, {Weiss}, {Behmard}, {Dai}, {Dalba}, {Giacalone}, {Hill},
  {Lubin}, {Mayo}, {Mo{\v{c}}nik}, {Polanski}, {Rosenthal}, {Scarsdale},
  {Turtelboom}, {Ricker}, {Vanderspek}, {Latham}, {Seager}, {Winn}, {Jenkins},
  {Quinn}, {Guerrero}, {Collins}, {Ciardi}, {Shporer}, {Goeke}, {Levine},
  {Ting}, {Bieryla}, {Collins}, {Kielkopf}, {Barkaoui}, {Benni},
  {Esparza-Borges}, {Conti}, {Hooton}, {Kagetani}, {Laloum}, {Marino},
  {Massey}, {Murgas}, {Papini}, {Schwarz}, {Srdoc}, {Stockdale}, {Wang},
  {Wittrock}, \& {Zou}}]{2022AJ....163..297C}
{Chontos}, A., {Murphy}, J. M.~A., {MacDougall}, M.~G., {et~al.} 2022, \aj,
  163, 297

\bibitem[{{Claret, A.}(2017)}]{Claret2017}
{Claret, A.} 2017, A\&A, 600, A30

\bibitem[{{Cloutier} {et~al.}(2020){Cloutier}, {Eastman}, {Rodriguez},
  {Astudillo-Defru}, {Bonfils}, {Mortier}, {Watson}, {Stalport}, {Pinamonti},
  {Lienhard}, {Harutyunyan}, {Damasso}, {Latham}, {Collins}, {Massey}, {Irwin},
  {Winters}, {Charbonneau}, {Ziegler}, {Matthews}, {Crossfield}, {Kreidberg},
  {Quinn}, {Ricker}, {Vanderspek}, {Seager}, {Winn}, {Jenkins}, {Vezie},
  {Udry}, {Twicken}, {Tenenbaum}, {Sozzetti}, {S{\'e}gransan}, {Schlieder},
  {Sasselov}, {Santos}, {Rice}, {Rackham}, {Poretti}, {Piotto}, {Phillips},
  {Pepe}, {Molinari}, {Mignon}, {Micela}, {Melo}, {de Medeiros}, {Mayor},
  {Matson}, {Martinez Fiorenzano}, {Mann}, {Magazz{\'u}}, {Lovis},
  {L{\'o}pez-Morales}, {Lopez}, {Lissauer}, {L{\'e}pine}, {Law}, {Kielkopf},
  {Johnson}, {Jensen}, {Howell}, {Gonzales}, {Ghedina}, {Forveille},
  {Figueira}, {Dumusque}, {Dressing}, {Doyon}, {D{\'\i}az}, {Fabrizio},
  {Delfosse}, {Cosentino}, {Conti}, {Collins}, {Cameron}, {Ciardi}, {Caldwell},
  {Burke}, {Buchhave}, {Brice{\~n}o}, {Boyd}, {Bouchy}, {Beichman}, {Artigau},
  \& {Almenara}}]{2020AJ....160....3C}
{Cloutier}, R., {Eastman}, J.~D., {Rodriguez}, J.~E., {et~al.} 2020, \aj, 160,
  3

\bibitem[{{Cosentino} {et~al.}(2012){Cosentino}, {Lovis}, {Pepe}, {Collier
  Cameron}, {Latham}, {Molinari}, {Udry}, {Bezawada}, {Black}, {Born},
  {Buchschacher}, {Charbonneau}, {Figueira}, {Fleury}, {Galli}, {Gallie},
  {Gao}, {Ghedina}, {Gonzalez}, {Gonzalez}, {Guerra}, {Henry}, {Horne},
  {Hughes}, {Kelly}, {Lodi}, {Lunney}, {Maire}, {Mayor}, {Micela}, {Ordway},
  {Peacock}, {Phillips}, {Piotto}, {Pollacco}, {Queloz}, {Rice}, {Riverol},
  {Riverol}, {San Juan}, {Sasselov}, {Segransan}, {Sozzetti}, {Sosnowska},
  {Stobie}, {Szentgyorgyi}, {Vick}, \& {Weber}}]{Cosentino2012}
{Cosentino}, R., {Lovis}, C., {Pepe}, F., {et~al.} 2012, Society of
  Photo-Optical Instrumentation Engineers (SPIE) Conference Series, Vol. 8446,
  {Harps-N: the new planet hunter at TNG}, 84461V

\bibitem[{{Covino} {et~al.}(2013){Covino}, {Esposito}, {Barbieri}, {Mancini},
  {Nascimbeni}, {Claudi}, {Desidera}, {Gratton}, {Lanza}, {Sozzetti}, {Biazzo},
  {Affer}, {Gandolfi}, {Munari}, {Pagano}, {Bonomo}, {Collier Cameron},
  {H{\'e}brard}, {Maggio}, {Messina}, {Micela}, {Molinari}, {Pepe}, {Piotto},
  {Ribas}, {Santos}, {Southworth}, {Shkolnik}, {Triaud}, {Bedin}, {Benatti},
  {Boccato}, {Bonavita}, {Borsa}, {Borsato}, {Brown}, {Carolo}, {Ciceri},
  {Cosentino}, {Damasso}, {Faedi}, {Mart{\'\i}nez Fiorenzano}, {Latham},
  {Lovis}, {Mordasini}, {Nikolov}, {Poretti}, {Rainer}, {Rebolo L{\'o}pez},
  {Scandariato}, {Silvotti}, {Smareglia}, {Alcal{\'a}}, {Cunial}, {Di
  Fabrizio}, {Di Mauro}, {Giacobbe}, {Granata}, {Harutyunyan}, {Knapic},
  {Lattanzi}, {Leto}, {Lodato}, {Malavolta}, {Marzari}, {Molinaro},
  {Nardiello}, {Pedani}, {Prisinzano}, \& {Turrini}}]{Covino2013}
{Covino}, E., {Esposito}, M., {Barbieri}, M., {et~al.} 2013, \aap, 554, A28

\bibitem[{{Cumming} {et~al.}(2008){Cumming}, {Butler}, {Marcy}, {Vogt},
  {Wright}, \& {Fischer}}]{2008PASP..120..531C}
{Cumming}, A., {Butler}, R.~P., {Marcy}, G.~W., {et~al.} 2008, \pasp, 120, 531

\bibitem[{{Cutri} {et~al.}(2021){Cutri}, {Wright}, {Conrow}, {Fowler},
  {Eisenhardt}, {Grillmair}, {Kirkpatrick}, {Masci}, {McCallon}, {Wheelock},
  {Fajardo-Acosta}, {Yan}, {Benford}, {Harbut}, {Jarrett}, {Lake}, {Leisawitz},
  {Ressler}, {Stanford}, {Tsai}, {Liu}, {Helou}, {Mainzer}, {Gettngs},
  {Gonzalez}, {Hoffman}, {Marsh}, {Padgett}, {Skrutskie}, {Beck}, {Papin}, \&
  {Wittman}}]{allwise2013}
{Cutri}, R.~M., {Wright}, E.~L., {Conrow}, T., {et~al.} 2021, VizieR Online
  Data Catalog, II/328

\bibitem[{Delrez {et~al.}(2021)Delrez, Ehrenreich, Alibert, Bonfanti, Borsato,
  Fossati, Hooton, Hoyer, Pozuelos, Salmon, Sulis, Wilson, Adibekyan, Bourrier,
  Brandeker, Charnoz, Deline, Guterman, Haldemann, Hara, Oshagh, Sousa,
  Grootel, Alonso, Anglada-Escud{\'{e}}, B{\'{a}}rczy, Barrado, Barros,
  Baumjohann, Beck, Bekkelien, Benz, Billot, Bonfils, Broeg, Cabrera, Cameron,
  Davies, Deleuil, Delisle, Demangeon, Demory, Erikson, Fortier, Fridlund,
  Futyan, Gandolfi, Mu{\~{n}}oz, Gillon, Guedel, Heng, Kiss, Laskar, des
  Etangs, Lendl, Lovis, Maxted, Nascimbeni, Olofsson, Osborn, Pagano,
  Pall{\'{e}}, Piotto, Pollacco, Queloz, Rauer, Ragazzoni, Ribas, Santos,
  Scandariato, S{\'{e}}gransan, Simon, Smith, Steller, Szab{\'{o}}, Thomas,
  Udry, \& Walton}]{Delrez2021}
Delrez, L., Ehrenreich, D., Alibert, Y., {et~al.} 2021, Nature Astronomy, 5,
  775

\bibitem[{{Diamond-Lowe} {et~al.}(2018){Diamond-Lowe}, {Berta-Thompson},
  {Charbonneau}, \& {Kempton}}]{2018AJ....156...42D}
{Diamond-Lowe}, H., {Berta-Thompson}, Z., {Charbonneau}, D., \& {Kempton}, E.
  M.~R. 2018, \aj, 156, 42

\bibitem[{{Dotter}(2016)}]{2016ApJS..222....8D}
{Dotter}, A. 2016, \apjs, 222, 8

\bibitem[{{Doyle} {et~al.}(2014){Doyle}, {Davies}, {Smalley}, {Chaplin}, \&
  {Elsworth}}]{Doyleetal2014}
{Doyle}, A.~P., {Davies}, G.~R., {Smalley}, B., {Chaplin}, W.~J., \&
  {Elsworth}, Y. 2014, \mnras, 444, 3592

\bibitem[{{Eastman}(2017)}]{2017ascl.soft10003E}
{Eastman}, J. 2017, {EXOFASTv2: Generalized publication-quality exoplanet
  modeling code}

\bibitem[{{Eastman} {et~al.}(2013){Eastman}, {Gaudi}, \& {Agol}}]{Eastman2012}
{Eastman}, J., {Gaudi}, B.~S., \& {Agol}, E. 2013, \pasp, 125, 83

\bibitem[{{Eastman} {et~al.}(2019){Eastman}, {Rodriguez}, {Agol}, {Stassun},
  {Beatty}, {Vanderburg}, {Gaudi}, {Collins}, \& {Luger}}]{2019arXiv190709480E}
{Eastman}, J.~D., {Rodriguez}, J.~E., {Agol}, E., {et~al.} 2019, arXiv
  e-prints, arXiv:1907.09480

\bibitem[{{Espinoza}(2018)}]{Espinoza2018}
{Espinoza}, N. 2018, Research Notes of the American Astronomical Society, 2,
  209

\bibitem[{Espinoza {et~al.}(2016)Espinoza, Brahm, Jord{\'{a}}n, Jenkins, Rojas,
  Jofr{\'{e}}, Mädler, Rabus, Chanam{\'{e}}, Pantoja, Soto, Morzinski, Males,
  Ward-Duong, \& Close}]{Espinoza_2016}
Espinoza, N., Brahm, R., Jord{\'{a}}n, A., {et~al.} 2016, The Astrophysical
  Journal, 830, 43

\bibitem[{Espinoza {et~al.}(2019)Espinoza, Kossakowski, \&
  Brahm}]{Espinoza2019}
Espinoza, N., Kossakowski, D., \& Brahm, R. 2019, Monthly Notices of the Royal
  Astronomical Society, 490

\bibitem[{{Foreman-Mackey} {et~al.}(2017){Foreman-Mackey}, {Agol},
  {Ambikasaran}, \& {Angus}}]{foreman}
{Foreman-Mackey}, D., {Agol}, E., {Ambikasaran}, S., \& {Angus}, R. 2017, \aj,
  154, 220

\bibitem[{{Fressin} {et~al.}(2013){Fressin}, {Torres}, {Charbonneau}, {Bryson},
  {Christiansen}, {Dressing}, {Jenkins}, {Walkowicz}, \&
  {Batalha}}]{2013ApJ...766...81F}
{Fressin}, F., {Torres}, G., {Charbonneau}, D., {et~al.} 2013, \apj, 766, 81

\bibitem[{{Fulton} \& {Petigura}(2018)}]{2018AJ....156..264F}
{Fulton}, B.~J. \& {Petigura}, E.~A. 2018, \aj, 156, 264

\bibitem[{{Fulton} {et~al.}(2018){Fulton}, {Petigura}, {Blunt}, \&
  {Sinukoff}}]{Fulton2018}
{Fulton}, B.~J., {Petigura}, E.~A., {Blunt}, S., \& {Sinukoff}, E. 2018, \pasp,
  130, 044504

\bibitem[{{Gaia Collaboration} {et~al.}(2021){Gaia Collaboration}, {Brown},
  {Vallenari}, {Prusti}, {de Bruijne}, {Babusiaux}, {Biermann}, {Creevey},
  {Evans}, {Eyer}, {Hutton}, {Jansen}, {Jordi}, {Klioner}, {Lammers},
  {Lindegren}, {Luri}, {Mignard}, {Panem}, {Pourbaix}, {Randich}, {Sartoretti},
  {Soubiran}, {Walton}, {Arenou}, {Bailer-Jones}, {Bastian}, {Cropper},
  {Drimmel}, {Katz}, {Lattanzi}, {van Leeuwen}, {Bakker}, {Cacciari},
  {Casta{\~n}eda}, {De Angeli}, {Ducourant}, {Fabricius}, {Fouesneau},
  {Fr{\'e}mat}, {Guerra}, {Guerrier}, {Guiraud}, {Jean-Antoine Piccolo},
  {Masana}, {Messineo}, {Mowlavi}, {Nicolas}, {Nienartowicz}, {Pailler},
  {Panuzzo}, {Riclet}, {Roux}, {Seabroke}, {Sordo}, {Tanga}, {Th{\'e}venin},
  {Gracia-Abril}, {Portell}, {Teyssier}, {Altmann}, {Andrae}, {Bellas-Velidis},
  {Benson}, {Berthier}, {Blomme}, {Brugaletta}, {Burgess}, {Busso}, {Carry},
  {Cellino}, {Cheek}, {Clementini}, {Damerdji}, {Davidson}, {Delchambre},
  {Dell'Oro}, {Fern{\'a}ndez-Hern{\'a}ndez}, {Galluccio}, {Garc{\'\i}a-Lario},
  {Garcia-Reinaldos}, {Gonz{\'a}lez-N{\'u}{\~n}ez}, {Gosset}, {Haigron},
  {Halbwachs}, {Hambly}, {Harrison}, {Hatzidimitriou}, {Heiter},
  {Hern{\'a}ndez}, {Hestroffer}, {Hodgkin}, {Holl}, {Jan{\ss}en}, {Jevardat de
  Fombelle}, {Jordan}, {Krone-Martins}, {Lanzafame}, {L{\"o}ffler}, {Lorca},
  {Manteiga}, {Marchal}, {Marrese}, {Moitinho}, {Mora}, {Muinonen}, {Osborne},
  {Pancino}, {Pauwels}, {Petit}, {Recio-Blanco}, {Richards}, {Riello},
  {Rimoldini}, {Robin}, {Roegiers}, {Rybizki}, {Sarro}, {Siopis}, {Smith},
  {Sozzetti}, {Ulla}, {Utrilla}, {van Leeuwen}, {van Reeven}, {Abbas}, {Abreu
  Aramburu}, {Accart}, {Aerts}, {Aguado}, {Ajaj}, {Altavilla}, {{\'A}lvarez},
  {{\'A}lvarez Cid-Fuentes}, {Alves}, {Anderson}, {Anglada Varela}, {Antoja},
  {Audard}, {Baines}, {Baker}, {Balaguer-N{\'u}{\~n}ez}, {Balbinot}, {Balog},
  {Barache}, {Barbato}, {Barros}, {Barstow}, {Bartolom{\'e}}, {Bassilana},
  {Bauchet}, {Baudesson-Stella}, {Becciani}, {Bellazzini}, {Bernet}, {Bertone},
  {Bianchi}, {Blanco-Cuaresma}, {Boch}, {Bombrun}, {Bossini}, {Bouquillon},
  {Bragaglia}, {Bramante}, {Breedt}, {Bressan}, {Brouillet}, {Bucciarelli},
  {Burlacu}, {Busonero}, {Butkevich}, {Buzzi}, {Caffau}, {Cancelliere},
  {C{\'a}novas}, {Cantat-Gaudin}, {Carballo}, {Carlucci}, {Carnerero},
  {Carrasco}, {Casamiquela}, {Castellani}, {Castro-Ginard}, {Castro Sampol},
  {Chaoul}, {Charlot}, {Chemin}, {Chiavassa}, {Cioni}, {Comoretto}, {Cooper},
  {Cornez}, {Cowell}, {Crifo}, {Crosta}, {Crowley}, {Dafonte}, {Dapergolas},
  {David}, {David}, {de Laverny}, {De Luise}, {De March}, {De Ridder}, {de
  Souza}, {de Teodoro}, {de Torres}, {del Peloso}, {del Pozo}, {Delbo},
  {Delgado}, {Delgado}, {Delisle}, {Di Matteo}, {Diakite}, {Diener},
  {Distefano}, {Dolding}, {Eappachen}, {Edvardsson}, {Enke}, {Esquej}, {Fabre},
  {Fabrizio}, {Faigler}, {Fedorets}, {Fernique}, {Fienga}, {Figueras},
  {Fouron}, {Fragkoudi}, {Fraile}, {Franke}, {Gai}, {Garabato},
  {Garcia-Gutierrez}, {Garc{\'\i}a-Torres}, {Garofalo}, {Gavras}, {Gerlach},
  {Geyer}, {Giacobbe}, {Gilmore}, {Girona}, {Giuffrida}, {Gomel}, {Gomez},
  {Gonzalez-Santamaria}, {Gonz{\'a}lez-Vidal}, {Granvik},
  {Guti{\'e}rrez-S{\'a}nchez}, {Guy}, {Hauser}, {Haywood}, {Helmi}, {Hidalgo},
  {Hilger}, {H{\l}adczuk}, {Hobbs}, {Holland}, {Huckle}, {Jasniewicz},
  {Jonker}, {Juaristi Campillo}, {Julbe}, {Karbevska}, {Kervella}, {Khanna},
  {Kochoska}, {Kontizas}, {Kordopatis}, {Korn}, {Kostrzewa-Rutkowska},
  {Kruszy{\'n}ska}, {Lambert}, {Lanza}, {Lasne}, {Le Campion}, {Le Fustec},
  {Lebreton}, {Lebzelter}, {Leccia}, {Leclerc}, {Lecoeur-Taibi}, {Liao},
  {Licata}, {Lindstr{\o}m}, {Lister}, {Livanou}, {Lobel}, {Madrero Pardo},
  {Managau}, {Mann}, {Marchant}, {Marconi}, {Marcos Santos}, {Marinoni},
  {Marocco}, {Marshall}, {Martin Polo}, {Mart{\'\i}n-Fleitas}, {Masip},
  {Massari}, {Mastrobuono-Battisti}, {Mazeh}, {McMillan}, {Messina},
  {Michalik}, {Millar}, {Mints}, {Molina}, {Molinaro}, {Moln{\'a}r},
  {Montegriffo}, {Mor}, {Morbidelli}, {Morel}, {Morris}, {Mulone}, {Munoz},
  {Muraveva}, {Murphy}, {Musella}, {Noval}, {Ord{\'e}novic}, {Orr{\`u}},
  {Osinde}, {Pagani}, {Pagano}, {Palaversa}, {Palicio}, {Panahi}, {Pawlak},
  {Pe{\~n}alosa Esteller}, {Penttil{\"a}}, {Piersimoni}, {Pineau}, {Plachy},
  {Plum}, {Poggio}, {Poretti}, {Poujoulet}, {Pr{\v{s}}a}, {Pulone}, {Racero},
  {Ragaini}, {Rainer}, {Raiteri}, {Rambaux}, {Ramos}, {Ramos-Lerate}, {Re
  Fiorentin}, {Regibo}, {Reyl{\'e}}, {Ripepi}, {Riva}, {Rixon}, {Robichon},
  {Robin}, {Roelens}, {Rohrbasser}, {Romero-G{\'o}mez}, {Rowell}, {Royer},
  {Rybicki}, {Sadowski}, {Sagrist{\`a} Sell{\'e}s}, {Sahlmann}, {Salgado},
  {Salguero}, {Samaras}, {Sanchez Gimenez}, {Sanna}, {Santove{\~n}a},
  {Sarasso}, {Schultheis}, {Sciacca}, {Segol}, {Segovia}, {S{\'e}gransan},
  {Semeux}, {Shahaf}, {Siddiqui}, {Siebert}, {Siltala}, {Slezak}, {Smart},
  {Solano}, {Solitro}, {Souami}, {Souchay}, {Spagna}, {Spoto}, {Steele},
  {Steidelm{\"u}ller}, {Stephenson}, {S{\"u}veges}, {Szabados}, {Szegedi-Elek},
  {Taris}, {Tauran}, {Taylor}, {Teixeira}, {Thuillot}, {Tonello}, {Torra},
  {Torra}, {Turon}, {Unger}, {Vaillant}, {van Dillen}, {Vanel}, {Vecchiato},
  {Viala}, {Vicente}, {Voutsinas}, {Weiler}, {Wevers}, {Wyrzykowski}, {Yoldas},
  {Yvard}, {Zhao}, {Zorec}, {Zucker}, {Zurbach}, \&
  {Zwitter}}]{2021A&A...649A...1G}
{Gaia Collaboration}, {Brown}, A.~G.~A., {Vallenari}, A., {et~al.} 2021, \aap,
  649, A1

\bibitem[{{Gandolfi} {et~al.}(2018){Gandolfi}, {Barrag{\'a}n}, {Livingston},
  {Fridlund}, {Justesen}, {Redfield}, {Fossati}, {Mathur}, {Grziwa}, {Cabrera},
  {Garc{\'\i}a}, {Persson}, {Van Eylen}, {Hatzes}, {Hidalgo}, {Albrecht},
  {Bugnet}, {Cochran}, {Csizmadia}, {Deeg}, {Eigm{\"u}ller}, {Endl}, {Erikson},
  {Esposito}, {Guenther}, {Korth}, {Luque}, {Monta{\~n}es Rodr{\'\i}guez},
  {Nespral}, {Nowak}, {P{\"a}tzold}, \& {Prieto-Arranz}}]{2018A&A...619L..10G}
{Gandolfi}, D., {Barrag{\'a}n}, O., {Livingston}, J.~H., {et~al.} 2018, \aap,
  619, L10

\bibitem[{{Girardi} {et~al.}(2012){Girardi}, {Barbieri}, {Groenewegen},
  {Marigo}, {Bressan}, {Rocha-Pinto}, {Santiago}, {Camargo}, \& {da
  Costa}}]{girardi12}
{Girardi}, L., {Barbieri}, M., {Groenewegen}, M. A.~T., {et~al.} 2012,
  Astrophysics and Space Science Proceedings, 26, 165

\bibitem[{{Gomes da Silva} {et~al.}(2018){Gomes da Silva}, {Figueira},
  {Santos}, \& {Faria}}]{actin}
{Gomes da Silva}, J., {Figueira}, P., {Santos}, N., \& {Faria}, J. 2018, The
  Journal of Open Source Software, 3, 667

\bibitem[{Gray \& Corbally(2009)}]{Gray2009}
Gray, R.~O. \& Corbally, C. 2009, Stellar Spectral Classification (Princeton
  University Press)

\bibitem[{{Guerrero} {et~al.}(2021){Guerrero}, {Seager}, {Huang}, {Vanderburg},
  {Garcia Soto}, {Mireles}, {Hesse}, {Fong}, {Glidden}, {Shporer}, {Latham},
  {Collins}, {Quinn}, {Burt}, {Dragomir}, {Crossfield}, {Vanderspek},
  {Fausnaugh}, {Burke}, {Ricker}, {Daylan}, {Essack}, {G{\"u}nther}, {Osborn},
  {Pepper}, {Rowden}, {Sha}, {Villanueva}, {Yahalomi}, {Yu}, {Ballard},
  {Batalha}, {Berardo}, {Chontos}, {Dittmann}, {Esquerdo}, {Mikal-Evans},
  {Jayaraman}, {Krishnamurthy}, {Louie}, {Mehrle}, {Niraula}, {Rackham},
  {Rodriguez}, {Rowden}, {Sousa-Silva}, {Watanabe}, {Wong}, {Zhan},
  {Zivanovic}, {Christiansen}, {Ciardi}, {Swain}, {Lund}, {Mullally},
  {Fleming}, {Rodriguez}, {Boyd}, {Quintana}, {Barclay}, {Col{\'o}n},
  {Rinehart}, {Schlieder}, {Clampin}, {Jenkins}, {Twicken}, {Caldwell},
  {Coughlin}, {Henze}, {Lissauer}, {Morris}, {Rose}, {Smith}, {Tenenbaum},
  {Ting}, {Wohler}, {Bakos}, {Bean}, {Berta-Thompson}, {Bieryla}, {Bouma},
  {Buchhave}, {Butler}, {Charbonneau}, {Doty}, {Ge}, {Holman}, {Howard},
  {Kaltenegger}, {Kane}, {Kjeldsen}, {Kreidberg}, {Lin}, {Minsky}, {Narita},
  {Paegert}, {P{\'a}l}, {Palle}, {Sasselov}, {Spencer}, {Sozzetti}, {Stassun},
  {Torres}, {Udry}, \& {Winn}}]{Guerrero2021}
{Guerrero}, N.~M., {Seager}, S., {Huang}, C.~X., {et~al.} 2021, \apjs, 254, 39

\bibitem[{{Henden} {et~al.}(2015){Henden}, {Levine}, {Terrell}, \&
  {Welch}}]{henden2015}
{Henden}, A.~A., {Levine}, S., {Terrell}, D., \& {Welch}, D.~L. 2015, in
  American Astronomical Society Meeting Abstracts, Vol. 225, American
  Astronomical Society Meeting Abstracts \#225, 336.16

\bibitem[{{H{\o}g} {et~al.}(2000){H{\o}g}, {Fabricius}, {Makarov}, {Urban},
  {Corbin}, {Wycoff}, {Bastian}, {Schwekendiek}, \& {Wicenec}}]{hog}
{H{\o}g}, E., {Fabricius}, C., {Makarov}, V.~V., {et~al.} 2000, \aap, 355, L27

\bibitem[{{Hormuth} {et~al.}(2008){Hormuth}, {Brandner}, {Hippler}, \&
  {Henning}}]{hormuth08}
{Hormuth}, F., {Brandner}, W., {Hippler}, S., \& {Henning}, T. 2008, Journal of
  Physics Conference Series, 131, 012051

\bibitem[{{Howard} {et~al.}(2012){Howard}, {Marcy}, {Bryson}, {Jenkins},
  {Rowe}, {Batalha}, {Borucki}, {Koch}, {Dunham}, {Gautier}, {Van Cleve},
  {Cochran}, {Latham}, {Lissauer}, {Torres}, {Brown}, {Gilliland}, {Buchhave},
  {Caldwell}, {Christensen-Dalsgaard}, {Ciardi}, {Fressin}, {Haas}, {Howell},
  {Kjeldsen}, {Seager}, {Rogers}, {Sasselov}, {Steffen}, {Basri},
  {Charbonneau}, {Christiansen}, {Clarke}, {Dupree}, {Fabrycky}, {Fischer},
  {Ford}, {Fortney}, {Tarter}, {Girouard}, {Holman}, {Johnson}, {Klaus},
  {Machalek}, {Moorhead}, {Morehead}, {Ragozzine}, {Tenenbaum}, {Twicken},
  {Quinn}, {Isaacson}, {Shporer}, {Lucas}, {Walkowicz}, {Welsh}, {Boss},
  {Devore}, {Gould}, {Smith}, {Morris}, {Prsa}, {Morton}, {Still}, {Thompson},
  {Mullally}, {Endl}, \& {MacQueen}}]{2012ApJS..201...15H}
{Howard}, A.~W., {Marcy}, G.~W., {Bryson}, S.~T., {et~al.} 2012, \apjs, 201, 15

\bibitem[{{Howard} {et~al.}(2010){Howard}, {Marcy}, {Johnson}, {Fischer},
  {Wright}, {Isaacson}, {Valenti}, {Anderson}, {Lin}, \&
  {Ida}}]{2010Sci...330..653H}
{Howard}, A.~W., {Marcy}, G.~W., {Johnson}, J.~A., {et~al.} 2010, Science, 330,
  653

\bibitem[{{Howell} {et~al.}(2014){Howell}, {Sobeck}, {Haas}, {Still},
  {Barclay}, {Mullally}, {Troeltzsch}, {Aigrain}, {Bryson}, {Caldwell},
  {Chaplin}, {Cochran}, {Huber}, {Marcy}, {Miglio}, {Najita}, {Smith},
  {Twicken}, \& {Fortney}}]{Howell2014}
{Howell}, S.~B., {Sobeck}, C., {Haas}, M., {et~al.} 2014, \pasp, 126, 398

\bibitem[{{Hunter} {et~al.}(2012){Hunter}, {Macgregor}, {Szabo}, {Wellington},
  \& {Bellgard}}]{YABI}
{Hunter}, A., {Macgregor}, A.~B., {Szabo}, T., {Wellington}, C., \& {Bellgard},
  M.~I. 2012, Source Code for Biology and Medicine, 7, 1

\bibitem[{{Jenkins}(2002)}]{jenkins2002}
{Jenkins}, J.~M. 2002, \apj, 575, 493

\bibitem[{{Jenkins} {et~al.}(2010){Jenkins}, {Chandrasekaran}, {McCauliff},
  {Caldwell}, {Tenenbaum}, {Li}, {Klaus}, {Cote}, \& {Middour}}]{jenkins2010}
{Jenkins}, J.~M., {Chandrasekaran}, H., {McCauliff}, S.~D., {et~al.} 2010, in
  Society of Photo-Optical Instrumentation Engineers (SPIE) Conference Series,
  Vol. 7740, Software and Cyberinfrastructure for Astronomy, ed. N.~M.
  {Radziwill} \& A.~{Bridger}, 77400D

\bibitem[{{Jenkins} {et~al.}(2020){Jenkins}, {Tenenbaum}, {Seader}, {Burke},
  {McCauliff}, {Smith}, {Twicken}, \& {Chandrasekaran}}]{Jenkins2020b}
{Jenkins}, J.~M., {Tenenbaum}, P., {Seader}, S., {et~al.} 2020, {Kepler Data
  Processing Handbook: Transiting Planet Search}, Kepler Science Document
  KSCI-19081-003, id. 9. Edited by Jon M. Jenkins.

\bibitem[{{Jenkins} {et~al.}(2016){Jenkins}, {Twicken}, {McCauliff},
  {Campbell}, {Sanderfer}, {Lung}, {Mansouri-Samani}, {Girouard}, {Tenenbaum},
  {Klaus}, {Smith}, {Caldwell}, {Chacon}, {Henze}, {Heiges}, {Latham},
  {Morgan}, {Swade}, {Rinehart}, \& {Vanderspek}}]{Jenkins2016}
{Jenkins}, J.~M., {Twicken}, J.~D., {McCauliff}, S., {et~al.} 2016, in Society
  of Photo-Optical Instrumentation Engineers (SPIE) Conference Series, Vol.
  9913, Software and Cyberinfrastructure for Astronomy IV, ed. G.~{Chiozzi} \&
  J.~C. {Guzman}, 99133E

\bibitem[{Kane {et~al.}(2020)Kane, Yal{\c{c}}{\i}nkaya, Osborn, Dalba, Nielsen,
  Vanderburg, Mo{\v{c}}nik, Hinkel, Ostberg, Esmer, Udry, Fetherolf, Özgür
  Ba{\c{s}}türk, Ricker, Vanderspek, Latham, Seager, Winn, Jenkins, Allart,
  Bailey, Bean, Bouchy, Butler, Campante, Carter, Daylan, Deleuil, Diaz,
  Dumusque, Ehrenreich, Horner, Howard, Isaacson, Jones, Kristiansen, Lovis,
  Marcy, Marmier, O'Toole, Pepe, Ragozzine, S{\'{e}}gransan, Tinney, Turnbull,
  Wittenmyer, Wright, \& Wright}]{Kane2020}
Kane, S.~R., Yal{\c{c}}{\i}nkaya, S., Osborn, H.~P., {et~al.} 2020, The
  Astronomical Journal, 160, 129

\bibitem[{Kass \& Raftery(1995)}]{Kass1995}
Kass, R.~E. \& Raftery, A.~E. 1995, Journal of the American Statistical
  Association, 90, 773

\bibitem[{Kempton {et~al.}(2018)Kempton, Bean, Louie, Deming, Koll, Mansfield,
  Christiansen, López-Morales, Swain, Zellem, Ballard, Barclay, Barstow,
  Batalha, Beatty, Berta-Thompson, Birkby, Buchhave, Charbonneau, Cowan,
  Crossfield, Val-Borro, Doyon, Dragomir, Gaidos, Heng, Hu, Kane, Kreidberg,
  Mallonn, Morley, Narita, Nascimbeni, Pallé, Quintana, Rauscher, Seager,
  Shkolnik, Sing, Sozzetti, Stassun, Valenti, \& Essen}]{Kempton2018}
Kempton, E. M.-R., Bean, J.~L., Louie, D.~R., {et~al.} 2018, Publications of
  the Astronomical Society of the Pacific, 130, 114401

\bibitem[{{Kipping}(2013)}]{Kipping2013}
{Kipping}, D.~M. 2013, \mnras, 435, 2152

\bibitem[{{Knutson} {et~al.}(2014){Knutson}, {Benneke}, {Deming}, \&
  {Homeier}}]{2014Natur.505...66K}
{Knutson}, H.~A., {Benneke}, B., {Deming}, D., \& {Homeier}, D. 2014, \nat,
  505, 66

\bibitem[{{Kreidberg}(2015)}]{Kreidberg2015}
{Kreidberg}, L. 2015, \pasp, 127, 1161

\bibitem[{{Kreidberg} {et~al.}(2014){Kreidberg}, {Bean}, {D{\'e}sert},
  {Benneke}, {Deming}, {Stevenson}, {Seager}, {Berta-Thompson}, {Seifahrt}, \&
  {Homeier}}]{2014Natur.505...69K}
{Kreidberg}, L., {Bean}, J.~L., {D{\'e}sert}, J.-M., {et~al.} 2014, \nat, 505,
  69

\bibitem[{Lacedelli {et~al.}(2020)Lacedelli, Malavolta, Borsato, Piotto,
  Nardiello, Mortier, Stalport, Cameron, Poretti, Buchhave, L{\'{o}
  }pez-Morales, Nascimbeni, Wilson, Udry, Latham, Bonomo, Damasso, Dumusque,
  Jenkins, Lovis, Rice, Sasselov, Winn, Andreuzzi, Cosentino, Charbonneau,
  Fabrizio, Fiorenzano, Ghedina, Harutyunyan, Lienhard, Micela, Molinari,
  Pagano, Pepe, Phillips, Pinamonti, Ricker, Scandariato, Sozzetti, \&
  Watson}]{Lacedelli2020}
Lacedelli, G., Malavolta, L., Borsato, L., {et~al.} 2020, Monthly Notices of
  the Royal Astronomical Society, 501, 4148

\bibitem[{Lacedelli {et~al.}(2022)Lacedelli, Wilson, Malavolta, Hooton,
  Cameron, Alibert, Mortier, Bonfanti, Haywood, Hoyer, Piotto, Bekkelien,
  Vanderburg, Benz, Dumusque, Deline, L{\'{o} }pez-Morales, Borsato, Rice,
  Fossati, Latham, Brandeker, Poretti, Sousa, Sozzetti, Salmon, Burke, Grootel,
  Fausnaugh, Adibekyan, Huang, Osborn, Mustill, Pall{\'{e}}, Bourrier,
  Nascimbeni, Alonso, Anglada, B{\'{a}}rczy, y~Navascues, Barros, Baumjohann,
  Beck, Beck, Billot, Bonfils, Broeg, Buchhave, Cabrera, Charnoz, Cosentino,
  Csizmadia, Davies, Deleuil, Delrez, Demangeon, Demory, Ehrenreich, Erikson,
  Esparza-Borges, Flor{\'{e}}n, Fortier, Fridlund, Futyan, Gandolfi, Ghedina,
  Gillon, Güdel, Guterman, Harutyunyan, Heng, Isaak, Jenkins, Kiss, Laskar,
  des Etangs, Lendl, Lovis, Magrin, Marafatto, Fiorenzano, Maxted, Mayor,
  Micela, Molinari, Murgas, Narita, Olofsson, Ottensamer, Pagano, Pasetti,
  Pedani, Pepe, Peter, Phillips, Pollacco, Queloz, Ragazzoni, Rando, Ratti,
  Rauer, Ribas, Santos, Sasselov, Scandariato, Seager, S{\'{e}}gransan,
  Serrano, Simon, Smith, Steinberger, Steller, Szab{\'{o}}, Thomas, Twicken,
  Udry, Walton, \& Winn}]{Lacedelli2022}
Lacedelli, G., Wilson, T.~G., Malavolta, L., {et~al.} 2022, Monthly Notices of
  the Royal Astronomical Society, 511, 4551

\bibitem[{Li {et~al.}(2019)Li, Tenenbaum, Twicken, Burke, Jenkins, Quintana,
  Rowe, \& Seader}]{Li2019}
Li, J., Tenenbaum, P., Twicken, J.~D., {et~al.} 2019, Publications of the
  Astronomical Society of the Pacific, 131, 1

\bibitem[{{Lightkurve Collaboration} {et~al.}(2018){Lightkurve Collaboration},
  {Cardoso}, {Hedges}, {Gully-Santiago}, {Saunders}, {Cody}, {Barclay}, {Hall},
  {Sagear}, {Turtelboom}, {Zhang}, {Tzanidakis}, {Mighell}, {Coughlin}, {Bell},
  {Berta-Thompson}, {Williams}, {Dotson}, \& {Barentsen}}]{lightkurve}
{Lightkurve Collaboration}, {Cardoso}, J.~V.~d.~M., {Hedges}, C., {et~al.}
  2018, {Lightkurve: Kepler and TESS time series analysis in Python},
  Astrophysics Source Code Library

\bibitem[{{Lillo-Box} {et~al.}(2012){Lillo-Box}, {Barrado}, \&
  {Bouy}}]{lillo-box12}
{Lillo-Box}, J., {Barrado}, D., \& {Bouy}, H. 2012, \aap, 546, A10

\bibitem[{{Lillo-Box} {et~al.}(2014){Lillo-Box}, {Barrado}, \&
  {Bouy}}]{lillo-box14b}
{Lillo-Box}, J., {Barrado}, D., \& {Bouy}, H. 2014, \aap, 566, A103

\bibitem[{{Lind} {et~al.}(2009){Lind}, {Asplund}, \& {Barklem}}]{lindetal2009}
{Lind}, K., {Asplund}, M., \& {Barklem}, P.~S. 2009, \aap, 503, 541L

\bibitem[{Lissauer {et~al.}(2013)Lissauer, Jontof-Hutter, Rowe, Fabrycky,
  Lopez, Agol, Marcy, Deck, Fischer, Fortney, Howell, Isaacson, Jenkins, Kolbl,
  Sasselov, Short, \& Welsh}]{Lissauer2013}
Lissauer, J.~J., Jontof-Hutter, D., Rowe, J.~F., {et~al.} 2013, The
  Astrophysical Journal, 770, 131

\bibitem[{{Lozovsky} {et~al.}(2018){Lozovsky}, {Helled}, {Dorn}, \&
  {Venturini}}]{2018ApJ...866...49L}
{Lozovsky}, M., {Helled}, R., {Dorn}, C., \& {Venturini}, J. 2018, \apj, 866,
  49

\bibitem[{{Maxted} {et~al.}(2011){Maxted}, {Anderson}, {Collier Cameron},
  {Hellier}, {Queloz}, {Smalley}, {Street}, {Triaud}, {West}, {Gillon},
  {Lister}, {Pepe}, {Pollacco}, {S{\'e}gransan}, {Smith}, \&
  {Udry}}]{2011PASP..123..547M}
{Maxted}, P.~F.~L., {Anderson}, D.~R., {Collier Cameron}, A., {et~al.} 2011,
  \pasp, 123, 547

\bibitem[{{Mayor} {et~al.}(2011){Mayor}, {Marmier}, {Lovis}, {Udry},
  {S{\'e}gransan}, {Pepe}, {Benz}, {Bertaux}, {Bouchy}, {Dumusque}, {Lo Curto},
  {Mordasini}, {Queloz}, \& {Santos}}]{2011arXiv1109.2497M}
{Mayor}, M., {Marmier}, M., {Lovis}, C., {et~al.} 2011, arXiv e-prints,
  arXiv:1109.2497

\bibitem[{{Miller-Ricci} {et~al.}(2009){Miller-Ricci}, {Seager}, \&
  {Sasselov}}]{2009ApJ...690.1056M}
{Miller-Ricci}, E., {Seager}, S., \& {Sasselov}, D. 2009, \apj, 690, 1056

\bibitem[{Modirrousta-Galian {et~al.}(2020)Modirrousta-Galian, Locci, \&
  Micela}]{Modirrousta2020}
Modirrousta-Galian, D., Locci, D., \& Micela, G. 2020, The Astrophysical
  Journal, 891, 158

\bibitem[{{Morris} {et~al.}(2020){Morris}, {Twicken}, {Smith}, {Clarke},
  {Jenkins}, {Bryson}, {Girouard}, \& {Klaus}}]{Morris2020}
{Morris}, R.~L., {Twicken}, J.~D., {Smith}, J.~C., {et~al.} 2020, {Kepler Data
  Processing Handbook: Photometric Analysis}, Kepler Science Document
  KSCI-19081-003, id. 6. Edited by Jon M. Jenkins.

\bibitem[{Nardiello {et~al.}(2019)Nardiello, Borsato, Piotto, Colombo,
  Manthopoulou, Bedin, Granata, Lacedelli, Libralato, Malavolta, Montalto, \&
  Nascimbeni}]{Nardiello2019}
Nardiello, D., Borsato, L., Piotto, G., {et~al.} 2019, Monthly Notices of the
  Royal Astronomical Society, 490, 3806

\bibitem[{{Pasquini} {et~al.}(2009){Pasquini}, {Biazzo}, {Bonifacio},
  {Randich}, \& {Bedin}}]{Pasquinietal2008}
{Pasquini}, L., {Biazzo}, K., {Bonifacio}, P., {Randich}, S., \& {Bedin}, L.~R.
  2009, \aap, 489, 677

\bibitem[{{Perger, M.} {et~al.}(2017){Perger, M.}, {Garc\'{\i}a-Piquer, A.},
  {Ribas, I.}, {Morales, J. C.}, {Affer, L.}, {Micela, G.}, {Damasso, M.},
  {Su\'arez-Mascare\~no, A.}, {Gonz\'alez-Hern\'andez, J. I.}, {Rebolo, R.},
  {Herrero, E.}, {Rosich, A.}, {Lafarga, M.}, {Bignamini, A.}, {Sozzetti, A.},
  {Claudi, R.}, {Cosentino, R.}, {Molinari, E.}, {Maldonado, J.}, {Maggio, A.},
  {Lanza, A. F.}, {Poretti, E.}, {Pagano, I.}, {Desidera, S.}, {Gratton, R.},
  {Piotto, G.}, {Bonomo, A. S.}, {Martinez Fiorenzano, A. F.}, {Giacobbe, P.},
  {Malavolta, L.}, {Nascimbeni, V.}, {Rainer, M.}, \& {Scandariato,
  G.}}]{perger2017}
{Perger, M.}, {Garc\'{\i}a-Piquer, A.}, {Ribas, I.}, {et~al.} 2017, A\&A, 598,
  A26

\bibitem[{Perruchot {et~al.}(2008)Perruchot, Kohler, Bouchy, Richaud, Richaud,
  Moreaux, Merzougui, Sottile, Hill, Knispel, Regal, Meunier, Ilovaisky,
  Coroller, Gillet, Schmitt, Pepe, Fleury, Sosnowska, Vors, Mégevand, Blanc,
  Carol, Point, Laloge, \& Brunel}]{Perruchot2008}
Perruchot, S., Kohler, D., Bouchy, F., {et~al.} 2008, in Ground-based and
  Airborne Instrumentation for Astronomy II, ed. I.~S. McLean \& M.~M. Casali,
  Vol. 7014, International Society for Optics and Photonics (SPIE), 235 -- 246

\bibitem[{{Petigura} {et~al.}(2017){Petigura}, {Howard}, {Marcy}, {Johnson},
  {Isaacson}, {Cargile}, {Hebb}, {Fulton}, {Weiss}, {Morton}, {Winn}, {Rogers},
  {Sinukoff}, {Hirsch}, \& {Crossfield}}]{2017AJ....154..107P}
{Petigura}, E.~A., {Howard}, A.~W., {Marcy}, G.~W., {et~al.} 2017, \aj, 154,
  107

\bibitem[{{Petigura} {et~al.}(2013){Petigura}, {Marcy}, \&
  {Howard}}]{2013ApJ...770...69P}
{Petigura}, E.~A., {Marcy}, G.~W., \& {Howard}, A.~W. 2013, \apj, 770, 69

\bibitem[{{Pino} {et~al.}(2020){Pino}, {D{\'e}sert}, {Brogi}, {Malavolta},
  {Wyttenbach}, {Line}, {Hoeijmakers}, {Fossati}, {Bonomo}, {Nascimbeni},
  {Panwar}, {Affer}, {Benatti}, {Biazzo}, {Bignamini}, {Borsa}, {Carleo},
  {Claudi}, {Cosentino}, {Covino}, {Damasso}, {Desidera}, {Giacobbe},
  {Harutyunyan}, {Lanza}, {Leto}, {Maggio}, {Maldonado}, {Mancini}, {Micela},
  {Molinari}, {Pagano}, {Piotto}, {Poretti}, {Rainer}, {Scandariato},
  {Sozzetti}, {Allart}, {Borsato}, {Bruno}, {Di Fabrizio}, {Ehrenreich},
  {Fiorenzano}, {Frustagli}, {Lavie}, {Lovis}, {Magazz{\`u}}, {Nardiello},
  {Pedani}, \& {Smareglia}}]{2020ApJ...894L..27P}
{Pino}, L., {D{\'e}sert}, J.-M., {Brogi}, M., {et~al.} 2020, \apjl, 894, L27

\bibitem[{{Pollacco} {et~al.}(2006){Pollacco}, {Skillen}, {Collier Cameron},
  {Christian}, {Hellier}, {Irwin}, {Lister}, {Street}, {West}, {Anderson},
  {Clarkson}, {Deeg}, {Enoch}, {Evans}, {Fitzsimmons}, {Haswell}, {Hodgkin},
  {Horne}, {Kane}, {Keenan}, {Maxted}, {Norton}, {Osborne}, {Parley}, {Ryans},
  {Smalley}, {Wheatley}, \& {Wilson}}]{2006PASP..118.1407P}
{Pollacco}, D.~L., {Skillen}, I., {Collier Cameron}, A., {et~al.} 2006, \pasp,
  118, 1407

\bibitem[{{Poretti} {et~al.}(2016){Poretti}, {Boccato}, {Claudi}, {Cosentino},
  {Covino}, {Desidera}, {Gratton}, {Lanza}, {Maggio}, {Micela}, {Molinari},
  {Pagano}, {Piotto}, {Smareglia}, {Sozzetti}, \& {GAPS
  Collaboration}}]{Poretti2015}
{Poretti}, E., {Boccato}, C., {Claudi}, R., {et~al.} 2016, \memsai, 87, 141

\bibitem[{Price-Whelan {et~al.}(2018)Price-Whelan, Sipőcz, Günther, Lim,
  Crawford, Conseil, Shupe, Craig, Dencheva, Ginsburg, VanderPlas, Bradley,
  Pérez-Suárez, de~Val-Borro, Aldcroft, Cruz, Robitaille, Tollerud, Ardelean,
  Babej, Bach, Bachetti, Bakanov, Bamford, Barentsen, Barmby, Baumbach, Berry,
  Biscani, Boquien, Bostroem, Bouma, Brammer, Bray, Breytenbach, Buddelmeijer,
  Burke, Calderone, Rodríguez, Cara, Cardoso, Cheedella, Copin, Corrales,
  Crichton, D’Avella, Deil, Depagne, Dietrich, Donath, Droettboom, Earl,
  Erben, Fabbro, Ferreira, Finethy, Fox, Garrison, Gibbons, Goldstein, Gommers,
  Greco, Greenfield, Groener, Grollier, Hagen, Hirst, Homeier, Horton,
  Hosseinzadeh, Hu, Hunkeler, Ivezić, Jain, Jenness, Kanarek, Kendrew, Kern,
  Kerzendorf, Khvalko, King, Kirkby, Kulkarni, Kumar, Lee, Lenz, Littlefair,
  Ma, Macleod, Mastropietro, McCully, Montagnac, Morris, Mueller, Mumford,
  Muna, Murphy, Nelson, Nguyen, Ninan, Nöthe, Ogaz, Oh, Parejko, Parley,
  Pascual, Patil, Patil, Plunkett, Prochaska, Rastogi, Janga, Sabater,
  Sakurikar, Seifert, Sherbert, Sherwood-Taylor, Shih, Sick, Silbiger,
  Singanamalla, Singer, Sladen, Sooley, Sornarajah, Streicher, Teuben, Thomas,
  Tremblay, Turner, Terrón, van Kerkwijk, de~la Vega, Watkins, Weaver,
  Whitmore, Woillez, \& Zabalza}]{Price2018}
Price-Whelan, A.~M., Sipőcz, B.~M., Günther, H.~M., {et~al.} 2018, The
  Astronomical Journal, 156, 123

\bibitem[{Prusti {et~al.}(2016)Prusti, de~Bruijne, Brown, Vallenari, Babusiaux,
  Bailer-Jones, Bastian, Biermann, Evans, Eyer, Jansen, Jordi, Klioner,
  Lammers, Lindegren, Luri, Mignard, Milligan, Panem, Poinsignon, Pourbaix,
  Randich, Sarri, Sartoretti, Siddiqui, Soubiran, Valette, van Leeuwen, Walton,
  Aerts, Arenou, Cropper, Drimmel, Høg, Katz, Lattanzi, O’Mullane, Grebel,
  Holland, Huc, Passot, Bramante, Cacciari, Castañeda, Chaoul, Cheek, Angeli,
  Fabricius, Guerra, Hernández, Jean-Antoine-Piccolo, Masana, Messineo,
  Mowlavi, Nienartowicz, Ordóñez-Blanco, Panuzzo, Portell, Richards, Riello,
  Seabroke, Tanga, Thévenin, Torra, Els, Gracia-Abril, Comoretto,
  Garcia-Reinaldos, Lock, Mercier, Altmann, Andrae, Astraatmadja,
  Bellas-Velidis, Benson, Berthier, Blomme, Busso, Carry, Cellino, Clementini,
  Cowell, Creevey, Cuypers, Davidson, Ridder, de~Torres, Delchambre,
  Dell’Oro, Ducourant, Frémat, García-Torres, Gosset, Halbwachs, Hambly,
  Harrison, Hauser, Hestroffer, Hodgkin, Huckle, Hutton, Jasniewicz, Jordan,
  Kontizas, Korn, Lanzafame, Manteiga, Moitinho, Muinonen, Osinde, Pancino,
  Pauwels, Petit, Recio-Blanco, Robin, Sarro, Siopis, Smith, Smith, Sozzetti,
  Thuillot, van Reeven, Viala, Abbas, Aramburu, Accart, Aguado, Allan, Allasia,
  Altavilla, Álvarez, Alves, Anderson, Andrei, Varela, Antiche, Antoja,
  Antón, Arcay, Atzei, Ayache, Bach, Baker, Balaguer-Núñez, Barache, Barata,
  Barbier, Barblan, Baroni, y~Navascués, Barros, Barstow, Becciani,
  Bellazzini, Bellei, García, Belokurov, Bendjoya, Berihuete, Bianchi,
  Bienaymé, Billebaud, Blagorodnova, Blanco-Cuaresma, Boch, Bombrun,
  Borrachero, Bouquillon, Bourda, Bouy, Bragaglia, Breddels, Brouillet,
  Brüsemeister, Bucciarelli, Budnik, Burgess, Burgon, Burlacu, Busonero,
  Buzzi, Caffau, Cambras, Campbell, Cancelliere, Cantat-Gaudin, Carlucci,
  Carrasco, Castellani, Charlot, Charnas, Charvet, Chassat, Chiavassa, Clotet,
  Cocozza, Collins, Collins, Costigan, Crifo, Cross, Crosta, Crowley, Dafonte,
  Damerdji, Dapergolas, David, David, Cat, de~Felice, de~Laverny, Luise, March,
  de~Martino, de~Souza, Debosscher, del Pozo, Delbo, Delgado, Delgado,
  di~Marco, Matteo, Diakite, Distefano, Dolding, Anjos, Drazinos, Durán,
  Dzigan, Ecale, Edvardsson, Enke, Erdmann, Escolar, Espina, Evans, Bontemps,
  Fabre, Fabrizio, Faigler, Falcão, Casas, Faye, Federici, Fedorets,
  Fernández-Hernández, Fernique, Fienga, Figueras, Filippi, Findeisen, Fonti,
  Fouesneau, Fraile, Fraser, Fuchs, Furnell, Gai, Galleti, Galluccio, Garabato,
  García-Sedano, Garé, Garofalo, Garralda, Gavras, Gerssen, Geyer, Gilmore,
  Girona, Giuffrida, Gomes, González-Marcos, González-Núñez,
  González-Vidal, Granvik, Guerrier, Guillout, Guiraud, Gúrpide,
  Gutiérrez-Sánchez, Guy, Haigron, Hatzidimitriou, Haywood, Heiter, Helmi,
  Hobbs, Hofmann, Holl, Holland, Hunt, Hypki, Icardi, Irwin, de~Fombelle,
  Jofré, Jonker, Jorissen, Julbe, Karampelas, Kochoska, Kohley, Kolenberg,
  Kontizas, Koposov, Kordopatis, Koubsky, Kowalczyk, Krone-Martins,
  Kudryashova, Kull, Bachchan, Lacoste-Seris, Lanza, Lavigne, Poncin-Lafitte,
  Lebreton, Lebzelter, Leccia, Leclerc, Lecoeur-Taibi, Lemaitre, Lenhardt,
  Leroux, Liao, Licata, Lindstrøm, Lister, Livanou, Lobel, Löffler, López,
  Lopez-Lozano, Lorenz, Loureiro, MacDonald, Fernandes, Managau, Mann,
  Mantelet, Marchal, Marchant, Marconi, Marie, Marinoni, Marrese, Marschalkó,
  Marshall, Martín-Fleitas, Martino, Mary, Matijevič, Mazeh, McMillan,
  Messina, Mestre, Michalik, Millar, Miranda, Molina, Molinaro, Molinaro,
  Molnár, Moniez, Montegriffo, Monteiro, Mor, Mora, Morbidelli, Morel,
  Morgenthaler, Morley, Morris, Mulone, Muraveva, Musella, Narbonne, Nelemans,
  Nicastro, Noval, Ordénovic, Ordieres-Meré, Osborne, Pagani, Pagano,
  Pailler, Palacin, Palaversa, Parsons, Paulsen, Pecoraro, Pedrosa,
  Pentikäinen, Pereira, Pichon, Piersimoni, Pineau, Plachy, Plum, Poujoulet,
  Prša, Pulone, Ragaini, Rago, Rambaux, Ramos-Lerate, Ranalli, Rauw, Read,
  Regibo, Renk, Reylé, Ribeiro, Rimoldini, Ripepi, Riva, Rixon, Roelens,
  Romero-Gómez, Rowell, Royer, Rudolph, Ruiz-Dern, Sadowski, Sellés,
  Sahlmann, Salgado, Salguero, Sarasso, Savietto, Schnorhk, Schultheis,
  Sciacca, Segol, Segovia, Segransan, Serpell, Shih, Smareglia, Smart, Smith,
  Solano, Solitro, Sordo, Nieto, Souchay, Spagna, Spoto, Stampa, Steele,
  Steidelmüller, Stephenson, Stoev, Suess, Süveges, Surdej, Szabados,
  Szegedi-Elek, Tapiador, Taris, Tauran, Taylor, Teixeira, Terrett, Tingley,
  Trager, Turon, Ulla, Utrilla, Valentini, van Elteren, Hemelryck, van Leeuwen,
  Varadi, Vecchiato, Veljanoski, Via, Vicente, Vogt, Voss, Votruba, Voutsinas,
  Walmsley, Weiler, Weingrill, Werner, Wevers, Whitehead, Wyrzykowski, Yoldas,
  Žerjal, Zucker, Zurbach, Zwitter, Alecu, Allen, Prieto, Amorim,
  Anglada-Escudé, Arsenijevic, Azaz, Balm, Beck, Bernstein, Bigot, Bijaoui,
  Blasco, Bonfigli, Bono, Boudreault, Bressan, Brown, Brunet, Bunclark,
  Buonanno, Butkevich, Carret, Carrion, Chemin, Chéreau, Corcione, Darmigny,
  de~Boer, de~Teodoro, de~Zeeuw, Luche, Domingues, Dubath, Fodor, Frézouls,
  Fries, Fustes, Fyfe, Gallardo, Gallegos, Gardiol, Gebran, Gomboc, Gómez,
  Grux, Gueguen, Heyrovsky, Hoar, Iannicola, Parache, Janotto, Joliet,
  Jonckheere, Keil, Kim, Klagyivik, Klar, Knude, Kochukhov, Kolka, Kos, Kutka,
  Lainey, LeBouquin, Liu, Loreggia, Makarov, Marseille, Martayan,
  Martinez-Rubi, Massart, Meynadier, Mignot, Munari, Nguyen, Nordlander,
  Ocvirk, O’Flaherty, Sanz, Ortiz, Osorio, Oszkiewicz, Ouzounis, Palmer,
  Park, Pasquato, Peltzer, Peralta, Péturaud, Pieniluoma, Pigozzi, Poels,
  Prat, Prod’homme, Raison, Rebordao, Risquez, Rocca-Volmerange, Rosen,
  Ruiz-Fuertes, Russo, Sembay, Vizcaino, Short, Siebert, Silva, Sinachopoulos,
  Slezak, Soffel, Sosnowska, Straižys, ter Linden, Terrell, Theil, Tiede,
  Troisi, Tsalmantza, Tur, Vaccari, Vachier, Valles, Hamme, Veltz, Virtanen,
  Wallut, Wichmann, Wilkinson, Ziaeepour, \& Zschocke}]{Prusti2016}
Prusti, T., de~Bruijne, J. H.~J., Brown, A. G.~A., {et~al.} 2016, \aap, 595, A1

\bibitem[{{Ramsay} {et~al.}(2021){Ramsay}, {Cirasuolo}, {Amico}, {Bezawada},
  {Caillier}, {Derie}, {Dorn}, {Egner}, {George}, {Gont{\'e}}, {Hammersley},
  {Haupt}, {Ives}, {Jakob}, {Kerber}, {Mainieri}, {Manescau}, {Oberti},
  {Peroux}, {Pfuhl}, {Seemann}, {Siebenmorgen}, {Schmid}, {Vernet}, \& {ESO ELT
  Programme and Follow-Up Team}}]{Ramsay2021}
{Ramsay}, S., {Cirasuolo}, M., {Amico}, P., {et~al.} 2021, The Messenger, 182,
  3

\bibitem[{Ricker {et~al.}(2014)Ricker, Winn, Vanderspek, Latham, Bakos, Bean,
  Berta-Thompson, Brown, Buchhave, Butler, Butler, Chaplin, Charbonneau,
  Christensen-Dalsgaard, Clampin, Deming, Doty, Lee, Dressing, Dunham, Endl,
  Fressin, Ge, Henning, Holman, Howard, Ida, Jenkins, Jernigan, Johnson,
  Kaltenegger, Kawai, Kjeldsen, Laughlin, Levine, Lin, Lissauer, MacQueen,
  Marcy, McCullough, Morton, Narita, Paegert, Palle, Pepe, Pepper, Quirrenbach,
  Rinehart, Sasselov, Sato, Seager, Sozzetti, Stassun, Sullivan, Szentgyorgyi,
  Torres, Udry, \& Villasenor}]{Ricker2014}
Ricker, G.~R., Winn, J.~N., Vanderspek, R., {et~al.} 2014, Journal of
  Astronomical Telescopes, Instruments, and Systems, 1

\bibitem[{Scarsdale {et~al.}(2021)Scarsdale, Murphy, Batalha, Crossfield,
  Dressing, Fulton, Howard, Huber, Isaacson, Kane, Petigura, Robertson, Roy,
  Weiss, Beard, Behmard, Chontos, Christiansen, Ciardi, Claytor, Collins,
  Collins, Dai, Dalba, Dragomir, Fetherolf, Fukui, Giacalone, Gonzales, Hill,
  Hirsch, Jensen, Kosiarek, de~Leon, Lubin, Lund, Luque, Mayo, Mo{\v{c} }nik,
  Mori, Narita, Nowak, Pall{\'{e}}, Rabus, Rosenthal, Rubenzahl, Schlieder,
  Shporer, Stassun, Twicken, Wang, Yahalomi, Jenkins, Latham, Ricker, Seager,
  Vanderspek, \& Winn}]{Scarsdale2021}
Scarsdale, N., Murphy, J. M.~A., Batalha, N.~M., {et~al.} 2021, The
  Astronomical Journal, 162, 215

\bibitem[{{Schlafly} \& {Finkbeiner}(2011)}]{2011ApJ...737..103S}
{Schlafly}, E.~F. \& {Finkbeiner}, D.~P. 2011, \apj, 737, 103

\bibitem[{{Schlegel} {et~al.}(1998){Schlegel}, {Finkbeiner}, \&
  {Davis}}]{1998ApJ...500..525S}
{Schlegel}, D.~J., {Finkbeiner}, D.~P., \& {Davis}, M. 1998, \apj, 500, 525

\bibitem[{{Skidmore} {et~al.}(2015){Skidmore}, {TMT International Science
  Development Teams}, \& {Science Advisory Committee}}]{Skidmore2015}
{Skidmore}, W., {TMT International Science Development Teams}, \& {Science
  Advisory Committee}, T. 2015, Research in Astronomy and Astrophysics, 15,
  1945

\bibitem[{Skrutskie {et~al.}(2006)Skrutskie, Cutri, Stiening, Weinberg,
  Schneider, Carpenter, Beichman, Capps, Chester, Elias, Huchra, Liebert,
  Lonsdale, Monet, Price, Seitzer, Jarrett, Kirkpatrick, Gizis, Howard, Evans,
  Fowler, Fullmer, Hurt, Light, Kopan, Marsh, McCallon, Tam, Dyk, \&
  Wheelock}]{Skrutskie2006}
Skrutskie, M.~F., Cutri, R.~M., Stiening, R., {et~al.} 2006, The Astronomical
  Journal, 131

\bibitem[{Smith {et~al.}(2012)Smith, Stumpe, Cleve, Jenkins, Barclay, Fanelli,
  Girouard, Kolodziejczak, McCauliff, Morris, \& Twicken}]{Smith2012}
Smith, J.~C., Stumpe, M.~C., Cleve, J. E.~V., {et~al.} 2012

\bibitem[{{Sneden}(1973)}]{Sneden1973}
{Sneden}, C. 1973, \aj, 184, 839

\bibitem[{{Snellen} {et~al.}(2010){Snellen}, {de Kok}, {de Mooij}, \&
  {Albrecht}}]{2010Natur.465.1049S}
{Snellen}, I. A.~G., {de Kok}, R.~J., {de Mooij}, E. J.~W., \& {Albrecht}, S.
  2010, \nat, 465, 1049

\bibitem[{Southworth(2010)}]{Southworth2010}
Southworth, J. 2010, Monthly Notices of the Royal Astronomical Society, 408,
  1689

\bibitem[{{Southworth}(2011)}]{2011MNRAS.417.2166S}
{Southworth}, J. 2011, \mnras, 417, 2166

\bibitem[{{Speagle}(2020)}]{Speagle2019}
{Speagle}, J.~S. 2020, \mnras, 493, 3132

\bibitem[{Spearman(1904)}]{Spearman1904}
Spearman, C. 1904, The American Journal of Psychology, 15, 72

\bibitem[{Stassun {et~al.}(2019)Stassun, Oelkers, Paegert, Torres, Pepper,
  De~Lee, Collins, Latham, Muirhead, Chittidi, Rojas-Ayala, Fleming, Rose,
  Tenenbaum, Ting, Kane, Barclay, Bean, Brassuer, Charbonneau, Ge, Lissauer,
  Mann, McLean, Mullally, Narita, Plavchan, Ricker, Sasselov, Seager, Sharma,
  Shiao, Sozzetti, Stello, Vanderspek, Wallace, \& Winn}]{Stassun2019}
Stassun, K.~G., Oelkers, R.~J., Paegert, M., {et~al.} 2019, \aj, 158, 138

\bibitem[{{Stassun} {et~al.}(2018){Stassun}, {Oelkers}, {Pepper}, {Paegert},
  {De Lee}, {Torres}, {Latham}, {Charpinet}, {Dressing}, {Huber}, {Kane},
  {L{\'e}pine}, {Mann}, {Muirhead}, {Rojas-Ayala}, {Silvotti}, {Fleming},
  {Levine}, \& {Plavchan}}]{Stassun2018}
{Stassun}, K.~G., {Oelkers}, R.~J., {Pepper}, J., {et~al.} 2018, \aj, 156, 102

\bibitem[{{Strehl}(1902)}]{strehl1902}
{Strehl}, K. 1902, Astronomische Nachrichten, 158, 89

\bibitem[{{Stumpe} {et~al.}(2014){Stumpe}, {Smith}, {Catanzarite}, {Van Cleve},
  {Jenkins}, {Twicken}, \& {Girouard}}]{Stumpe2014}
{Stumpe}, M.~C., {Smith}, J.~C., {Catanzarite}, J.~H., {et~al.} 2014, \pasp,
  126, 100

\bibitem[{{Stumpe} {et~al.}(2012){Stumpe}, {Smith}, {Van Cleve}, {Twicken},
  {Barclay}, {Fanelli}, {Girouard}, {Jenkins}, {Kolodziejczak}, {McCauliff}, \&
  {Morris}}]{Stumpe2012}
{Stumpe}, M.~C., {Smith}, J.~C., {Van Cleve}, J.~E., {et~al.} 2012, \pasp, 124,
  985

\bibitem[{{Tinetti} {et~al.}(2021){Tinetti}, {Eccleston}, {Haswell}, {Lagage},
  {Leconte}, {L{\"u}ftinger}, {Micela}, {Min}, {Pilbratt}, {Puig}, {Swain},
  {Testi}, {Turrini}, {Vandenbussche}, {Rosa Zapatero Osorio}, {Aret},
  {Beaulieu}, {Buchhave}, {Ferus}, {Griffin}, {Guedel}, {Hartogh}, {Machado},
  {Malaguti}, {Pall{\'e}}, {Rataj}, {Ray}, {Ribas}, {Szab{\'o}}, {Tan},
  {Werner}, {Ratti}, {Scharmberg}, {Salvignol}, {Boudin}, {Halain}, {Haag},
  {Crouzet}, {Kohley}, {Symonds}, {Renk}, {Caldwell}, {Abreu}, {Alonso},
  {Amiaux}, {Berth{\'e}}, {Bishop}, {Bowles}, {Carmona}, {Coffey},
  {Colom{\'e}}, {Crook}, {D{\'e}sjonqueres}, {D{\'\i}az}, {Drummond},
  {Focardi}, {G{\'o}mez}, {Holmes}, {Krijger}, {Kovacs}, {Hunt}, {Machado},
  {Morgante}, {Ollivier}, {Ottensamer}, {Pace}, {Pagano}, {Pascale}, {Pearson},
  {M{\o}ller Pedersen}, {Pniel}, {Roose}, {Savini}, {Stamper}, {Szirovicza},
  {Szoke}, {Tosh}, {Vilardell}, {Barstow}, {Borsato}, {Casewell}, {Changeat},
  {Charnay}, {Civi{\v{s}}}, {Coud{\'e} du Foresto}, {Coustenis}, {Cowan},
  {Danielski}, {Demangeon}, {Drossart}, {Edwards}, {Gilli}, {Encrenaz}, {Kiss},
  {Kokori}, {Ikoma}, {Morales}, {Mendon{\c{c}}a}, {Moneti}, {Mugnai},
  {Garc{\'\i}a Mu{\~n}oz}, {Helled}, {Kama}, {Miguel}, {Nikolaou}, {Pagano},
  {Panic}, {Rengel}, {Rickman}, {Rocchetto}, {Sarkar}, {Selsis}, {Tennyson},
  {Tsiaras}, {Venot}, {Vida}, {Waldmann}, {Yurchenko}, {Szab{\'o}}, {Zellem},
  {Al-Refaie}, {Perez Alvarez}, {Anisman}, {Arhancet}, {Ateca}, {Baeyens},
  {Barnes}, {Bell}, {Benatti}, {Biazzo}, {B{\l}{\k{e}}cka}, {Bonomo}, {Bosch},
  {Bossini}, {Bourgalais}, {Brienza}, {Brucalassi}, {Bruno}, {Caines},
  {Calcutt}, {Campante}, {Canestrari}, {Cann}, {Casali}, {Casas}, {Cassone},
  {Cara}, {Carmona}, {Carone}, {Carrasco}, {Changeat}, {Chioetto},
  {Cortecchia}, {Czupalla}, {Chubb}, {Ciaravella}, {Claret}, {Claudi},
  {Codella}, {Garcia Comas}, {Cracchiolo}, {Cubillos}, {Da Peppo}, {Decin},
  {Dejabrun}, {Delgado-Mena}, {Di Giorgio}, {Diolaiti}, {Dorn}, {Doublier},
  {Doumayrou}, {Dransfield}, {Dumaye}, {Dunford}, {Jimenez Escobar}, {Van
  Eylen}, {Farina}, {Fedele}, {Fern{\'a}ndez}, {Fleury}, {Fonte}, {Fontignie},
  {Fossati}, {Funke}, {Galy}, {Garai}, {Garc{\'\i}a}, {Garc{\'\i}a-Rigo},
  {Garufi}, {Germano Sacco}, {Giacobbe}, {G{\'o}mez}, {Gonzalez},
  {Gonzalez-Galindo}, {Grassi}, {Griffith}, {Guarcello}, {Goujon}, {Gressier},
  {Grzegorczyk}, {Guillot}, {Guilluy}, {Hargrave}, {Hellin}, {Herrero},
  {Hills}, {Horeau}, {Ito}, {Jessen}, {Kabath}, {K{\'a}lm{\'a}n}, {Kawashima},
  {Kimura}, {Kn{\'\i}{\v{z}}ek}, {Kreidberg}, {Kruid}, {Kruijssen},
  {Kubel{\'\i}k}, {Lara}, {Lebonnois}, {Lee}, {Lefevre}, {Lichtenberg},
  {Locci}, {Lombini}, {Sanchez Lopez}, {Lorenzani}, {MacDonald}, {Magrini},
  {Maldonado}, {Marcq}, {Migliorini}, {Modirrousta-Galian}, {Molaverdikhani},
  {Molinari}, {Molli{\`e}re}, {Moreau}, {Morello}, {Morinaud}, {Morvan},
  {Moses}, {Mouzali}, {Nakhjiri}, {Naponiello}, {Narita}, {Nascimbeni},
  {Nikolaou}, {Noce}, {Oliva}, {Palladino}, {Papageorgiou}, {Parmentier},
  {Peres}, {P{\'e}rez}, {Perez-Hoyos}, {Perger}, {Cecchi Pestellini},
  {Petralia}, {Philippon}, {Piccialli}, {Pignatari}, {Piotto}, {Podio},
  {Polenta}, {Preti}, {Pribulla}, {Lopez Puertas}, {Rainer}, {Reess}, {Rimmer},
  {Robert}, {Rosich}, {Rossi}, {Rust}, {Saleh}, {Sanna}, {Schisano},
  {Schreiber}, {Schwartz}, {Scippa}, {Seli}, {Shibata}, {Simpson}, {Shorttle},
  {Skaf}, {Skup}, {Sobiecki}, {Sousa}, {Sozzetti}, {{\v{S}}poner}, {Steiger},
  {Tanga}, {Tackley}, {Taylor}, {Tecza}, {Terenzi}, {Tremblin}, {Tozzi},
  {Triaud}, {Trompet}, {Tsai}, {Tsantaki}, {Valencia}, {Carine Vandaele}, {Van
  der Swaelmen}, {Adibekyan}, {Vasisht}, {Vazan}, {Del Vecchio}, {Waltham},
  {Wawer}, {Widemann}, {Wolkenberg}, {Hou Yip}, {Yung}, {Zilinskas},
  {Zingales}, \& {Zuppella}}]{Tinetti2021}
{Tinetti}, G., {Eccleston}, P., {Haswell}, C., {et~al.} 2021, arXiv e-prints,
  arXiv:2104.04824

\bibitem[{{Trifonov}(2019)}]{Trifonov2019}
{Trifonov}, T. 2019, {The Exo-Striker: Transit and radial velocity interactive
  fitting tool for orbital analysis and N-body simulations}, Astrophysics
  Source Code Library, record ascl:1906.004

\bibitem[{{Turbet} {et~al.}(2020){Turbet}, {Bolmont}, {Ehrenreich}, {Gratier},
  {Leconte}, {Selsis}, {Hara}, \& {Lovis}}]{2020A&A...638A..41T}
{Turbet}, M., {Bolmont}, E., {Ehrenreich}, D., {et~al.} 2020, \aap, 638, A41

\bibitem[{{Twicken} {et~al.}(2018){Twicken}, {Catanzarite}, {Clarke},
  {Girouard}, {Jenkins}, {Klaus}, {Li}, {McCauliff}, {Seader}, {Tenenbaum},
  {Wohler}, {Bryson}, {Burke}, {Caldwell}, {Haas}, {Henze}, \&
  {Sanderfer}}]{Twicken2018}
{Twicken}, J.~D., {Catanzarite}, J.~H., {Clarke}, B.~D., {et~al.} 2018, \pasp,
  130, 064502

\bibitem[{Twicken {et~al.}(2010)Twicken, Clarke, Bryson, Tenenbaum, Wu,
  Jenkins, Girouard, \& Klaus}]{Twicken2010}
Twicken, J.~D., Clarke, B.~D., Bryson, S.~T., {et~al.} 2010, in Software and
  Cyberinfrastructure for Astronomy, ed. N.~M. Radziwill \& A.~Bridger, Vol.
  7740, International Society for Optics and Photonics (SPIE), 749 -- 760

\bibitem[{{Van Eylen} {et~al.}(2018){Van Eylen}, {Agentoft}, {Lundkvist},
  {Kjeldsen}, {Owen}, {Fulton}, {Petigura}, \& {Snellen}}]{Vaneylen2018}
{Van Eylen}, V., {Agentoft}, C., {Lundkvist}, M.~S., {et~al.} 2018, \mnras,
  479, 4786

\bibitem[{Vissapragada {et~al.}(2020)Vissapragada, Jontof-Hutter, Shporer,
  Knutson, Liu, Thorngren, Lee, Chachan, Mawet, Millar-Blanchaer, Nilsson,
  Tinyanont, Vasisht, \& Wright}]{Vissapragada2020}
Vissapragada, S., Jontof-Hutter, D., Shporer, A., {et~al.} 2020, The
  Astronomical Journal, 159, 108

\bibitem[{{Weiss} \& {Marcy}(2014)}]{2014ApJ...783L...6W}
{Weiss}, L.~M. \& {Marcy}, G.~W. 2014, \apjl, 783, L6

\bibitem[{{Winn} \& {Fabrycky}(2015)}]{Winn2015}
{Winn}, J.~N. \& {Fabrycky}, D.~C. 2015, \araa, 53, 409

\bibitem[{{Wright} {et~al.}(2012){Wright}, {Marcy}, {Howard}, {Johnson},
  {Morton}, \& {Fischer}}]{2012ApJ...753..160W}
{Wright}, J.~T., {Marcy}, G.~W., {Howard}, A.~W., {et~al.} 2012, \apj, 753, 160

\bibitem[{{Wyttenbach} {et~al.}(2015){Wyttenbach}, {Ehrenreich}, {Lovis},
  {Udry}, \& {Pepe}}]{2015A&A...577A..62W}
{Wyttenbach}, A., {Ehrenreich}, D., {Lovis}, C., {Udry}, S., \& {Pepe}, F.
  2015, \aap, 577, A62

\bibitem[{{Xie} {et~al.}(2016){Xie}, {Dong}, {Zhu}, {Huber}, {Zheng}, {De Cat},
  {Fu}, {Liu}, {Luo}, {Wu}, {Zhang}, {Zhang}, {Zhou}, {Cao}, {Hou}, {Wang}, \&
  {Zhang}}]{Xie2016}
{Xie}, J.-W., {Dong}, S., {Zhu}, Z., {et~al.} 2016, Proceedings of the National
  Academy of Science, 113, 11431

\bibitem[{{Zeng} {et~al.}(2019){Zeng}, {Jacobsen}, {Sasselov}, {Petaev},
  {Vanderburg}, {Lopez-Morales}, {Perez-Mercader}, {Mattsson}, {Li}, {Heising},
  {Bonomo}, {Damasso}, {Berger}, {Cao}, {Levi}, \&
  {Wordsworth}}]{2019PNAS..116.9723Z}
{Zeng}, L., {Jacobsen}, S.~B., {Sasselov}, D.~D., {et~al.} 2019, Proceedings of
  the National Academy of Science, 116, 9723

\end{thebibliography}

\appendix
\section{HARPS-N RV datapoints}\label{appendix:rv}
\begin{table*}
\centering
\caption[]{HARPS-N RV data points and activity indexes (used in Fig. \ref{fig:GLS_activity}) obtained with the TERRA reduction pipeline between June 8, 2020 and January 21, 2022. The four lines in bold highlight the RV data points that have been removed because they do not fit Chauvenet's criterion.}\label{tab:TERRA}
{\tiny\renewcommand{\arraystretch}{.8}
\resizebox{!}{.4\paperheight}{%
\begin{tabular}{lllllllllllll}
    \toprule
    $\mathrm{BJD_{\textsf{UTC}}}$ & RV & ±$1\upsigma_{\textsf{RV}}$ & FWHM & BIS$^{(\dagger)}$ & Exp.$^{(*)}$ & $\mathrm{S/N}$ & $I_{\rm Ca~{\sc ii}}$ & $I_{\rm H\upalpha06}$ & $I_{\rm He~{\sc i}}$ & $I_{\rm Na~{\sc i}}$ & $I_{\rm Ca~{\sc i}}$ & $I_{\rm H\upalpha16}$ \\ 
    $-2457000\,[days]$ & \multicolumn{2}{c}{$\mathrm{~~[m\,s^{-1}]}$} & --& --& $\mathrm{[sec]}$ & --& --& --& --& --& --& --\rule[-0.8ex]{0pt}{0pt} \\ 
    \hline \\
2008.69349316 & -7.95 & 1.95 & 7174.86 & -7.56 & 900 & 42.8 & 0.549907 & 0.084427 & 0.097028 & 0.180378 & 0.527652 & 0.683496 \\
2009.71160774 & -3.78 & 2.25 & 7181.28 & -6.09 & 900 & 34.3 & 0.555074 & 0.078803 & 0.097526 & 0.179596 & 0.522912 & 0.63963 \\
2026.66187880 & -4.51 & 2.31 & 7164.91 & 0.894 & 900 & 44.2 & 0.554022 & 0.082134 & 0.100171 & 0.187731 & 0.505988 & 0.687073 \\
2027.72062839 & 4.586 & 2.2 & 7153.54 & -4.7 & 900 & 36.4 & 0.558749 & 0.083288 & 0.107005 & 0.187897 & 0.530667 & 0.687811 \\
2028.69301741 & 3.085 & 2.06 & 7178.27 & -6.91 & 900 & 39.3 & 0.587418 & 0.078941 & 0.098429 & 0.181754 & 0.53219 & 0.657997 \\
2037.70925340 & -1.04 & 2.78 & 7190.11 & 1.317 & 900 & 27.1 & 0.588273 & 0.087717 & 0.10143 & 0.181427 & 0.533928 & 0.670313 \\
2038.71004788 & -6.86 & 2.35 & 7174.7 & -12.8 & 900 & 33.8 & 0.589738 & 0.08641 & 0.102117 & 0.18419 & 0.531931 & 0.671326 \\
2039.71645615 & -2.51 & 2.17 & 7176.79 & -5.84 & 900 & 40.5 & 0.590585 & 0.078021 & 0.099401 & 0.182083 & 0.530063 & 0.658114 \\
2040.70404346 & -0.98 & 2.65 & 7182.67 & -18.1 & 900 & 31.6 & 0.593109 & 0.085039 & 0.099033 & 0.182093 & 0.528559 & 0.650582 \\
2050.69789104 & -1.68 & 1.94 & 7172.47 & -1.01 & 900 & 40.2 & 0.594075 & 0.087503 & 0.099635 & 0.183098 & 0.534144 & 0.666092 \\
2051.69861363 & -4.63 & 1.99 & 7177.72 & -11.5 & 900 & 38.6 & 0.591485 & 0.076251 & 0.099171 & 0.182165 & 0.525447 & 0.654697 \\
2054.72404480 & -1.22 & 3.15 & 7162.98 & -4.59 & 900 & 30.1 & 0.564703 & 0.094168 & 0.100699 & 0.182831 & 0.523618 & 0.663779 \\
2068.61577511 & -3.21 & 6.73 & 7165.8 & -6.58 & 900 & 12.7 & 0.584758 & 0.089195 & 0.098718 & 0.179674 & 0.52911 & 0.641145 \\
2069.66437348 & 1.928 & 1.46 & 7167.86 & -1.31 & 900 & 52.5 & 0.562065 & 0.080459 & 0.103277 & 0.183022 & 0.530759 & 0.690607 \\
2070.70250746 & -3.5 & 1.57 & 7172.32 & -3.47 & 900 & 47.8 & 0.561804 & 0.078625 & 0.09964 & 0.18108 & 0.529827 & 0.668034 \\
2071.70780293 & -0.88 & 2.12 & 7195.44 & -4.27 & 900 & 39.4 & 0.589073 & 0.083154 & 0.099452 & 0.180534 & 0.52584 & 0.65915 \\
2072.70309705 & -2.2 & 1.86 & 7185.56 & -14.3 & 900 & 42.3 & 0.557986 & 0.082279 & 0.098576 & 0.180502 & 0.525855 & 0.654666 \\
2075.63970105 & -3.37 & 1.9 & 7169.71 & 0.109 & 900 & 45.6 & 0.594915 & 0.083487 & 0.100564 & 0.181682 & 0.527055 & 0.685465 \\
2076.71092255 & -6.68 & 2.31 & 7174.73 & -9.8 & 900 & 33.1 & 0.554061 & 0.07931 & 0.103664 & 0.184892 & 0.530864 & 0.665231 \\
2078.68129258 & -6.1 & 1.94 & 7175.62 & -16.9 & 900 & 44.2 & 0.588419 & 0.082137 & 0.100369 & 0.183009 & 0.534477 & 0.67444 \\
2079.69077171 & -4.53 & 1.95 & 7176.88 & -8.31 & 900 & 45.3 & 0.560252 & 0.079819 & 0.100953 & 0.183346 & 0.530948 & 0.673179 \\
2091.68437135 & 8.017 & 1.77 & 7185.69 & -8.84 & 900 & 41.1 & 0.549246 & 0.082682 & 0.099392 & 0.181514 & 0.526944 & 0.682402 \\
2092.62561704 & -1.48 & 2.43 & 7172.71 & -7.71 & 900 & 34.1 & 0.581979 & 0.086567 & 0.100783 & 0.183421 & 0.530712 & 0.666342 \\
2093.65087003 & 7.487 & 7.04 & 7164.43 & -16.6 & 900 & 13.4 & 0.564012 & 0.107488 & 0.101191 & 0.188208 & 0.50388 & 0.653611 \\
2094.60798244 & -1.1 & 1.7 & 7168.72 & -6.22 & 900 & 42.4 & 0.5533 & 0.084541 & 0.100049 & 0.184004 & 0.527844 & 0.70821 \\
2095.61169299 & -0.2 & 2.3 & 7152.3 & 3.952 & 900 & 36.4 & 0.585249 & 0.077758 & 0.105133 & 0.187703 & 0.525175 & 0.697218 \\
2096.62962752 & -0.29 & 1.86 & 7166.3 & -11.5 & 900 & 46.4 & 0.541437 & 0.080363 & 0.09865 & 0.182509 & 0.530522 & 0.677679 \\
2097.66412402 & -2.95 & 2.64 & 7180.27 & -4.25 & 900 & 33.7 & 0.55499 & 0.084606 & 0.102655 & 0.182256 & 0.527347 & 0.650539 \\
2099.63975353 & -8.99 & 2.53 & 7178.63 & -1 & 900 & 32.3 & 0.511441 & 0.083238 & 0.09935 & 0.183486 & 0.53274 & 0.664088 \\
2106.65854394 & -2.12 & 2.29 & 7175.27 & -17.8 & 900 & 34.3 & 0.542303 & 0.089511 & 0.101622 & 0.184597 & 0.529016 & 0.670419 \\
2110.71566068 & -6.18 & 3.41 & 7186.25 & 5.57 & 900 & 25.3 & 0.512902 & 0.082166 & 0.101099 & 0.184164 & 0.530842 & 0.67291 \\
2111.55269319 & -6.16 & 1.92 & 7162.99 & -2.12 & 900 & 41 & 0.525014 & 0.080415 & 0.100104 & 0.185605 & 0.529595 & 0.68975 \\
2112.58796961 & -7.39 & 1.88 & 7169.58 & -11.6 & 900 & 40 & 0.534984 & 0.083924 & 0.099656 & 0.182888 & 0.53337 & 0.680635 \\
2119.69014174 & 11.98 & 2.85 & 7190.32 & -9.13 & 900 & 25.8 & 0.545065 & 0.087481 & 0.098395 & 0.182445 & 0.534123 & 0.626241 \\
2120.67679558 & -2.05 & 2.87 & 7208.68 & -6.74 & 900 & 27.8 & 0.528515 & 0.086578 & 0.102032 & 0.186926 & 0.532501 & 0.627348 \\
2125.54442015 & -5.53 & 3.13 & 7169.97 & -3.52 & 900 & 28.1 & 0.555249 & 0.085601 & 0.101368 & 0.186582 & 0.523009 & 0.665638 \\
2126.56835804 & -6.72 & 2.67 & 7174.3 & -16.4 & 1200 & 32.1 & 0.567202 & 0.083452 & 0.099046 & 0.183561 & 0.527745 & 0.651399 \\
2127.63415799 & 0.931 & 2.18 & 7180.78 & -4.92 & 900 & 35.3 & 0.539365 & 0.085257 & 0.098857 & 0.18364 & 0.524582 & 0.655533 \\
2130.59628332 & -1.27 & 3.36 & 7187.06 & -7.56 & 900 & 23.7 & 0.576932 & 0.083983 & 0.098713 & 0.183139 & 0.531644 & 0.651165 \\
2134.62011272 & 0.717 & 3.58 & 7200.6 & -10.9 & 900 & 22.7 & 0.580358 & 0.091434 & 0.100845 & 0.184286 & 0.525551 & 0.661751 \\
2137.55086797 & -6.08 & 1.75 & 7184.04 & -1.32 & 900 & 39.9 & 0.587977 & 0.087623 & 0.097656 & 0.18263 & 0.528299 & 0.65772 \\
2153.56077137 & -4.39 & 7.54 & 7144.15 & -24 & 900 & 14.5 & 0.561764 & 0.077853 & 0.102889 & 0.189554 & 0.527182 & 0.670189 \\
2156.53992496 & -0.87 & 2.45 & 7194.74 & -1.33 & 900 & 34.9 & 0.596947 & 0.089431 & 0.103586 & 0.187334 & 0.531839 & 0.639791 \\
\textbf{2157.58172574} & \textbf{15.06} & \textbf{6.49} & \textbf{7189.21} & \textbf{8.093} & \textbf{900} & \textbf{15.9} & \textbf{0.554652} & \textbf{0.093102} & \textbf{0.098102} & \textbf{0.18062} & \textbf{0.525648} & \textbf{0.651384} \\
2169.31333487 & 0.713 & 5.22 & 7202.4 & -9.03 & 900 & 19.2 & 0.569926 & 0.086208 & 0.102926 & 0.186679 & 0.532988 & 0.706365 \\
2170.34375539 & -3.22 & 2.75 & 7181.57 & -5.96 & 900 & 26.3 & 0.573886 & 0.092071 & 0.106226 & 0.189905 & 0.524569 & 0.677237 \\
2171.31709456 & 5.22 & 2.1 & 7199.72 & -2.35 & 900 & 40 & 0.567555 & 0.080705 & 0.097921 & 0.182549 & 0.526484 & 0.643617 \\
2172.31291994 & -1.16 & 1.78 & 7183.03 & -13.1 & 900 & 51.7 & 0.602331 & 0.082135 & 0.100467 & 0.184348 & 0.521872 & 0.655749 \\
2189.39389821 & -5.12 & 2.18 & 7187.35 & 0.742 & 900 & 34.9 & 0.553612 & 0.087636 & 0.106345 & 0.186232 & 0.524683 & 0.636639 \\
2190.35888875 & -5.56 & 2.1 & 7181.28 & -1.38 & 900 & 37.8 & 0.553642 & 0.08599 & 0.10436 & 0.185399 & 0.52397 & 0.649955 \\
2192.33613311 & -1.58 & 1.8 & 7190.22 & -4.64 & 900 & 41.2 & 0.594328 & 0.086471 & 0.099245 & 0.182607 & 0.521139 & 0.661821 \\
2212.37354384 & 7.092 & 3.59 & 7203.08 & -9.36 & 900 & 23.9 & 0.593524 & 0.09262 & 0.102766 & 0.183774 & 0.521448 & 0.630614 \\
\textbf{2213.42063681} & \textbf{16.78} & \textbf{4.42} & \textbf{7211.7} & \textbf{3.168} & \textbf{900} & \textbf{19} & \textbf{0.596428} & \textbf{0.099179} & \textbf{0.101745} & \textbf{0.181668} & \textbf{0.520206} & \textbf{0.654342} \\
2216.40383735 & -8.94 & 2.19 & 7203.82 & -10.3 & 900 & 33.9 & 0.594565 & 0.089359 & 0.100295 & 0.182115 & 0.519823 & 0.629985 \\
2235.36719701 & 6.639 & 2.74 & 7177.77 & 0.334 & 900 & 25.5 & 0.595733 & 0.091527 & 0.101648 & 0.182908 & 0.520753 & 0.626604 \\
2236.31768047 & 5.56 & 2.2 & 7188.86 & -8.92 & 900 & 38.7 & 0.567033 & 0.087864 & 0.100285 & 0.182334 & 0.518575 & 0.610068 \\
2237.32364626 & 4.933 & 3.21 & 7181.9 & -17.9 & 1200 & 28.2 & 0.559051 & 0.086238 & 0.104913 & 0.186654 & 0.525845 & 0.664309 \\
\textbf{2239.31266431} & \textbf{21.6} & \textbf{2.85} & \textbf{7219.49} & \textbf{1.592} & \textbf{900} & \textbf{31.5} & \textbf{0.591982} & \textbf{0.088633} & \textbf{0.099598} & \textbf{0.180553} & \textbf{0.524177} & \textbf{0.630278} \\
2240.31312271 & 7.368 & 2.31 & 7208.34 & -2.51 & 900 & 37.6 & 0.594946 & 0.081703 & 0.10278 & 0.183971 & 0.522467 & 0.662409 \\
2244.31706817 & 6.052 & 2.84 & 7195.35 & 0.352 & 900 & 32.5 & 0.594812 & 0.084492 & 0.100398 & 0.18777 & 0.524697 & 0.655945 \\
2245.31980936 & 0.237 & 2.24 & 7209.07 & -5.47 & 900 & 34.9 & 0.589965 & 0.087524 & 0.103149 & 0.183372 & 0.519622 & 0.663637 \\
\textbf{2246.32103509} & \textbf{27.1} & \textbf{7.78} & \textbf{7248.46} & \textbf{34.51} & \textbf{1200} & \textbf{12.7} & \textbf{0.594927} & \textbf{0.090211} & \textbf{0.100742} & \textbf{0.184404} & \textbf{0.521228} & \textbf{0.74003} \\
2412.68868670 & -3.7 & 2.54 & 7163.71 & -3.37 & 900 & 32.8 & 0.550281 & 0.090724 & 0.100293 & 0.183955 & 0.523314 & 0.6709 \\
2413.66378329 & -1.5 & 2.16 & 7179.72 & -6.74 & 900 & 34.4 & 0.558588 & 0.088191 & 0.099293 & 0.1792 & 0.526702 & 0.652927 \\
2414.68886021 & 2.937 & 2.15 & 7166.9 & 4.331 & 900 & 37.1 & 0.555402 & 0.084723 & 0.098655 & 0.183102 & 0.530929 & 0.676626 \\
2416.63114629 & 2.289 & 1.95 & 7191.8 & -11.7 & 900 & 48 & 0.584258 & 0.083699 & 0.099042 & 0.180173 & 0.526418 & 0.68478 \\
2417.65137234 & 3.912 & 1.73 & 7191.95 & -8.14 & 900 & 40.8 & 0.583166 & 0.087257 & 0.098779 & 0.18156 & 0.52857 & 0.664774 \\
2418.65486069 & 4.286 & 3.33 & 7195.12 & -13.2 & 900 & 21.6 & 0.587824 & 0.10659 & 0.10043 & 0.180994 & 0.498526 & 0.662161 \\
2427.73181821 & 1.618 & 2.29 & 7191.43 & -0.36 & 900 & 34.2 & 0.595541 & 0.089463 & 0.100339 & 0.183087 & 0.530979 & 0.649579 \\
2428.66018094 & 0.03 & 3.07 & 7187.2 & -11.9 & 900 & 28.1 & 0.587265 & 0.089537 & 0.098082 & 0.181868 & 0.526651 & 0.668995 \\
2430.66988978 & -0.96 & 2.26 & 7186.32 & -3.5 & 900 & 36 & 0.558113 & 0.085708 & 0.102817 & 0.184679 & 0.529491 & 0.668573 \\
2431.71360583 & -5.64 & 2.64 & 7190.54 & -22.7 & 900 & 32.1 & 0.550955 & 0.07889 & 0.100986 & 0.181203 & 0.529026 & 0.662741 \\
2443.66087057 & 8.667 & 3.5 & 7188.34 & -4.35 & 900 & 25.2 & 0.590039 & 0.092999 & 0.096614 & 0.179042 & 0.526005 & 0.696157 \\
2444.57982389 & 9.299 & 3.32 & 7180.37 & -2.39 & 900 & 26.8 & 0.589859 & 0.091057 & 0.098188 & 0.179557 & 0.522807 & 0.666047 \\
2445.59632825 & 10.81 & 1.86 & 7178.02 & -8.87 & 900 & 36.9 & 0.557645 & 0.085087 & 0.10075 & 0.181234 & 0.52886 & 0.678235 \\
2446.60873425 & 5.874 & 1.87 & 7182.03 & 2.719 & 900 & 45 & 0.552972 & 0.085681 & 0.101935 & 0.183638 & 0.525628 & 0.682197 \\
2447.61736597 & 1.855 & 2.1 & 7181.21 & -2 & 900 & 39.9 & 0.585974 & 0.086557 & 0.101268 & 0.182021 & 0.529423 & 0.6903 \\
2448.59166621 & 2.524 & 1.79 & 7183.26 & -6.53 & 900 & 46.5 & 0.550409 & 0.081722 & 0.099201 & 0.181717 & 0.528627 & 0.676518 \\
2449.61666267 & -2.34 & 1.85 & 7181.85 & -9.02 & 900 & 46.7 & 0.55726 & 0.081361 & 0.100242 & 0.183494 & 0.527072 & 0.693437 \\
2453.58138876 & -5.57 & 2.17 & 7179.71 & -10.5 & 900 & 38.1 & 0.576 & 0.087524 & 0.101564 & 0.182982 & 0.529699 & 0.666481 \\
2454.66732062 & -2.11 & 5 & 7157.53 & -13.6 & 900 & 16.4 & 0.534792 & 0.105417 & 0.106316 & 0.188685 & 0.531533 & 0.672621 \\
2455.68136136 & 1.242 & 2.34 & 7201.87 & -15 & 900 & 37.6 & 0.559782 & 0.085698 & 0.097922 & 0.182038 & 0.527513 & 0.642527 \\
2456.70358427 & 1.194 & 1.71 & 7193.3 & -2.59 & 900 & 39.1 & 0.539261 & 0.083335 & 0.100678 & 0.18192 & 0.526882 & 0.654302 \\
2457.65444875 & 0.232 & 1.53 & 7186.93 & -4.04 & 900 & 45.9 & 0.544456 & 0.082158 & 0.103781 & 0.185289 & 0.52979 & 0.678245 \\
2458.62416725 & 1.797 & 2.1 & 7191.62 & -2.73 & 900 & 42.5 & 0.571849 & 0.083688 & 0.10082 & 0.181891 & 0.525159 & 0.670374 \\
2459.58868774 & 0 & 2.42 & 7166.31 & -9.47 & 900 & 32 & 0.534143 & 0.084143 & 0.101018 & 0.183983 & 0.531303 & 0.671843 \\
2460.68008744 & 2.859 & 2 & 7199.43 & -10 & 900 & 39 & 0.567005 & 0.085291 & 0.09952 & 0.182178 & 0.53004 & 0.641539 \\
2461.61247483 & -3.43 & 2.01 & 7177.71 & -14.6 & 900 & 39.8 & 0.531283 & 0.081882 & 0.101482 & 0.18371 & 0.532312 & 0.662684 \\
2462.59699290 & -1.49 & 2.41 & 7193.46 & -10.2 & 900 & 34 & 0.53251 & 0.091757 & 0.1007 & 0.182059 & 0.53047 & 0.678952 \\
2464.59087617 & 5.604 & 5.45 & 7198.52 & -24.4 & 900 & 17.8 & 0.536789 & 0.095036 & 0.099161 & 0.183787 & 0.528055 & 0.686225 \\
2465.55285533 & 1.953 & 1.95 & 7189.61 & -10.6 & 900 & 37.5 & 0.525385 & 0.081987 & 0.102391 & 0.184078 & 0.528833 & 0.659363 \\
2472.64437339 & 9.088 & 3.09 & 7206.96 & -15.6 & 900 & 28.8 & 0.582423 & 0.088182 & 0.102267 & 0.18471 & 0.526636 & 0.671635 \\
2473.58211801 & -2.46 & 3.71 & 7182.04 & -22.5 & 900 & 24 & 0.548713 & 0.090743 & 0.100974 & 0.184532 & 0.522132 & 0.670676 \\
2475.57083577 & 3.505 & 2.35 & 7184.54 & -16.3 & 1800 & 32.2 & 0.580897 & 0.092011 & 0.101234 & 0.182594 & 0.528772 & 0.656172 \\
2476.56282582 & 7.827 & 1.91 & 7191.68 & -6.25 & 900 & 44 & 0.556952 & 0.083328 & 0.09885 & 0.181388 & 0.523729 & 0.660628 \\
2477.54247641 & 5.777 & 1.95 & 7190.45 & -13.2 & 900 & 40.1 & 0.584375 & 0.084401 & 0.105974 & 0.188686 & 0.528026 & 0.684554 \\
2478.53938306 & 3.887 & 2.42 & 7199.21 & 1.209 & 900 & 33.4 & 0.542519 & 0.091376 & 0.099644 & 0.180595 & 0.522094 & 0.662569 \\
2479.53060547 & 0.796 & 2.4 & 7200.54 & -6.9 & 900 & 35.5 & 0.578876 & 0.086324 & 0.099936 & 0.181075 & 0.527574 & 0.654195 \\
2481.49382164 & 3.548 & 2.82 & 7186.71 & -6.58 & 900 & 31.2 & 0.53899 & 0.091801 & 0.101227 & 0.183502 & 0.527366 & 0.676158 \\
2513.44982261 & 1.106 & 2.07 & 7176.01 & -3.44 & 900 & 41.6 & 0.561363 & 0.088328 & 0.100405 & 0.183709 & 0.523996 & 0.678603 \\
2513.50218232 & -4.57 & 2.69 & 7172.76 & -4.86 & 900 & 33.2 & 0.566953 & 0.08974 & 0.10596 & 0.188059 & 0.524511 & 0.666513 \\
2515.41917706 & -2.34 & 1.93 & 7184.92 & -5.97 & 900 & 41.7 & 0.564153 & 0.089897 & 0.101247 & 0.182411 & 0.526736 & 0.673459 \\
2515.44453511 & -3.02 & 1.76 & 7178.72 & -6.73 & 900 & 49.2 & 0.567076 & 0.088259 & 0.098628 & 0.181831 & 0.520853 & 0.679953 \\
2516.53077561 & -4.93 & 2.92 & 7190.8 & 1.972 & 900 & 31.9 & 0.602149 & 0.092247 & 0.099896 & 0.182444 & 0.527717 & 0.666103 \\
2565.36649887 & 0.845 & 1.64 & 7187.23 & -3.3 & 900 & 54.5 & 0.596593 & 0.088658 & 0.09841 & 0.180042 & 0.522949 & 0.675468 \\
2566.34542939 & -5.47 & 2.15 & 7179.09 & -6.14 & 900 & 45.4 & 0.601739 & 0.091592 & 0.099575 & 0.18055 & 0.525212 & 0.678527 \\
2575.40497038 & 3.93 & 2.12 & 7181.99 & -8.36 & 900 & 40.1 & 0.566264 & 0.087701 & 0.098051 & 0.182169 & 0.523518 & 0.671318 \\
2579.38524573 & 5.059 & 2.18 & 7191.98 & -5.62 & 900 & 38.3 & 0.562049 & 0.089366 & 0.104764 & 0.183719 & 0.525329 & 0.674917 \\
2580.41266409 & 10.09 & 5.97 & 7187.22 & 4.876 & 900 & 19.9 & 0.572011 & 0.107006 & 0.107628 & 0.185394 & 0.52275 & 0.687077 \\
2584.33980519 & 0.821 & 2.95 & 7183.34 & 0.433 & 900 & 29.2 & 0.56581 & 0.092026 & 0.104575 & 0.185035 & 0.514433 & 0.687819 \\
2588.34483394 & 1.991 & 5.05 & 7181.58 & 9.001 & 900 & 19.2 & 0.560173 & 0.094073 & 0.110324 & 0.186892 & 0.524911 & 0.680451 \\
2601.31675212 & 5.918 & 2.03 & 7196.44 & -6.76 & 900 & 45.2 & 0.565656 & 0.087291 & 0.10094 & 0.17769 & 0.523886 & 0.667702 \rule[-0.8ex]{0pt}{0pt} \\

    \bottomrule
\end{tabular}}}
\tablefoot{$^{(*)}$~Duration of each individual exposure. $^{(\dagger)}$~Bisector spans; error bars are twice those of RVs.}
\end{table*}


\section{Priors and posteriors}
\label{appendix:prior}

\begin{table}
\centering
\caption{Prior volume for the parameters of the one-planet model fit of Sect.~\ref{sec:rv} processed with \texttt{juliet}. $\mathcal{U}(a,\:b)$ indicates a uniform distribution between $a$ and $b$; $\mathcal{L}(a,\:b)$ a log-normal distribution, $\mathcal{N}(a,\:b)$ a normal distribution and $\mathcal{T}(a,\:b)$ a truncated normal distribution (where lower possible value equals zero) with mean $a$ and standard deviation~$b$.}\label{tab:prior1p}
\renewcommand{\arraystretch}{1.2}\begin{tabular}{llc}
    \toprule
    \multicolumn{2}{l}{Parameter} &  Prior distribution \\ 
    \hline
    \multicolumn{2}{l}{\footnotesize{Keplerian Parameters:}} & \rule{0pt}{2.6ex} \\ 
    $\rho_{\star}$ & $\mathrm{[kg/m^3]}$   & $\mathcal{N}(1300,\:100)$ \\
    $T_{0,b}$ & $\mathrm{[BJD]}$    & $\mathcal{N}(2458745.921,\:0.003)$ \\
    $P_b$ & $\mathrm{[days]}$   & $\mathcal{N}(12.998,\:0.002)$ \\
    $e_{b}^{*}$ &  & $0$ \\
    $\omega_b^{*}$ &  & $90$ \\
    
    \multicolumn{2}{l}{\footnotesize{Transit Parameters:}} & \rule{0pt}{2.6ex} \\ 
    $R_{\rm p}/R_{\rm \star}$ &     & $\mathcal{U}(0.0,\:1.0)$ \\
    $D$ &    & $1.0$ \\
    $q_1$ &     & $\mathcal{N}(0.31,\:0.30)$ \\
    $q_2$ &     & $\mathcal{N}(0.25,\:0.10)$ \\

    \multicolumn{2}{l}{\footnotesize{Light curve GP Hyperparameters:}} & \rule{0pt}{2.6ex} \\ 
    $\sigma_{\textsf{TESS}}$ & $\mathrm{[ppt]}$  & $\mathcal{L}(10^{-3},\:10)$ \\
    $\rho_{\textsf{TESS}}$ & $\mathrm{[days]}$  & $\mathcal{L}(10^{-1},\:10)$ \\
    
    \multicolumn{2}{l}{\footnotesize{RV parameters:}} & \rule{0pt}{2.6ex} \\
    $K_b$ & $\mathrm{[m/s]}$   & $\mathcal{U}(0.0,\:10.0)$ \\
    $\sigma_{\textsf{HARPS-N}}$ & $\mathrm{[m\,s^{-1}]}$   & $\mathcal{U}(0,\:10)$ \\
    $A$ & $\mathrm{[m\,s^{-1}\,days^{-1}]}$   & $\mathcal{U}(-1,\:1)$ \\
    $B$ & $\mathrm{[m\,s^{-1}]}$   & $\mathcal{U}(-20,\:20)$ \\
    
    \bottomrule
\end{tabular}
\tablefoot{$^{(*)}$~In the case of non-null eccentricity, the priors were set as follows: $(\sqrt{e}\,\sin\omega, \sqrt{e}\,\cos\omega)$ in $\mathcal{U}(-1.0,\:1.0)$.}
\end{table}

\begin{table}
\centering
\caption{Prior volume for the parameters of the two-planet model fit of Sect.~\ref{sec:rv} processed with \texttt{juliet}.}\label{tab:prior2p}
\renewcommand{\arraystretch}{1.2}\begin{tabular}{llc}
    \toprule
    \multicolumn{2}{l}{Parameter} &  Prior distribution \\ 
    \hline
    \multicolumn{2}{l}{\footnotesize{Keplerian Parameters:}} & \rule{0pt}{2.6ex} \\ 
    $\rho_{\star}$ & $\mathrm{[kg/m^3]}$   & $\mathcal{N}(1300,\:100)$ \\
    $T_{0,b}$ & $\mathrm{[BJD]}$    & $\mathcal{N}(2458745.921,\:0.003)$ \\
    $P_b$ & $\mathrm{[days]}$   & $\mathcal{N}(12.998,\:0.002)$ \\
    $T_{0,c}$ & $\mathrm{[BJD]}$    & $\mathcal{N}(2458740,\:2458790)$ \\
    $P_c$ & $\mathrm{[days]}$   & $\mathcal{U}(1,\:100)$ \\
    $(e_b, e_c)^*$ &  & $0$ \\
    $(\omega_b, \omega_c)^*$ &  & $90$ \\
    
    \multicolumn{2}{l}{\footnotesize{Transit Parameters:}} & \rule{0pt}{2.6ex} \\ 
    $R_{\rm p}/R_{\rm \star}$ &     & $\mathcal{U}(0.0,\:1.0)$ \\
    $D$ &    & $1.0$ \\
    $q_1$ &     & $\mathcal{N}(0.31,\:0.30)$ \\
    $q_2$ &     & $\mathcal{N}(0.25,\:0.10)$ \\

    \multicolumn{2}{l}{\footnotesize{Light curve GP Hyperparameters:}} & \rule{0pt}{2.6ex} \\ 
    $\sigma_{\textsf{TESS}}$ & $\mathrm{[ppt]}$  & $\mathcal{L}(10^{-3},\:10)$ \\
    $\rho_{\textsf{TESS}}$ & $\mathrm{[days]}$  & $\mathcal{L}(10^{-1},\:10)$ \\
    
    \multicolumn{2}{l}{\footnotesize{RV parameters:}} \rule{0pt}{2.6ex} \\
    $K_b$ & $\mathrm{[m/s]}$   & $\mathcal{U}(0.0,\:10.0)$ \\
    $K_c$ & $\mathrm{[m/s]}$   & $\mathcal{U}(0,\:10)$ \\
    $\sigma_{\textsf{HARPS-N}}$ & $\mathrm{[m\,s^{-1}]}$   & $\mathcal{U}(0,\:10)$ \\
    $A$ & $\mathrm{[m\,s^{-1}\,days^{-1}]}$   & $\mathcal{U}(-1,\:1)$ \\
    $B$ & $\mathrm{[m\,s^{-1}]}$   & $\mathcal{U}(-20,\:20)$ \\
    
    \bottomrule
\end{tabular}
\tablefoot{$^{(*)}$~In the case of non-null eccentricity, the priors were set as follows: $(\sqrt{e}\,\sin\omega, \sqrt{e}\,\cos\omega)$ in $\mathcal{U}(-1.0,\:1.0)$.}
\end{table}

\begin{table}
\centering
\caption{Posteriors result for the parameters of the two-planet eccentric model fit of Sect.~\ref{sec:rv} processed with \texttt{juliet}.}\label{tab:posterior2p}
\renewcommand{\arraystretch}{1.2}\begin{tabular}{llc}
    \toprule
    \multicolumn{2}{l}{Parameter} &  Value ($\pm 1\sigma$) \\ 
    \hline
    \multicolumn{2}{l}{\footnotesize{Keplerian Parameters:}} & \rule{0pt}{2.6ex} \\ 
    $\rho_{\star}$ & $\mathrm{[kg/m^3]}$   & $1312_{-68}^{+55}$ \\
    $a_b/R_{\star}$ & & $22.72_{-0.40}^{+0.31}$ \\
    $a_c/R_{\star}$ & & $39.05_{-0.73}^{+0.50}$ \\
    $T_{0,b}$ & $\mathrm{[BJD]}$    & $2458745.9205_{-0.0011}^{+0.0012}$ \\
    $P_b$ & $\mathrm{[days]}$   & $12.9972\pm0.0006$ \\
    $T_{0,c}$ & $\mathrm{[BJD]}$    & $2458776.6_{-4.6}^{+4.5}$ \\
    $P_c$ & $\mathrm{[days]}$   & $29.29_{-0.20}^{+0.21}$ \\
    
    \multicolumn{2}{l}{\footnotesize{Transit Parameters:}} & \rule{0pt}{2.6ex} \\ 
    $R_{\rm p_b}/R_{\rm \star}$ &     & $0.0356\pm_{-0.0005}^{+0.0007}$ \\
    $q_1$ &     & $0.28^{+0.11}_{-0.08}$ \\
    $q_2$ &     & $0.30^{+0.05}_{-0.05}$ \\
    $b_b$ & & $0.19_{-0.10}^{+0.11}$ \\
    $i_b$ & [deg] & $89.52^{+0.26}_{-0.28}$ \\ 

    \multicolumn{2}{l}{\footnotesize{Light curve GP Hyperparameters:}} & \rule{0pt}{2.6ex} \\ 
    $\sigma_{\textsf{TESS}}$ & $\mathrm{[ppt]}$  & $0.19_{-0.02}^{+0.03}$ \\
    $\rho_{\textsf{TESS}}$ & $\mathrm{[days]}$  & $0.76_{-0.15}^{+0.19}$ \\
    
    \multicolumn{2}{l}{\footnotesize{RV parameters:}} \rule{0pt}{2.6ex} \\
    $K_b$ & $\mathrm{[m/s]}$   & $2.47_{-0.46}^{+0.50}$ \\
    $K_c$ & $\mathrm{[m/s]}$   & $2.36_{-0.40}^{+0.42}$ \\
    $\sigma_{\textsf{HARPS-N}}$ & $\mathrm{[m\,s^{-1}]}$ & $2.93_{-0.32}^{+0.35}$ \\
    $A$ & $\mathrm{[m\,s^{-1}\,days^{-1}]}$   & $0.0110\pm0.0015$ \\
    $B$ & $\mathrm{[m\,s^{-1}]}$   & $-9.1\pm1.3$ \\
    
    \bottomrule
\end{tabular}
\end{table}

\section{Corner plots}
\label{appendix:posteriors}


\begin{figure*}
    \centering
    \includegraphics[width=\textwidth]{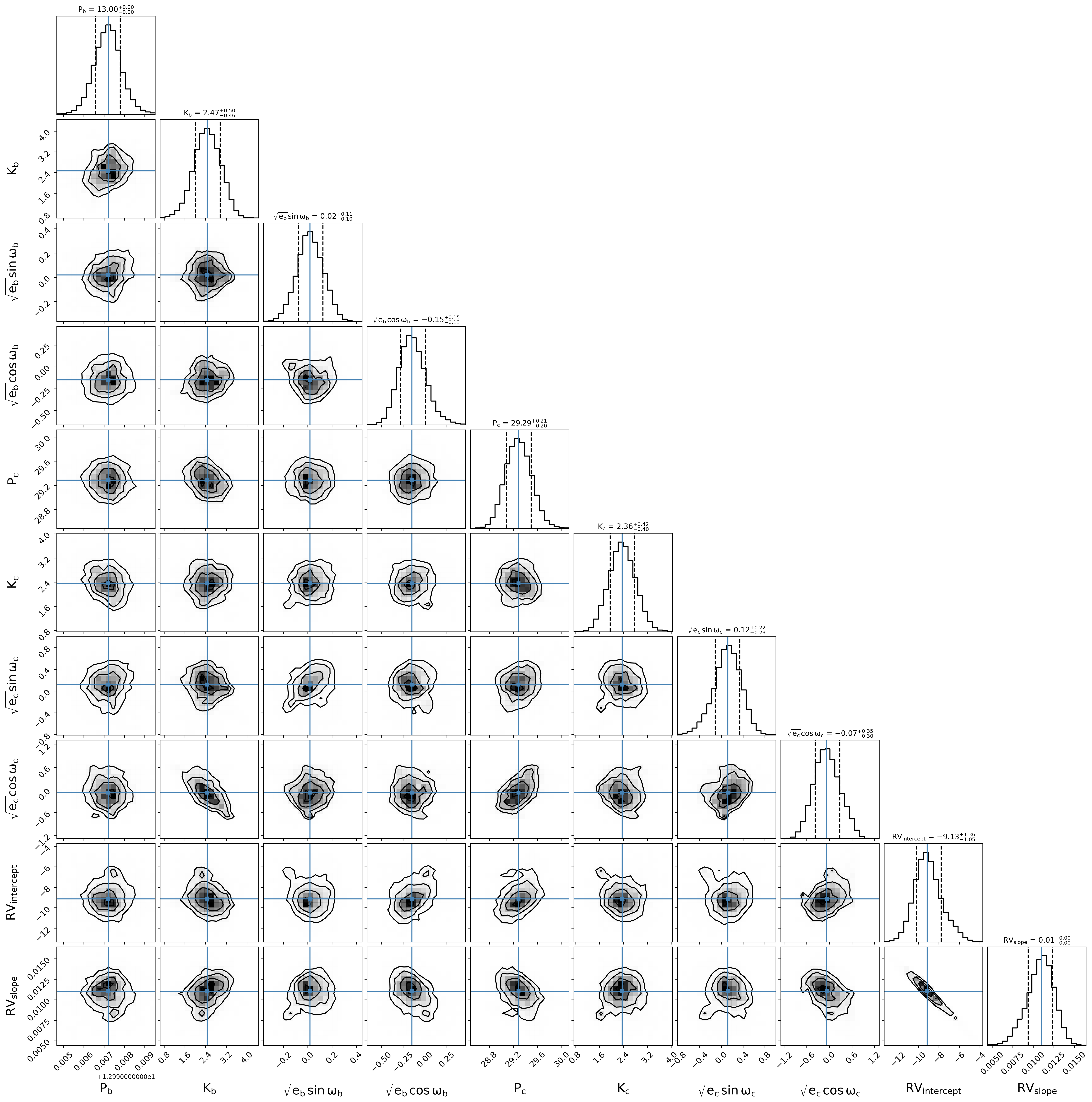}
    \caption{Corner plot for the posterior distribution of the joint transit and RV analysis of Sect.~\ref{sec:joint_analysis} in the case of 2 planets, elaborated with \texttt{juliet}.}
    \label{fig:corner2p}
\end{figure*}

\section{Additional plots}
\label{appendix:addplots}


\begin{figure*}
   \centering
   \includegraphics[width=0.9\textwidth]{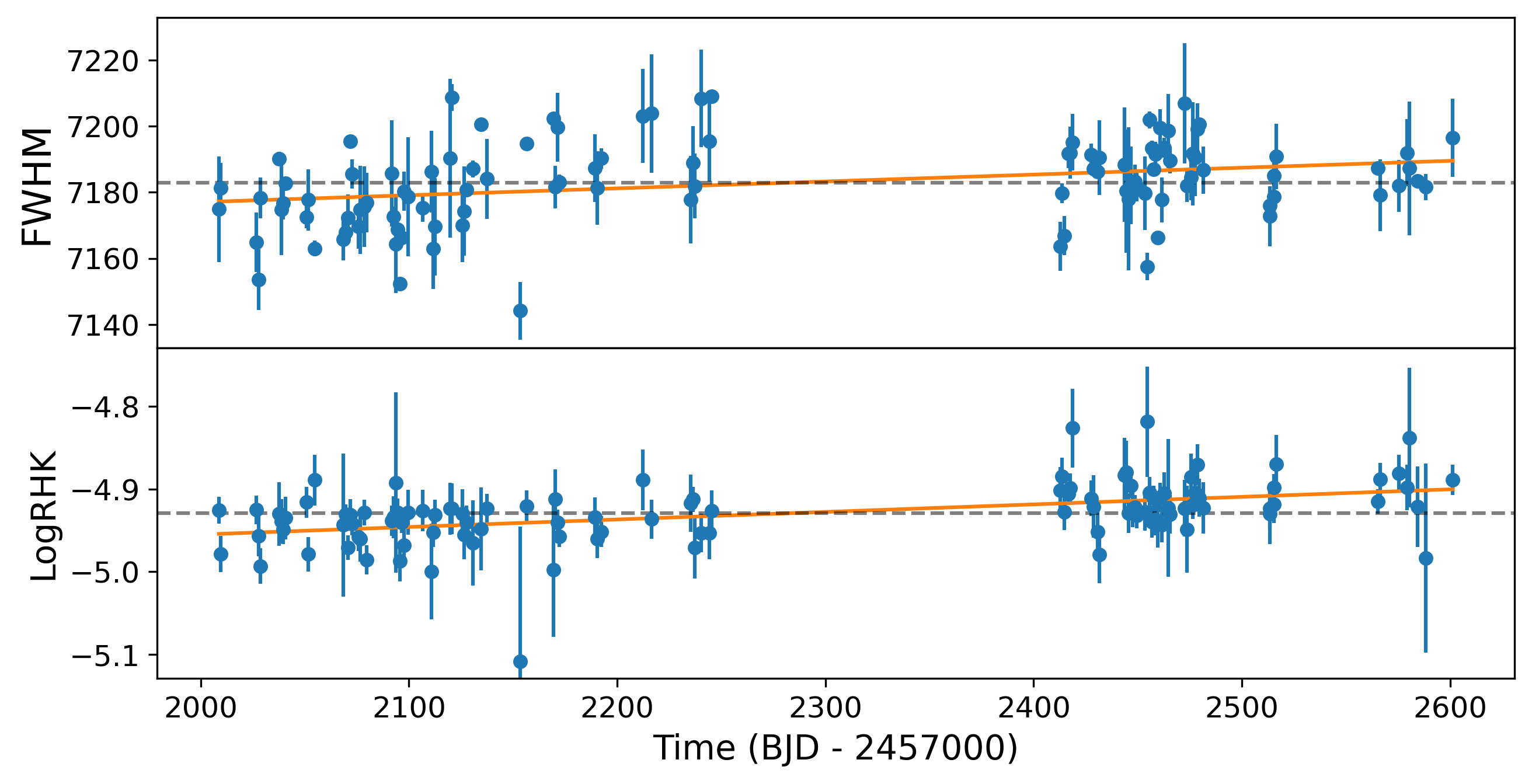}
      \caption{The FWHM and $\log R^{\prime}_{\rm HK}$ are plotted over time respectively in the upper and lower panel, along with their linear trends (orange line) and average value (dashed gray line).}
         \label{fig:fwhmrhk}
\end{figure*}

\begin{figure*}
   \centering
   \includegraphics[width=0.8\textwidth]{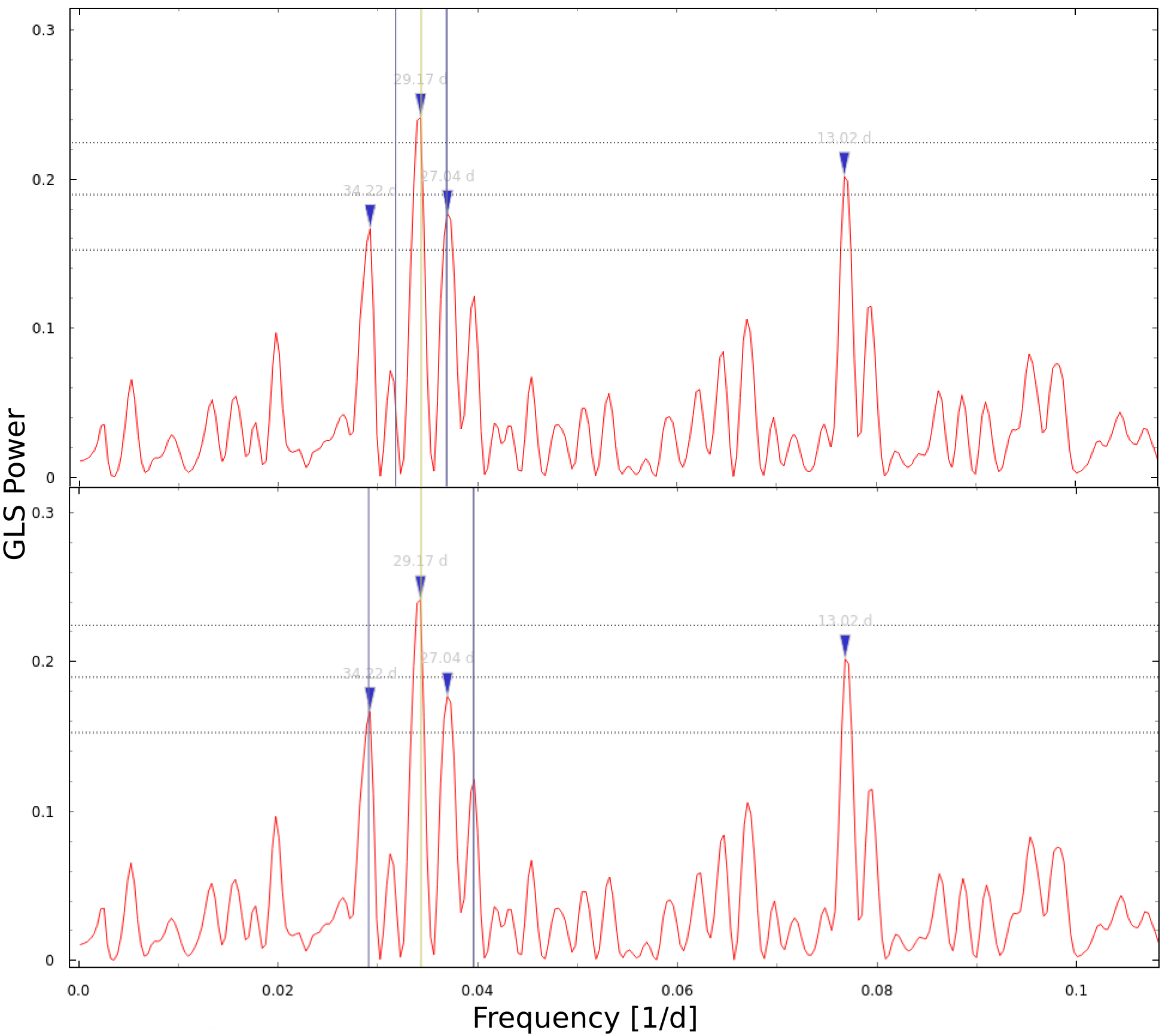}
      \caption{Close-up look of the RV GLS periodogram, executed with the publicly available tool Exo-Striker (\citealp{Trifonov2019}; {\tt https://github.com/3fon3fonov/exostriker}) after the removal of a linear trend. The two vertical blue lines, around the 29-days signal (indicated by a vertical yellow line), show the main peak aliases due to the two highest frequencies of the window function respectively in the upper and bottom panel. The three horizontal dotted lines represent the $10\%$, $1\%$ and $0.1\%$ FAP levels.}
         \label{fig:aliases}
\end{figure*}

\begin{figure*}
   \centering
   \includegraphics[width=0.7\textwidth]{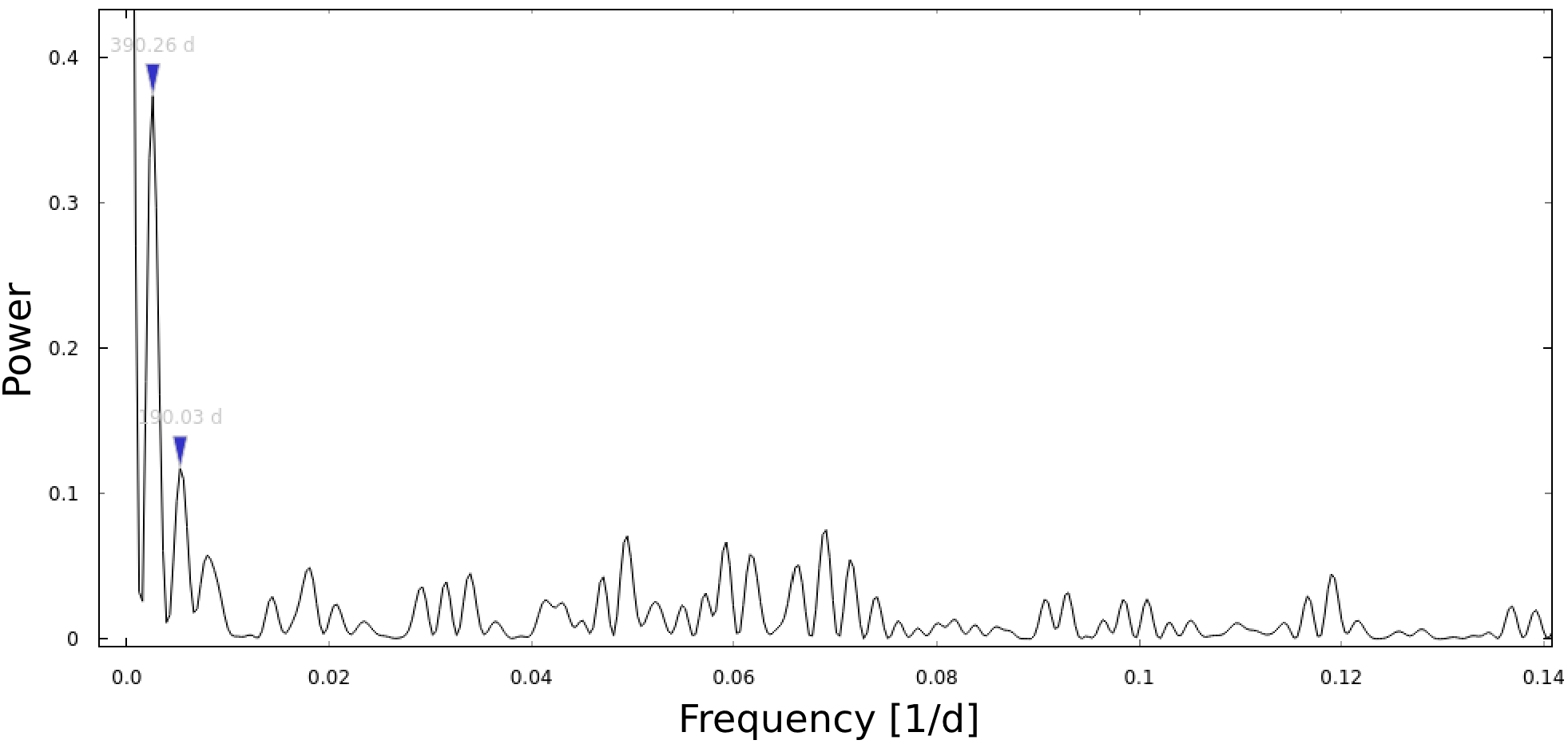}
      \caption{Window function of the HARPS-N RV measurements, as evaluated with Exo-Striker. The two highest peaks, excluding the 1-day peak and frequencies close to zero,  are indicated by the respective labels.}
         \label{fig:window}
\end{figure*}

\begin{figure*}
    \centering
    \includegraphics[width=0.8\textwidth]{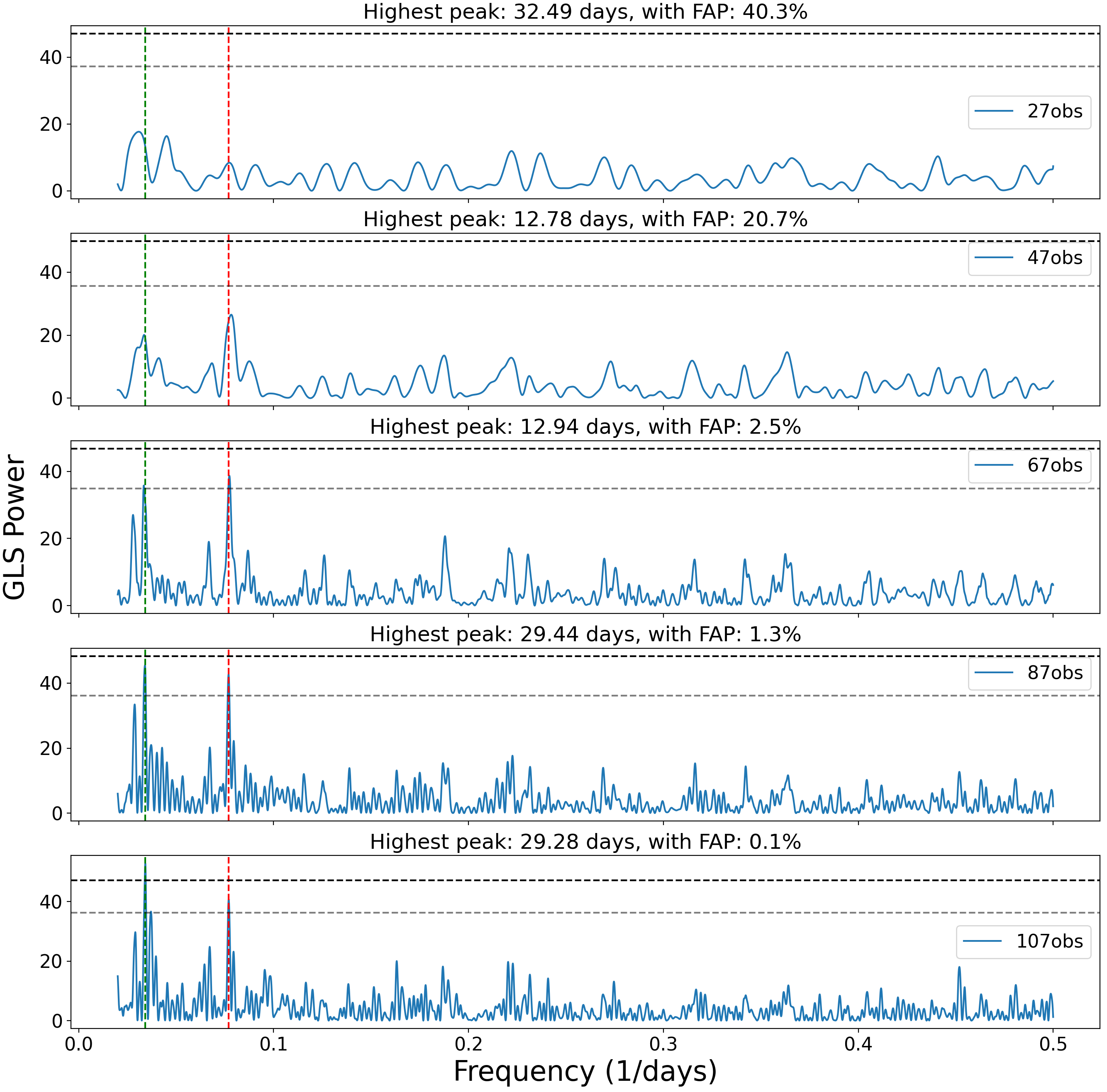}
    \caption{Unnormalized GLS power for different number of HARPS-N observations. The power of the 29-day signal increases with more observations. The red and green dashed vertical lines indicate respectively TOI-1422\,b and TOI-1422\,c orbital periods, while the horizontal dashed lines signal respectively the 10\% and 1\% confidence levels (evaluated with the bootstrap method).}
    \label{fig:power_obs}
\end{figure*}

\begin{figure*}
   \centering
   \includegraphics[width=0.65\textwidth]{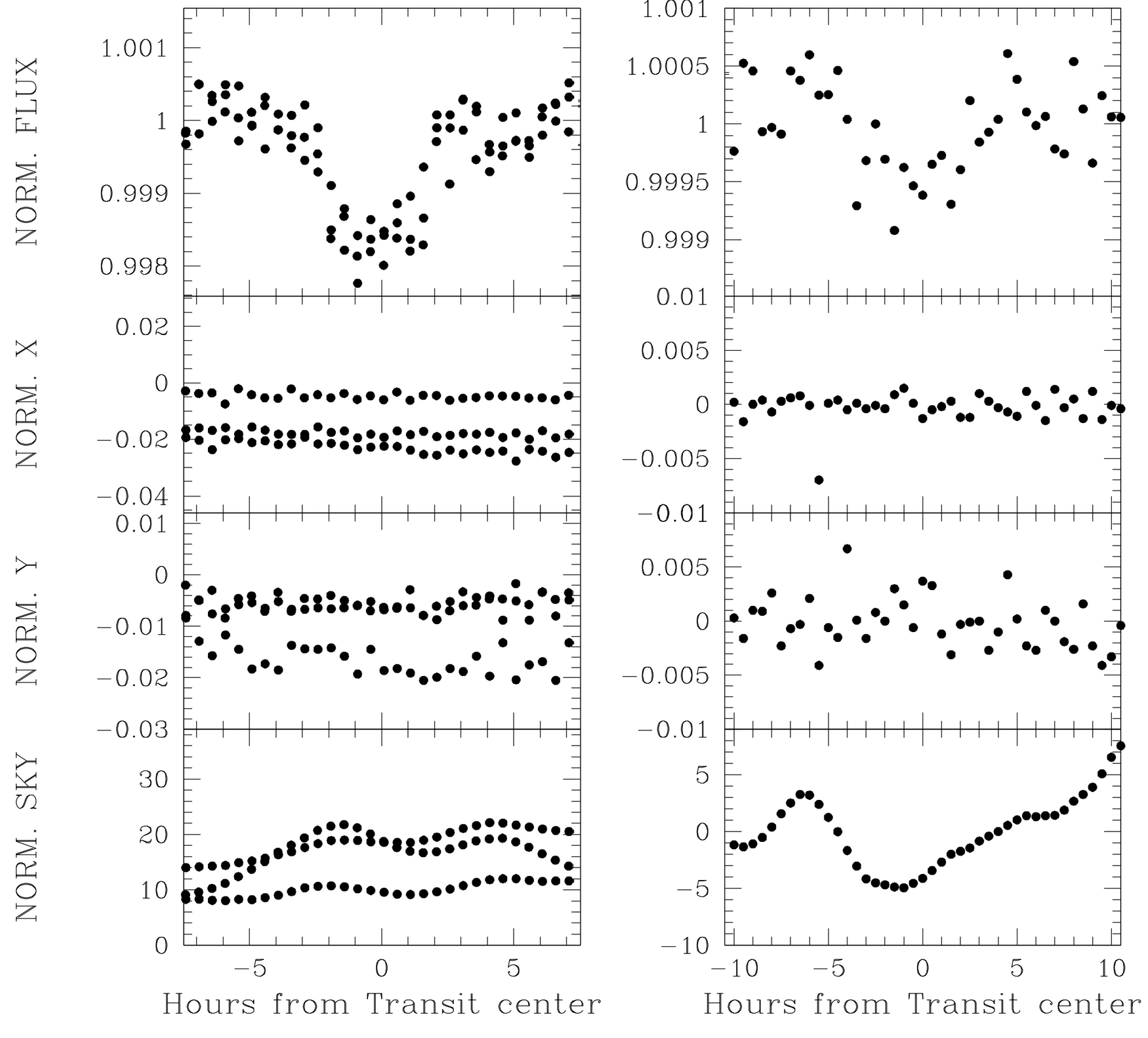}
      \caption{TOI-1422\,b transits, as seen with PATHOS, folded on the first row of the left column and the single transit event on the right one, with X/Y and the sky background on the following rows, showing no correlation with the transits.}
         \label{fig:systematics}
\end{figure*}
\begin{figure*}
   \centering
   \includegraphics[width=0.65\textwidth]{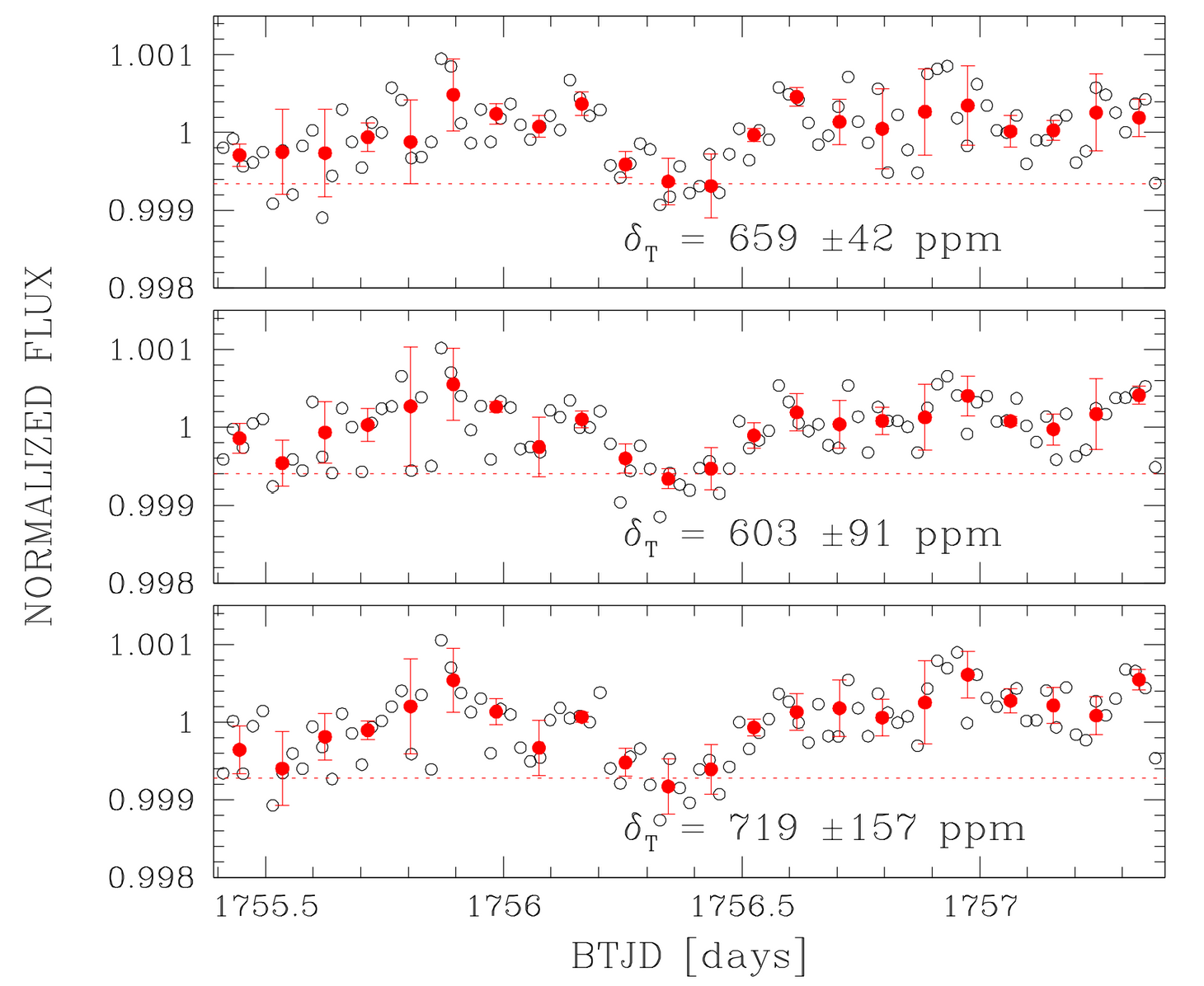}
      \caption{Single transit depth from PATHOS in different apertures, with the three rows showing the transit depth at aperture radius 2, 3 and 4 pixels respectively.}
         \label{fig:singledepth}
\end{figure*}


\end{document}